\documentclass[12pt,thmsa]{article}

\input{tcilatex}
\usepackage{graphicx}
\begin{document}

\setcounter{page}{0} \topmargin 0pt \oddsidemargin 5mm \renewcommand{%
\thefootnote}{\fnsymbol{footnote}} \newpage \setcounter{page}{0} 
\begin{titlepage}
\begin{flushright}
Berlin Sfb288 Preprint  \\
hep-th/9907125\\
\end{flushright}
\vspace{0.5cm}
\begin{center}
{\Large {\bf On the universal Representation of the Scattering Matrix of
Affine Toda Field Theory} }

\vspace{0.8cm}
{\large  A. Fring, C. Korff and B.J. Schulz }

\vspace{0.5cm}
{\em Institut f\"ur Theoretische Physik,
Freie Universit\"at Berlin\\ 
Arnimallee 14, D-14195 Berlin, Germany }
\end{center}
\vspace{0.2cm}
 
\renewcommand{\thefootnote}{\arabic{footnote}}
\setcounter{footnote}{0}

\begin{abstract}
By exploiting the properties of q-deformed Coxeter elements, the scattering matrices of 
affine Toda field theories with real coupling constant related to any dual pair of simple Lie 
algebras may be expressed in a completely generic way. We discuss the governing 
equations for  the existence of bound states, i.e. the  fusing rules, in terms 
of q-deformed Coxeter elements, twisted q-deformed Coxeter elements and 
undeformed Coxeter elements. We establish the precise relation between these different 
formulations and study their solutions. The generalized  S-matrix bootstrap equations are shown to be 
equivalent to the fusing rules. The relation between  different  versions of fusing rules and 
quantum conserved quantities, which result as nullvectors of  a doubly q-deformed Cartan
like matrix, is presented. The properties of this matrix together with the so-called  
combined bootstrap equations are utilised in order to derive generic integral representations 
for the scattering matrix in terms of quantities of either of the two dual algebras. We present 
extensive case-by-case data, in particular on the orbits generated by  the various Coxeter 
elements.
\medskip
\par\noindent
PACS numbers: 11.10Kk, 11.55.Ds
\end{abstract}
\vfill{ \hspace*{-9mm}
\begin{tabular}{l}
\rule{6 cm}{0.05 mm}\\
Fring@physik.fu-berlin.de \\
Korff@physik.fu-berlin.de \\
Schulzb@physik.fu-berlin.de
\end{tabular}}
\end{titlepage}
\newpage


\section{Introduction}

The perturbation of 1+1 dimensional conformal field theories\footnote{%
There exist earlier considerations of field theories in 1+1 dimensions which
focus on the aspect of conformal invariance, e.g. \cite{SCH}. However, the
key feature, i.e. the role played by the Virasoro algebra, which lead to a
more universal formulation and allowed to find their solution was first
realised and exploited in \cite{BPZ}.} \cite{BPZ} in a suitable way leads to
massive quantum field theories which possess a rich underlying structure.
Soon after the seminal paper by Zamolodchikov \cite{Zper} a decade ago on
the perturbation of the Ising model, it was realized \cite{PT} that most of
these massive theories are closely related to affine Toda field theories 
\cite{ATFT}, either in a ``minimal'' sense or with the coupling constant
included. On the base of case-by-case studies for various algebras several
explicit scattering matrices were constructed thereafter \cite{TodaS}. For
the simply laced algebras (ADE) this series of investigations culminated
with the formulation of universal formulae which encompass all these
algebras at once \cite{PD,FO}. The universal nature of these representations
for the scattering matrices allowed also to establish the equivalence
between the bootstrap equations and a classical fusing rule \cite{PD}
formulated with the orbits generated by Coxeter elements of the related
algebra \cite{FLO}. Furthermore the fusing rule is closely linked to the
quantum conservation laws. The origin for the structural interrelation
between the classical and the quantum field theory is the fact that for the
simply laced theories all masses of the theory renormalise with an overall
factor \cite{TodaS}. It is the breakdown of this property for theories
related to a non-simply laced algebra which constituted the main obstacle in
the construction of consistent scattering matrices on the base of the
boostrap principle. Once again numerous candidates were proposed on the base
of case-by-case studies \cite{G2,nons,Donons,Khast}, but it remained a
challenge to find a closed universal representation similar to the simply
laced case for these theories, until Oota recently \cite{Oota} succeeded.

The main conceptual breakthrough towards this goal was the proposal by Dorey 
\cite{PT}, that one may regard these theories in a dual sense,
mathematically in a Lie algebraic way and physically equivalent to this in
the strong-weak duality sense in the coupling constant and the
generalization of the bootstrap principle \cite{nons} by Corrigan, Dorey and
Sasaki. From this point of view affine Toda theories constitute some
concrete simple examples for the Olive-Montonen duality \cite{OM}.
Technically it was also very important to express the scattering matrices in
the adequate building blocks \cite{Donons}. Chari and Pressley \cite{CP}
succeeded thereafter to work out in detail the suggested \cite{PT} fusing
rules in terms of the two dual algebras which reproduced precisely the
allowed fusing processes. Oota \cite{Oota} suggested to re-formulate these
fusing rules in terms of q-deformed Coxeter transformations of either of the
two dual Lie algebras. Viewing matters in the latter fashion allows to link
the fusing rules to the scattering matrices and find closed universal
representations.

One of the purposes of this paper is to precisely establish and derive the
interrelation between the different versions of the fusing rules. We further
demonstrate that these fusing rules are equivalent to the S-matrix bootstrap
equations. Numerous identities which were hitherto only claimed on the base
of case-by-case analysis are rigorously derived. We manifest the relation
between quantum conserved quantities and the various versions of the fusing
rules. We derive a set of equations, which we refer to as combined bootstrap
equations, and exploit them systematically to derive generic integral
representations for the scattering matrix.

Our manuscript is organized as follows: We first develop the mathematics
needed and apply it thereafter in the physical context. In section 2 we
define two different q-deformed Coxeter elements related to two Lie algebras
dual to each other. We derive some of their properties which we need later
on in the physical context. In particular their action in the root space and
inner product relations. In section 3 we formulate several equivalent
versions of the fusing rule, study their different solutions and establish
their relation to quantum conserved quantities. In section 4 we apply our
results to a universal formula for the scattering matrices of affine Toda
field theories in terms of basic building blocks consisting of specific
combinations of hyperbolic functions, whose powers may be obtained from
q-deformed quantities of either of the two dual algebras. An alternative
formula for the scattering matrix in form of an integral representation is
derived in section 5. We exploit the properties of matrices $M$ and $\hat{N}$
related to the untwisted and twisted algebra, respectively, and establish
their equality. In section 6 we reduce the expressions for the scattering
matrix to the simply laced case. In section 7 we provide a case-by-case
analysis for all non-simply laced algebras. Our conclusions are stated in
section 8.

\section{q-deformed Coxeter Elements of dual Pairs}

Adopting the standard notation of \cite{Kac}, we let $X_{n}^{(1)}$ be a
simple simply laced Lie algebra of rank $n$ endowed with a Dynkin diagram
automorphism $\omega $ of order $l$. Employing this automorphism to fix a
subalgebra in $X_{n}^{(1)}$ we obtain the twisted Lie algebra $\hat{X}%
_{n}^{(l)}$ of rank $r$. Changing the orientation of the arrows of the
Dynkin diagram related to this twisted Lie algebra $\hat{X}_{n}^{(l)}$, that
is interchanging long and short roots, produces a Dynkin diagram related to
a Lie algebra $X_{r}^{(1)}$. Two Lie algebras which are related by this map
are referred to as dual pair $(X_{r}^{(1)},\hat{X}_{n}^{(l)})$. Simply laced
Lie algebras are self-dual in this sense.

Before we move on to the q-deformed case we shall collect a few well known
facts in order to define our notations. To each simple root $\alpha _{i}$ of 
$X_{r}^{(1)}$ or $\hat{\alpha}_{i}$ of $\hat{X}_{n}^{(l)}$ a reflection on
the hyperplane through the origin orthogonal to $\alpha _{i}$ or $\hat{\alpha%
}_{i}$ may be associated 
\begin{equation}
\sigma _{i}(x)=x-2\frac{x\cdot \alpha _{i}}{\alpha _{i}^{2}}\alpha
_{i}\,\qquad \text{or\qquad }\hat{\sigma}_{i}(x)=x-2\frac{x\cdot \hat{\alpha}%
_{i}}{\hat{\alpha}_{i}^{2}}\hat{\alpha}_{i}\,.  \label{SWR}
\end{equation}
Note that there is no sum over $i$ implied here on the r.h.s. These are the
Weyl reflections constituting the Weyl group which are used to construct the
so-called Coxeter- and twisted Coxeter element 
\begin{equation}
\sigma =\prod\limits_{i=1}^{r}\sigma _{i}\qquad \text{and\qquad }\hat{\sigma}%
=\prod\limits_{i=1}^{r}\hat{\sigma}_{i}\omega  \label{WR}
\end{equation}
for $X_{r}^{(1)}$ and $\hat{X}_{n}^{(l)}$, respectively. The latter
definition is originally due to Springer \cite{Springer}. We also note here
that these elements are not unique and only defined up to conjugation. There
are several Coxeter numbers (see e.g. \cite{Kac}), whose intimate relations
we wish to exploit. Expressing the highest root of $X_{r}^{(1)}$ as $\psi
=\sum_{i=1}^{r}n_{i}\alpha _{i}$, the corresponding Coxeter- and the dual
Coxeter numbers are defined as 
\begin{equation}
h=1+\sum\limits_{i=1}^{r}n_{i}\qquad \text{and\qquad }h^{\vee
}=1+\sum\limits_{i=1}^{r}n_{i}^{\vee }\,\,.
\end{equation}
The so-called marks $n_{i}$ (or Kac labels) and co-marks $n_{i}^{\vee }$ are
related by $n_{i}^{\vee }=n_{i}\alpha _{i}^{2}/2$. Since dual algebras are
obtained from each other by the interchange of roots and co-roots, i.e. $%
\alpha _{i}\rightarrow 2\alpha _{i}/\alpha _{i}^{2}$, one deduces easily
that 
\begin{equation}
h=\hat{h}^{\vee }\qquad \text{and\qquad }h^{\vee }=\hat{h}\,\,,
\end{equation}
where $\hat{h}^{\vee }$,$\hat{h}$ are the Coxeter numbers of $\hat{X}%
_{n}^{(l)}$. The order of the Coxeter elements read 
\begin{equation}
\sigma ^{h}=1\qquad \text{and\qquad }\hat{\sigma}^{H}=1
\end{equation}
where $H$ is the $l$-th Coxeter number of $\hat{X}_{n}^{(l)}$, i.e. $H=l\hat{%
h}$.

Following now essentially Oota \cite{Oota} the definitions of the Coxeter
elements (\ref{WR}) can be generalized by introducing a q-deformation.

\subsection{q-deformed Coxeter Element of $X_{r}^{(1)}$}

\subsubsection{Definitions}

Using the standard notation $[n]_{q}=(q^{n}-q^{-n})/(q^{1}-q^{-1})$ 
for q-deformed integers, we define the action of the q-deformed Weyl
reflection $\sigma _{i}^{q}$ on a simple root $\alpha _{i}$ as 
\begin{equation}
\sigma _{i}^{q}(\alpha _{j}):=\alpha _{j}-\left( 2\delta _{ij}-\left[
I_{ji}\right] _{q}\right) \alpha _{i}\,\,.  \label{Weyl}
\end{equation}
Here $I$ denotes the incidence matrix, i.e. twice the unit matrix minus the
Cartan matrix $K_{ij}=2\alpha _{i}\cdot \alpha _{j}/\alpha _{j}^{2}$,
related to the simply laced Lie algebra $X_{r}^{(1)}$. We easily verify the
usual properties of a reflection $(\sigma _{i}^{q})^{2}=1$. For the time
being we assume the deformation parameter $q$ to be completely generic, that
is some complex number which is not a root of unity. In some later
applications we will specify $q$ to be a root of unity and also introduce a
particular parameterization $q(\beta ),$ where $\beta $ is a coupling
constant. In that situation the ``classical'' limit $q\rightarrow 1$
corresponds to the vanishing of the coupling constant.

Since in general Weyl reflections do not commute, Coxeter elements, i.e. the
products of all Weyl reflections related to simple roots, only form a
conjugacy class. However, by introducing a particular ordering amongst the
simple roots, one is able to define the Coxeter element uniquely. For this
purpose we partition the set of simple roots, denoted by $\Delta $, into two
disjoint sets of roots, say $\Delta _{\pm }$, by associating the values $%
c_{i}=\pm 1$ to the vertices $i$ of the Dynkin diagram of $X_{r}^{(1)}$, in
such a way that no two vertices related to the same set are linked together.
Then it clearly holds by (\ref{Weyl}) that two reflections related to simple
roots belonging to the same colour set commute, 
\begin{equation}
\left[ \sigma _{i}^{q},\sigma _{j}^{q}\right] =0\qquad \qquad \text{for\quad
\thinspace \thinspace }c_{i}=c_{j}\,\,.  \label{comm}
\end{equation}
Consequently the two special elements 
\begin{equation}
\sigma _{\pm }^{q}:=\prod\limits_{\alpha _{i}\in \Delta _{\pm }}\sigma
_{i}^{q}\,\,,  \label{spm}
\end{equation}
are uniquely defined, having obviously the property $(\sigma _{\pm
}^{q})^{2}=1$. For reasons which become more apparent below, it is
convenient to introduce the simple root times its colour value as a separate
quantity $\gamma _{i}:=c_{i}\alpha _{i}$. Then, the action of the
reflections on these elements is easily worked out. With the help of (\ref
{Weyl}), (\ref{comm}) and (\ref{spm}) we obtain 
\begin{equation}
\sigma _{c_{i}}^{q}(\gamma _{i})=-\gamma _{i}\quad \text{and}\quad \sigma
_{-c_{i}}^{q}(\gamma _{i})=\gamma _{i}-\sum\limits_{\alpha _{j}\in \Delta
_{-c_{i}}}\left[ I_{ij}\right] _{q}\gamma _{j}\,\,.  \label{acti}
\end{equation}
Here we introduced the notation $\sigma _{c_{i}}^{q}$, meaning that it takes
the values $\sigma _{+}^{q}$ or $\sigma _{-}^{q}$ when $c_{i}=1$ or $%
c_{i}=-1 $, respectively. Denoting now by $\alpha _{s}\in \Delta _{s}$ and $%
\alpha _{l}\in \Delta _{l}$ the short and the long roots, respectively, we
define some integers 
\begin{equation}
t_{i}=\left\{ 
\begin{array}{l}
1\qquad \qquad \qquad \,\,\text{for\thinspace \thinspace \thinspace }\alpha
_{i}\in \Delta _{s} \\ 
\alpha _{l}^{2}/\alpha _{s}^{2}\qquad \qquad \text{for\thinspace \thinspace
\thinspace }\alpha _{i}\in \Delta _{l}
\end{array}
\right.  \label{symm}
\end{equation}
which symmetrize the incidence matrix 
\begin{equation}
I_{ij}t_{j}=I_{ji}t_{i}\,\,.  \label{deft}
\end{equation}
The ratio $\alpha _{l}^{2}/\alpha _{s}^{2}$ is indeed an integer, which
follows directly from the definition of the Cartan matrix. In fact it equals 
$l$ ($1,2$ or $3$), the highest order of the Dynkin diagram automorphism of
the algebra $X_{n}^{(1)}$. The occurrence of quantities of $X_{n}^{(1)}$,
despite the fact that we are discussing $X_{r}^{(1)}$, is a feature we will
encounter more frequently in the course of our discussion and indicates the
close interrelation between the two dual algebras. We employ the
symmetrizers (\ref{symm}) to introduce the map 
\begin{equation}
\tau (\gamma _{i}):=q^{t_{i}}\gamma _{i}\,\,.  \label{tau}
\end{equation}
We have now assembled all the ingredients in order to define the q-deformed
Coxeter element 
\begin{equation}
\sigma _{q}:=\sigma _{-}^{q}\,\tau \,\sigma _{+}^{q}\,\tau \,\,.
\label{qCox}
\end{equation}
Having eliminated the ambiguity in the ordering of the q-deformed Weyl
reflections within $\,\sigma _{\pm }^{q}$, the only matter left to
convention with regard to the q-deformed Coxeter element is the ordering of
the four maps in (\ref{qCox}) and the two possible choices for the colour
values we attribute to the vertices of the Dynkin diagram. The former
ambiguity is fixed by the choice in (\ref{qCox}) and the latter by choosing
the unique vertex of the short root which is connected to a long root as $%
c_{i}=-1$. Note also that $\lim_{q\rightarrow 1}\sigma _{q}=\sigma $, that
is in the ``classical'' limit we recover the usual Coxeter element (\ref{WR}%
) from the q-deformed Coxeter element (\ref{qCox}).

\subsubsection{Action of $\sigma _{q}$ in the Root Space}

There are several properties of the q-deformed Coxeter element which we wish
to exploit in the context of the scattering matrix of affine Toda field
theories. First we state the identities 
\begin{eqnarray}
\sigma _{q^{-1}}^{x} &=&\tau ^{\frac{1+c_{i}}{2}}\sigma _{c_{i}}^{q}\tau ^{%
\frac{1+c_{i}}{2}}\,\,\sigma _{q}^{-x}\,\,\tau ^{-\frac{1+c_{i}}{2}}\sigma
_{c_{i}}^{q}\tau ^{-\frac{1+c_{i}}{2}}\,\,  \label{q-q} \\
&=&\tau ^{\frac{1-c_{j}}{2}}\sigma _{-c_{j}}^{q}\tau ^{\frac{1-c_{j}}{2}%
}\,\,\sigma _{q}^{-x+\frac{c_{i}+c_{j}}{2}}\,\,\tau ^{-\frac{1+c_{i}}{2}%
}\sigma _{c_{i}}^{q}\tau ^{-\frac{1+c_{i}}{2}}\,\,,  \label{q-q2}
\end{eqnarray}
which follow immediately by noting that under the interchange of $q$ and $%
q^{-1}$ the elements $\sigma _{\pm }^{q}$ remain invariant and $\tau
\rightarrow \tau ^{-1}$. In fact the r.h.s. of these equations correspond to
several equations which are combined to one by including the colour values $%
c_{i}$ and $c_{j}$ in the way we need them. Obviously, $(\sigma
_{q})^{-1}=\tau ^{-1}\sigma _{+}^{q}\,\tau \,^{-1}\sigma _{-}^{q}$ is the
inverse q-deformed Coxeter element\footnote{%
We differ here from the definition of the inverse in \cite{Oota}.}.

We further need to know the action of $\sigma _{q}$ on the simple roots.
From (\ref{acti}), (\ref{tau}) and (\ref{qCox}) we obtain 
\begin{equation}
\sigma _{q}(\alpha _{i})+q^{2t_{i}}\alpha _{i}=\!\!\!\!\!\sum\limits_{\alpha
_{j}\in \Delta _{-c_{i}}}\!\!\!\!\!q^{\frac{3+c_{i}}{2}t_{i}+\frac{1+c_{j}}{2%
}t_{j}}\left[ I_{ij}\right] _{q}\gamma _{j}+\frac{c_{i}-1}{2}\sum\limits\Sb %
\alpha _{j}\in \Delta _{+}  \\ \alpha _{l}\in \Delta _{-}  \endSb %
\!\!\!q^{t_{i}+t_{j}}\left[ I_{ij}\right] _{q}\left[ I_{jl}\right]
_{q}\gamma _{l}  \label{sigal}
\end{equation}
and also the crucial identity 
\begin{equation}
\left( q^{-c_{i}t_{i}}(\sigma _{q})^{c_{i}}+q^{c_{i}t_{i}}\right) (\gamma
_{i})=\sum\limits_{\alpha _{j}\in \Delta _{-c_{i}}}q^{\frac{1+c_{i}}{2}t_{i}-%
\frac{1+c_{j}}{2}t_{j}}\left[ I_{ij}\right] _{q}\gamma _{j}\,\,.
\label{bootp}
\end{equation}
Acting now successively with $\sigma _{q}$ on $\gamma _{i}$ and the
multiplication with powers of $q$ will create an orbit which we denote by $%
\Omega _{i}^{q}$, i.e. for $x,y$ being arbitrary integers a typical element
in $\Omega _{i}^{q}$ reads $q^{x}\sigma _{q}^{y}(\gamma _{i})$. The
periodicity of these orbits reads 
\begin{equation}
q^{-2H}\left( \sigma _{q}\right) ^{h}=1\,\,.  \label{quasi}
\end{equation}
We do not have a general proof of (\ref{quasi}), but it is confirmed on the
base of a case-by-case analysis in section 7 as may be seen from the data
presented. To each orbit $\Omega _{i}^{q}$ we associate a particle species.
The anti-particle is identified with the orbit in which we find the element 
\begin{equation}
-q^{-H+\frac{c_{\bar{\imath}}-c_{i}}{2}t_{i}}\,\sigma _{q}^{\frac{h}{2}+%
\frac{c_{i}-c_{\bar{\imath}}}{4}}\gamma _{i}\,\,=\gamma _{\bar{\imath}}\in
\Omega _{\bar{\imath}}^{q}\,\,.  \label{anti}
\end{equation}
The property $c_{i}c_{\bar{\imath}}=(-1)^{h}$ \cite{FO} ensures that the
power of the Coxeter element is always an integer. Conjugating $\bar{\imath}$
once more in (\ref{anti}) leads to (\ref{quasi}), when $t_{i}=t_{\bar{\imath}%
}$. For the non-simply laced algebras, the relation (\ref{anti}) reduces to 
\begin{equation}
\,\sigma _{q}^{\frac{h}{2}}\gamma _{i}\,\,=-q^{H}\gamma _{\bar{\imath}}\,\,,
\label{self}
\end{equation}
since in that case all particles are self-conjugate. The motivation of this
definition is analogue to the one known from the simply laced case \cite{FLO}%
. This means complex conjugating the field which creates the particle of
type $i$ in the classical theory corresponds to the creation of the
anti-particle $\bar{\imath}$, suggesting to associate $-\gamma _{i}$ to the
anti-particle. However, one should keep in mind that in this context the
classical theory is only known in the extreme weak or extreme strong limit
of the coupling constant. In the classical limit we recover the known
identity \cite{FO} for the simply laced case $\sigma ^{\frac{h}{2}+\frac{%
c_{i}-c_{\bar{\imath}}}{4}}\gamma _{i}\,\,=\gamma _{\bar{\imath}}$ which
relates particles and anti-particles.

\subsubsection{Inner Product Identities}

We introduce now the co-fundamental weights $\,\tilde{\lambda}_{i}$, related
to the fundamental weights $\lambda _{i}$ as $\,\tilde{\lambda}%
_{i}:=\,2\lambda _{i}/\alpha _{i}^{2}$, such that they constitute a dual
base to the simple roots, i.e. $\,\tilde{\lambda}_{i}\cdot \alpha
_{j}=\delta _{ij}$. In comparison to the non-deformed (simply laced) case,
it is important to note that $\sigma _{q}$ does in general not preserve the
inner product, i.e. $\tilde{\lambda}_{j}\cdot (\sigma _{q})^{x}\gamma
_{i}\neq (\sigma _{q})^{-x}\tilde{\lambda}_{j}\cdot \gamma _{i}$.

In view of (\ref{acti}), (\ref{tau}) and the orthogonality of roots and
co-fundamental weights we can write 
\begin{equation}
\tilde{\lambda}_{j}\cdot \sigma _{q}^{x}\gamma _{i}=\tilde{\lambda}_{j}\cdot
\sigma _{-c_{j}}^{q}\sigma _{q}^{x}\gamma _{i}=q^{-t_{j}}\,\tilde{\lambda}%
_{j}\cdot \tau \sigma _{q}^{x}\gamma _{i}\,\,.  \label{inner}
\end{equation}
Using now (\ref{acti}), (\ref{q-q2}) and exploiting (\ref{inner}) we derive 
\begin{eqnarray}
\tilde{\lambda}_{j}\cdot \sigma _{q}^{x}\gamma _{i} &=&\tilde{\lambda}%
_{j}\cdot \tau ^{\frac{c_{j}-1}{2}}\sigma _{-c_{j}}^{q}\tau ^{\frac{c_{j}-1}{%
2}}\,\,\sigma _{q^{-1}}^{-x+\frac{c_{i}+c_{j}}{2}}\,\,\tau ^{\frac{1+c_{i}}{2%
}}\sigma _{c_{i}}^{q}\tau ^{\frac{1+c_{i}}{2}}\gamma _{i} \\
&=&-q^{(c_{j}-1)t_{j}+(1+c_{i})t_{i}}\,\,\tilde{\lambda}_{j}\cdot \sigma
_{q^{-1}}^{-x+\frac{c_{i}+c_{j}}{2}}\,\,\gamma _{i}\,\,,
\end{eqnarray}
which may also be re-written as

\begin{equation}
q^{\frac{(1-c_{j})t_{j}-(1+c_{i})t_{i}}{2}}\,\tilde{\lambda}_{j}\cdot \sigma
_{q}^{x}\gamma _{i}+\,\,q^{2H+\frac{(c_{j}-1)t_{j}+(1+c_{i})t_{i}}{2}}\,%
\tilde{\lambda}_{j}\cdot \sigma _{q^{-1}}^{h-x+(c_{i}+c_{j})/2}\gamma
_{i}=0\,\,,  \label{kro}
\end{equation}
with the help of (\ref{quasi}).

As the last inner product identity we show 
\begin{equation}
(q^{2t_{j}}-1)(\,\tilde{\lambda}_{j}\cdot (\sigma _{q})^{x}\gamma
_{i})=(q^{2t_{i}}-1)(\,\tilde{\lambda}_{i}\cdot (\sigma _{q})^{x}\gamma
_{j})\,\,.  \label{Pari}
\end{equation}
We prove (\ref{Pari}) by induction and demonstrate therefore first that it
holds for $x=1$. With the help of (\ref{sigal}) we obtain 
\begin{eqnarray}
(q^{2t_{j}}-1)\sigma _{q}(\alpha _{i})\cdot \,\tilde{\lambda}_{j}
&=&(q^{2t_{j}}-1)\left( -\frac{1-c_{i}}{2}\sum\limits_{p\in \Delta
_{-c_{i}}}q^{t_{i}+t_{p}}\left[ I_{ip}\right] _{q}\left[ I_{pj}\right]
_{q}\right.  \nonumber \\
&&\left. -q^{2t_{i}}\delta _{ij}+c_{j}q^{\frac{3+c_{i}}{2}t_{i}+\frac{1+c_{j}%
}{2}t_{j}}\left[ I_{ij}\right] _{q}\delta _{c_{i},-c_{j}}\right) \,\,.
\label{iddd}
\end{eqnarray}
Noting that (\ref{deft}) also holds for the q-deformed quantities, i.e. $%
[I_{ij}]_{q}[t_{j}]_{q}=[I_{ji}]_{q}[t_{i}]_{q}$, it is easy to verify that
the r.h.s. of equation (\ref{iddd}) is symmetric in $i$ and $j$. Assuming
now relation (\ref{Pari}) to be valid for $x$, one deduces by the similar
reasoning as for the case $x=1$, that (\ref{Pari}) also holds for $x+1$ and
therefore for all integers $x$. This establishes (\ref{Pari}).

It should be mentioned that once we have the matrix representation of
section 5.1 the symmetry property (\ref{Pari}) follows more easily.

\subsection{q-deformed twisted Coxeter Element of $\hat{X}_{n}^{(l)}$}

\subsubsection{Definitions}

Let us now consider a Lie algebra $X_{n}^{(1)}$, whose associated Dynkin
diagram is endowed with an automorphism $\omega $ which acts on a simple
root $\alpha _{i}$ with length $l_{i}$, i.e. $\omega ^{l_{i}}\alpha
_{i}=\alpha _{i}$. The largest value of $l_{i}$ corresponds to $l$.
Sometimes we will also use the common notation $\omega \alpha _{i}=\alpha
_{\omega (i)}$. We may employ this automorphism to define the orbits $\Omega
_{i}^{\omega }$ by successive actions of $\omega $ on a simple root $\alpha
_{i}$. By selecting a representative of the orbit $\Omega _{i}^{\omega }$,
we can build up a set of roots, which we denote by $\hat{\alpha}_{i}\in \hat{%
\Delta}$. The algebra related to these roots is the twisted Lie algebra $%
\hat{X}_{n}^{(l)}$. To each of the $r$ elements in $\hat{\Delta}$ we
associate a particular particle species. We choose the conventions in such a
way that we may carry out a one-to-one correspondence between the two dual
algebras without renaming the particles, see section 7. The Weyl reflections
related to these representatives are now defined in the usual fashion as in (%
\ref{SWR}) 
\begin{equation}
\sigma _{i}(\alpha _{j})=\alpha _{j}-K_{ji}\alpha _{i}\,\,\,,  \label{Weylr}
\end{equation}
where $K$ denotes the Cartan matrix of $X_{n}^{(1)}.$ Analogously to the
non-twisted case, treated in the previous section, we can bi-colour the
Dynkin diagram related to $X_{n}^{(1)}$ and divide the set of
representatives into two sets $\hat{\Delta}_{-}$ and $\hat{\Delta}_{+}$.
Note that roots related by the automorphism $\omega $ possess naturally the
same colour value. Hence we may define uniquely the elements 
\begin{equation}
\hat{\sigma}_{\pm }:=\prod\limits_{\hat{\alpha}_{i}\in \hat{\Delta}_{\pm
}}\sigma _{i}\,\,\,.
\end{equation}
Besides the absence of the q-deformation, the difference between these
special elements of the Weyl group in comparison with the non-twisted case
is that the product runs only over the representatives. We define now the
integers 
\begin{equation}
\hat{t}_{i}=\left\{ 
\begin{array}{l}
1\qquad \qquad \text{for\thinspace \thinspace \thinspace }\alpha _{i}\in 
\hat{\Delta} \\ 
0\qquad \qquad \text{for\thinspace \thinspace \thinspace }\alpha _{i}\notin 
\hat{\Delta}
\end{array}
\right. \,\,.
\end{equation}
With the help of (\ref{Weylr}) we easily compute the action of $\hat{\sigma}%
_{\pm }$ on some $\gamma _{i}:=c_{i}\alpha _{i}$, where we stress that $%
\alpha _{i}$ is not necessarily a representative 
\begin{equation}
\hat{\sigma}_{c_{i}}\gamma _{i}=(-1)^{\hat{t}_{i}}\gamma _{i}\qquad \text{%
and\qquad }\hat{\sigma}_{-c_{i}}\gamma _{i}=\gamma _{i}-\sum\limits_{\hat{%
\alpha}_{j}\in \hat{\Delta}_{-c_{i}}}I_{ij}\hat{\gamma}_{j}\,\,\,.
\label{sigg}
\end{equation}
The incidence matrix $I$ is here related to $X_{n}^{(1)}$, but note that $%
1\leq i\leq n$ and $1\leq j\leq r$. In addition we introduce the map which
will serve as a q-deformation 
\begin{equation}
\hat{\tau}(\alpha _{i}):=q^{2\hat{t}_{i}}\alpha _{i}\,\,\,.  \label{thh}
\end{equation}
At last we are in the position to define, analogously to \cite{Oota}, the
q-deformed twisted Coxeter element as 
\begin{equation}
\hat{\sigma}_{q}:=\omega ^{-1}\,\hat{\sigma}_{-}\,\hat{\tau}\,\hat{\sigma}%
_{+}\,\,\,.  \label{sigt}
\end{equation}
Once again by means of the bi-colouration, we have achieved that $\hat{\sigma%
}_{q}$ is uniquely defined up to the ordering of the maps occurring in (\ref
{sigt}). For $q\rightarrow 1$ we obtain one of the standard twisted Coxeter
elements in the conjugacy class as originally introduced by Springer (\ref
{WR}). We will not elaborate here on the alternative characterization of the
twisted Coxeter element, which may be obtained from the folding of an affine
simply laced Dynkin diagram, see e.g. \cite{Kac,TO,KO}.

\subsubsection{Action of $\hat{\sigma}_{q}$ in the Root Space}

Introducing for convenience the quantities $\gamma _{i}^{\pm }:=\omega ^{%
\frac{\pm c_{i}-1}{2}}\gamma _{i}$, the action of $\hat{\sigma}_{q}$ on the
simple roots is computed to 
\begin{equation}
\hat{\sigma}_{q}(\hat{\gamma}_{i})+q^{2}\omega ^{-1}\hat{\gamma}%
_{i}=-\!\!\!\!\!\sum\limits_{\hat{\alpha}_{j}\in \hat{\Delta}%
_{-c_{i}}}\!\!\!\!\!q^{2}I_{ij}\omega ^{-1}\hat{\alpha}_{j}+\frac{1-c_{i}}{2}%
q^{2}\!\!\!\!\sum\limits\Sb \alpha _{j}\in \hat{\Delta}_{+}  \\ \alpha
_{l}\in \hat{\Delta}_{-}  \endSb \!\!\!I_{ij}I_{jl}\omega ^{-1}\hat{\gamma}%
_{l}
\end{equation}
and 
\begin{equation}
\gamma _{i}^{-}=(-q^{-2c_{i}})^{\hat{t}_{i}}\hat{\sigma}_{q}^{c_{i}}\left(
\gamma _{i}^{+}\right) +\sum\limits_{\alpha _{j}\in \Delta
_{-c_{i}}}\!\!\!\!\!I_{ij}\hat{\gamma}_{j}^{+}\,\,\,,  \label{tw}
\end{equation}
with the help of (\ref{sigg}), (\ref{thh}) and (\ref{sigt}).

Acting successively with $\hat{\sigma}_{q}$ and $q$ on the elements of $\hat{%
\Delta}$, we construct the orbits of the q-deformed twisted Coxeter element,
which we denote by $\hat{\Omega}_{i}^{q}$. The order of the q-deformed
twisted Coxeter element reads 
\begin{equation}
q^{-2h}\hat{\sigma}_{q}^{H}=1\,\,.  \label{quasi2}
\end{equation}
Thus in comparison with (\ref{quasi}) the roles of $h$ and $H$ are just
interchanged. Like in the non-twisted case we do not have a generic proof of
this periodicity property, but we have verified it case-by-case in section 7.

The anti-particle is identified with the orbit in which we find the element 
\begin{equation}
-q^{-h+\frac{c_{\bar{\imath}}-c_{i}}{2}\hat{t}_{i}}\,\hat{\sigma}_{q}^{\frac{%
H}{2}+\frac{c_{i}-c_{\bar{\imath}}}{4}(2-l_{i})}\hat{\gamma}_{i}^{+}\,\,=%
\hat{\gamma}_{\bar{\imath}}^{+}\in \hat{\Omega}_{\bar{\imath}}^{q}\,\,.
\label{antit}
\end{equation}
Conjugating $\bar{\imath}$ once more in (\ref{antit}) leads to (\ref{quasi2}%
), when $l_{i}=l_{\bar{\imath}}$. For the non-simply laced algebras, the
relation (\ref{anti}) reduces to 
\begin{equation}
\,\hat{\sigma}_{q}^{\frac{H}{2}}\hat{\gamma}_{i}^{+}\,\,=-q^{h}\hat{\gamma}_{%
\bar{\imath}}^{+}\,\,,
\end{equation}
since in that case all particles are self-conjugate. In the limit $%
q\rightarrow 1$ we obtain $\hat{\sigma}^{\frac{H}{2}+\frac{c_{i}-c_{\bar{%
\imath}}}{4}(2-l_{i})}\hat{\gamma}_{i}^{+}\,\,=\hat{\gamma}_{\bar{\imath}%
}^{+}$, which relates particles and anti-particles in twisted algebras.

\subsubsection{Inner Product Identities}

To each orbit $\hat{\Omega}_{i}^{q}$ we associate now a fundamental weight $%
\,\hat{\lambda}_{i}$ which is dual to all elements inside the $\omega $%
-orbit, i.e. 
\begin{equation}
\hat{\lambda}_{i}\cdot \sum\limits_{k=1}^{l_{j}}\omega ^{k}(\alpha
_{j})=\delta _{ij}\,\,\,,  \label{ort}
\end{equation}
for $\alpha _{i}$ being a root of $X_{n}^{(1)}$. With the help of (\ref{thh}%
), (\ref{sigg}) and the orthogonality relation (\ref{ort}) we derive easily 
\begin{equation}
\hat{\lambda}_{j}\cdot \hat{\sigma}_{q}^{x}\hat{\gamma}_{i}=\hat{\lambda}%
_{j}\cdot \hat{\sigma}_{-c_{j}}^{q}\hat{\sigma}_{q}^{x}\hat{\gamma}_{i}=q^{-2%
\hat{t}_{j}}\,\hat{\lambda}_{j}\cdot \hat{\tau}\hat{\sigma}_{q}^{x}\gamma
_{i}\,\,.  \label{innert}
\end{equation}
We also have the identities 
\begin{equation}
\hat{\lambda}_{j}\cdot \hat{\sigma}_{q}^{x}\hat{\gamma}_{i}^{+}=\hat{\lambda}%
_{i}\cdot \hat{\sigma}_{q}^{x+\frac{c_{j}-c_{i}}{2}+\frac{c_{i}-1}{2}l_{i}+%
\frac{1-c_{j}}{2}l_{j}}\hat{\gamma}_{j}^{+}  \label{paritt}
\end{equation}
and 
\begin{equation}
\hat{\lambda}_{j}\cdot \hat{\sigma}_{q}^{x}\hat{\gamma}%
_{i}^{+}=-q^{2h+c_{i}+c_{j}}\,\,\hat{\lambda}_{j}\cdot \hat{\sigma}%
_{q^{-1}}^{H-x+c_{i}+\frac{1-c_{i}}{2}l_{i}+\frac{c_{j}-1}{2}l_{j}}\hat{%
\gamma}_{i}^{+}  \label{mm}
\end{equation}
To prove these identities directly is much more involved as for the
equivalent relations in the untwisted case. We will therefore postpone the
proof until section 5.2. where we can exploit properties of a different
quantity which then implies the validity of (\ref{paritt}) and (\ref{mm}).

\section{The Fusing Rules}

We are now in the position to formulate the universal fusing rules. This may
be done either by exploiting the properties of the orbits of the q-deformed
Coxeter Element of $X_{r}^{(1)}$ or the q-deformed twisted Coxeter Element
of $\hat{X}_{n}^{(l)}$ similar to the approach of Oota \cite{Oota} or
alternatively in the spirit of Chari and Pressley \cite{CP} one may consider
the orbits of the non-deformed Coxeter Element of $X_{r}^{(1)}$and
simultaneously the non-deformed twisted Coxeter Element of $\hat{X}%
_{n}^{(l)} $. Additionally one may formulate the fusing rule in terms of the
quantum conserved quantities. We will discuss the solutions to these
different fusing rules and prove in general that they are in fact all
equivalent. We derive the precise quantitative relation between the relevant
quantities.

\subsection{The Fusing Rule in $\Omega ^{q}$}

\emph{The generalized\footnote{%
Usually we really refer to the three-point-coupling in the common sense,
i.e. related to the process $i+j\rightarrow \bar{k}$. The only exceptions
are the processes $2+2\rightarrow 2$ and $3+3\rightarrow 3$ in $%
(F_{4}^{(1)},E_{6}^{(2)})$, which are possible from the fusing rule point of
view. However, on the S-matrix bootstrap side these processes correspond to
third order poles. } three-point-coupling related to three particles of the
type }$i,j$\emph{\ and }$k$\emph{\ is non-vanishing, i.e. the process }$%
i+j\rightarrow \bar{k}$\emph{\ is possible, if and only if there exist
representatives of the q-deformed orbits }$\Omega _{i}^{q},$\emph{\ }$\Omega
_{j}^{q}$\emph{\ and }$\Omega _{k}^{q}$\emph{\ whose sum is zero. }

This means there should exist two triplets of integers $(\xi _{i},\xi
_{j},\xi _{k})$ and $(\zeta _{i},\zeta _{j},\zeta _{k})$ such that 
\begin{equation}
\sum\limits_{l=i,j,k}q^{\zeta _{l}}\,\sigma _{q}^{\xi _{l}}\,\gamma
_{l}=0\,\,.  \label{FU}
\end{equation}
Multiplying (\ref{FU}) by $q^{n}$ or $\sigma _{q}^{m}$ corresponds naturally
to the same process and we should therefore view the triplets as equivalence
classes\footnote{%
We shall see below that from a physical point of view this corresponds to a
simple shift in the bootstrap functional equations which involve the
scattering matrix.}. In this sense we regard two pairs of triplets as
equivalent if they may be constructed from each other by the displacements $%
\zeta _{l}\rightarrow \zeta _{l}+m$ or $\xi _{l}\rightarrow \xi _{l}+n$.
Similarly as in the simply laced case \cite{FO}, it will turn out to be
crucial that there exists a second solution to (\ref{FU}) 
\begin{equation}
\sum\limits_{l=i,j,k}q^{\zeta _{l}^{\prime }}\,\sigma _{q}^{\xi _{l}^{\prime
}}\,\gamma _{l}=0\,\,\,\,.  \label{FU2}
\end{equation}
The two solutions may not be obtained from each other by simple shifts, but
they are related as 
\begin{equation}
\xi _{l}^{\prime }=-\xi _{l}+\frac{c_{l}-1}{2}\quad \text{and\quad }\zeta
_{l}^{\prime }=-\zeta _{l}-(1+c_{l})t_{l}\,\,,\quad \quad l=i,j,k.
\label{FU22}
\end{equation}
Nonetheless, as an existence criterion for the fusing process, the variant (%
\ref{FU}) is sufficient, since the second solution may always be constructed
from the first as we now demonstrate. Changing $q$ to $q^{-1}$ in the fusing
rule (\ref{FU}) and using (\ref{q-q}) thereafter, we obtain 
\begin{equation}
\sum\limits_{l=i,j,k}q^{-\zeta _{l}}\,\tau ^{\frac{1+c_{l}}{2}}\sigma
_{c_{l}}^{q}\tau ^{\frac{1+c_{l}}{2}}\,\,\sigma _{q}^{-\xi _{l}}\,\,\tau ^{-%
\frac{1+c_{l}}{2}}\sigma _{c_{l}}^{q}\tau ^{-\frac{1+c_{l}}{2}}\,\,\gamma
_{l}=0\,\,.
\end{equation}
Acting on this equation with $\tau ^{-1}\sigma _{+}^{q}\tau ^{-1}$ yields (%
\ref{FU2}), with the help of (\ref{acti}), (\ref{tau}) and (\ref{qCox}).
What remains to be shown is that these two solutions are indeed
non-equivalent in the sense defined above. For this purpose we may take the
limit $q\rightarrow 1$ and note that the quantities $\xi _{l}$ and $\xi
_{l}^{\prime }$ are related to each other in the same way as in the simply
laced case. We may now simply refer to \cite{FO} for the proof of the
non-equivalence of this two triplets. This is sufficient to establish the
non-equivalence between the two solutions. In addition we shall demonstrate
below that there exists in fact no further non-equivalent solution.

\subsection{The Fusing Rule in $\hat{\Omega}^{q}$}

\emph{The generalized three-point-coupling related to three particles of the
type }$i,j$\emph{\ and }$k$\emph{\ is non-vanishing, i.e. the process }$%
i+j\rightarrow \bar{k}$\emph{\ is possible, if and only if there exist
representatives of the q-deformed orbits }$\omega ^{\frac{c_{i}-1}{2}}\hat{%
\Omega}_{i}^{q},$\emph{\ }$\omega ^{\frac{c_{j}-1}{2}}\hat{\Omega}_{j}^{q}$%
\emph{\ and }$\omega ^{\frac{c_{k}-1}{2}}\hat{\Omega}_{k}^{q}$\emph{\ whose
sum is zero. }

This means there should exist two triplets of integers $(\hat{\xi}_{i},\hat{%
\xi}_{j},\hat{\xi}_{k})$ and $(\hat{\zeta}_{i},\hat{\zeta}_{j},\hat{\zeta}%
_{k})$ such that 
\begin{equation}
\sum\limits_{l=i,j,k}\,\,q^{\hat{\zeta}_{l}}\hat{\sigma}^{\hat{\xi}_{l}}\,%
\hat{\gamma}_{l}^{+}=0\,\,\,.\,\,\,  \label{FUt}
\end{equation}
Equivalence$\,$ of two solutions is defined as in the previous section, i.e.
two triplets which are obtained by simple shifts of the type $\hat{\xi}%
_{l}\rightarrow \hat{\xi}_{l}+m$ and $\hat{\zeta}_{l}$ $\rightarrow \hat{%
\zeta}_{l}+n$ are considered equivalent to the original solution. However,
as in the non-twisted case, also (\ref{FUt}) always admits a second
non-equivalent solution 
\begin{equation}
\sum\limits_{l=i,j,k}q^{\hat{\zeta}_{l}^{\prime }}\,\,\hat{\sigma}_{q}^{\hat{%
\xi}_{l}^{\prime }}\,\hat{\gamma}_{l}^{+}=0\,\,\,.  \label{FU2t}
\end{equation}
The relations between the two solutions read 
\begin{equation}
\hat{\zeta}_{l}^{\prime }=-\hat{\zeta}_{l}+1-c_{l}\quad \text{and\quad }\hat{%
\xi}_{l}^{\prime }=-\hat{\xi}_{l}+\frac{1-c_{l}}{2}l_{l}+1+c_{l}\,,\quad
l=i,j,k.  \label{sect}
\end{equation}
As in the previous section the second solution may be constructed from the
first, and therefore the variant (\ref{FUt}) is sufficient as an existence
criterion.

\subsection{The Fusing Rule in $\Omega $ and $\hat{\Omega}$}

\emph{The generalized three-point-coupling related to three particles of the
type }$i,j$\emph{\ and }$k$\emph{\ is non-vanishing, i.e. the process }$%
i+j\rightarrow \bar{k}$\emph{\ is possible, if and only if there exist
representatives of the orbits }$\Omega _{i},\Omega _{j}$ $\emph{and}$ $%
\Omega _{k}$ \emph{whose sum is zero and if in addition there exist
representatives of the orbits }$\omega ^{\frac{c_{i}-1}{2}}\hat{\Omega}_{i},$%
\emph{\ }$\omega ^{\frac{c_{j}-1}{2}}\hat{\Omega}_{j}$\emph{\ and }$\omega ^{%
\frac{c_{k}-1}{2}}\hat{\Omega}_{k}$\emph{\ which also sum up to zero. }

Quantitatively this means there should exist two triplets of integers $(\xi
_{i},\xi _{j},\xi _{k})$ and $(\hat{\xi}_{i},\hat{\xi}_{j},\hat{\xi}_{k})$
such that 
\begin{equation}
\sum\limits_{l=i,j,k}\,\,\sigma ^{\xi _{l}}\,\gamma _{l}=0\,\qquad \text{%
and\qquad }\sum\limits_{l=i,j,k}\,\,\hat{\sigma}^{\hat{\xi}_{l}}\,\hat{\gamma%
}_{l}^{+}=0\,\,.  \label{FU3}
\end{equation}
The version (\ref{FU3}) of the fusing rule was first stated by Chari and
Pressley \cite{CP}, with the only difference that our $\hat{\sigma}$
corresponds to the inverse twisted Coxeter element in \cite{CP} and also $%
\hat{\gamma}_{l}^{+}$ is defined differently in their formulation. The
multiplication of the first equation in (\ref{FU3}) by powers of the Coxeter
element $\sigma $ and the second by powers of the twisted Coxeter element $%
\hat{\sigma}$ will produce further solutions, which we regard as equivalent.
Once again there exists a second non-equivalent solution 
\begin{equation}
\sum\limits_{l=i,j,k}\,\,\sigma ^{\xi _{l}^{\prime }}\,\gamma _{l}=0\,\qquad 
\text{and\qquad }\sum\limits_{l=i,j,k}\,\,\hat{\sigma}^{\hat{\xi}%
_{l}^{\prime }}\,\hat{\gamma}_{l}^{+}=0\,\,,  \label{FU32}
\end{equation}
which is related to the first by the relevant relations in (\ref{FU22}) and (%
\ref{sect}). The equations (\ref{FU3}) and (\ref{FU32}) may be obtained in
the limit $q\rightarrow 1$ from (\ref{FU}), (\ref{FU2}) and (\ref{FUt}), (%
\ref{FU2t}), respectively. Since we have already shown that neither the
triplet $(\xi _{i}^{\prime },\xi _{j}^{\prime },\xi _{k}^{\prime })$ may be
obtained from $(\xi _{i},\xi _{j},\xi _{k})$ by simple shifts nor $(\hat{\xi}%
_{i}^{\prime },\hat{\xi}_{j}^{\prime },\hat{\xi}_{k}^{\prime })$ from $(\hat{%
\xi}_{i},\hat{\xi}_{j},\hat{\xi}_{k})$ by the same means, we have
established the nonequivalence between the two solutions. It is also clear
from the preceding sections that we may construct the second solution always
from the first.

\subsection{The Fusing Rule and conserved Quantities}

Let $y(n)$ ($1\leq n\leq r$) be a vector\footnote{%
In fact we see below that this will be the nullvector of a particular matrix
as specified in equation (\ref{qqK}).} whose components are labeled by
particle types. In particular for $n=1$ we identify $y_{i}(1)$ with the
quantum mass $m_{i}$ of the particle of species $i$. Then we may formulate a
further variant of the fusing rule:

\emph{The generalized three-point-coupling related to three particles of the
type }$i,j$\emph{\ and }$k$\emph{\ is non-vanishing, i.e. the process }$%
i+j\rightarrow \bar{k}$\emph{\ is possible, if there exist two triplets of
integers }$(\eta _{i},\eta _{j},\eta _{k})$ \emph{and} $(\bar{\eta}_{i},\bar{%
\eta}_{j},\bar{\eta}_{k})$ \emph{such that} 
\begin{equation}
\sum\limits_{l=i,j,k}e^{s_{n}(\eta _{l}\theta _{h}+\bar{\eta}_{l}\theta
_{H})}\,y_{l}(n)=0\,\,.  \label{fumat}
\end{equation}
The $s_{n}$ ($1\leq n\leq r$) label the exponents of the algebra $%
X_{r}^{(1)} $ in increasing order. We further introduced the angles 
\begin{equation}
\theta _{h}:=\frac{i\pi (2-B)}{2h}\qquad \text{and\qquad }\theta _{H}:=\frac{%
i\pi B}{2H}\,,
\end{equation}
whose deeper origin becomes more apparent when we discuss the scattering
matrix in section 4. The coupling constant $\beta $ enters here the
expressions through the function $B=2H\beta ^{2}/(H\beta ^{2}+4\pi h)$ which
takes values between $0$ and $2$. Obviously, multiplying equation (\ref
{fumat}) by $e^{ms_{n}\eta _{l}\theta _{h}}$ and $e^{ks_{n}\bar{\eta}%
_{l}\theta _{H}}$, with $m,k$ being arbitrary integers, will also produce a
solution, which we regard as equivalent in the same spirit as in the
previous subsections. Likewise there exists a second non-equivalent solution 
\begin{equation}
\sum\limits_{l=i,j,k}\,e^{s_{n}(\eta _{l}^{\prime }\theta _{h}+\bar{\eta}%
_{l}^{\prime }\theta _{H})}\,y_{l}(n)=0\quad \quad \,\,,  \label{fumat2}
\end{equation}
related to the first simply as 
\begin{equation}
\eta _{l}^{\prime }=-\eta _{l}\qquad \text{and \qquad }\bar{\eta}%
_{l}^{\prime }=-\bar{\eta}_{l}\,\,.
\end{equation}
Clearly we can not construct (\ref{fumat}) from (\ref{fumat2}) by
multiplication of $e^{s_{n}m\eta _{l}\theta _{h}}$ and $e^{s_{n}k\bar{\eta}%
_{l}\theta _{H}}$ unless $\eta _{i}=\eta _{j}=\eta _{k}$ and $\bar{\eta}_{i}=%
\bar{\eta}_{j}=\bar{\eta}_{k}$. The latter fact would mean that $%
\sum_{l=i,j,k}\,y_{l}(n)=0$, $\,$which in particular for $n=1$ is impossible
since all quantities in the sum, the masses, are positive. We have therefore
established that the two solutions are indeed non-equivalent. However, one
solution may always be constructed from the other simply by replacing $%
s_{n}\rightarrow -s_{n}$ or complex conjugation of (\ref{fumat}) or (\ref
{fumat2}).

\medskip

\begin{picture}(240.00,200.00)(-50.00,0.00)
\put(188.00,182.00){$\delta_{ji}^{k+}$}
\put(236.00,106.00){$\delta_{kj}^{i+}$}
\put(38.00,106.00){$\delta_{ik}^{j-}$}
\put(92.00,10.00){$m_{\bar \imath}$}
\put(92.00,190.00){$m_{\bar \imath}$}
\put(18.00,40.00){$m_{\bar \jmath}$}
\put(18.00,160.00){$m_{\bar\jmath}$}
\put(158.00,10.00){$m_i$}
\put(158.00,190.00){$m_i$}
\put(230.00,40.00){$m_j$}
\put(230.00,160.00){$m_j$}
\put(130.00,90.00){$m_k$}
\qbezier(245.00,124.00)(227.00,121.00)(223.00,100.00)
\qbezier(172.00,185.00)(190.00,166.00)(214.00,176.00)
\qbezier(47.00,123.00)(58.00,116.00)(56.00,100.00)
\put(60.00,0.00){\line(2,1){200.00}}
\put(0.00,100.00){\line(3,-5){60.00}}
\put(60.00,200.00){\line(-3,-5){60.00}}
\put(260.00,100.00){\line(-2,1){200.00}}
\put(200.00,0.00){\line(3,5){60.00}}
\put(0.00,100.00){\line(2,-1){200.00}}
\put(10.00,100.00){\line(-1,0){10.00}}
\put(0.00,100.00){\line(1,0){10.00}}
\put(200.00,200.00){\line(-2,-1){200.00}}
\put(260.00,100.00){\line(-3,5){60.00}}
\put(10.00,100.00){\line(1,0){250.00}}
\end{picture}

{\small \noindent Figure 1: Mass triangles in the complex velocity plane.
The angles are defined as $i\delta _{jk}^{i\pm }=(\eta _{j}-\eta _{k})\theta
_{h}+(\bar{\eta}_{j}-\bar{\eta}_{k})\theta _{H}\pm i\pi .$ }

\medskip

Having obtained the fusing angles $\eta $ we may immediately compute
relations among the quantum conserved quantities. Combining (\ref{fumat})
and (\ref{fumat2}) we derive 
\begin{equation}
\frac{y_{i}(n)}{y_{j}(n)}=\frac{\sinh \left( s_{n}(\eta _{k}-\eta
_{j})\theta _{h}+s_{n}(\bar{\eta}_{k}-\bar{\eta}_{j})\theta _{H}\right) }{%
\sinh \left( s_{n}(\eta _{i}-\eta _{k})\theta _{h}+s_{n}(\bar{\eta}_{i}-\bar{%
\eta}_{k})\theta _{H}\right) }\,\,.
\end{equation}
We may interpret these relations in the complex velocity plane as explained
in \cite{FO}. In particular for $s_{1}=1$ we obtain the important ratios of
the quantum masses 
\begin{equation}
\frac{m_{i}}{m_{j}}=\frac{\sinh \left( (\eta _{k}-\eta _{j})\theta _{h}+(%
\bar{\eta}_{k}-\bar{\eta}_{j})\theta _{H}\right) }{\sinh \left( (\eta
_{i}-\eta _{k})\theta _{h}+(\bar{\eta}_{i}-\bar{\eta}_{k})\theta _{H}\right) 
}\,.  \label{massr}
\end{equation}
As the main difference to the simply laced case we note that the masses now
depend on the coupling constant. The relevant triangles are depicted in
figure 1. We will now be more specific on how to calculate the fusing angles
from Lie algebraic properties.

\subsection{Relations between the Fusing Rules}

The four versions of the fusing rules are all related to each other, meaning
that having one solution of one particular formulation of the fusing rule we
are able to construct all the other solutions. The precise relations read 
\begin{equation}
\eta _{l}=\xi _{l}^{\prime }-\xi _{l}=\frac{\hat{\zeta}_{l}-\hat{\zeta}%
_{l}^{\prime }}{2}\quad \text{and\quad }\bar{\eta}_{l}=\frac{\zeta
_{l}-\zeta _{l}^{\prime }}{2}=\hat{\xi}_{l}^{\prime }-\hat{\xi}_{l}\qquad 
\text{for }l=i,j,k.\quad \quad  \label{eta11}
\end{equation}
We see that the interchange of the two solutions of one version of the
fusing rule immediately demands that the two solutions of the other rules
should also be exchanged. In particular it follows that 
\begin{equation}
-2\xi _{l}=\hat{\zeta}_{l},\qquad \zeta _{l}=-2\hat{\xi}_{l}+\frac{1-c_{l}}{2%
}l_{l}\,\,-\frac{1+c_{l}}{2}t_{l}+1+c_{l}\quad \text{for }l=i,j,k.\,
\end{equation}
These relations do not only relate the fusing rule in $\Omega ^{q}$ and $%
\hat{\Omega}^{q}$ to each other, but they also provide the precise link
between the q-deformed and non-deformed versions of the fusing rule. It will
take until subsection 5.1. to have assembled all the ingredients for the
proof of (\ref{eta11}).

There is one last question which we should answer with regard to possible
solutions of the fusing rules: Are there any further non-equivalent
solutions to these equations? The answer is no. For the proof of this
statement we assume at this point that the rules are indeed equivalent, such
that it suffices to discuss only one version. We adopt the argumentation of 
\cite{FO} for this purpose. The only four triangles which we may construct
in the complex velocity plane from three sides with fixed modulus are the
ones depicted in figure 1. Hence there are no further possible angles,
meaning no additional non-equivalent solution to (\ref{fumat}) exist. By (%
\ref{eta11}) this fact is also established for all other versions of the
fusing rule we have stated.

Treating the fusing rule as a pure existence criterion for the possibility
of certain fusing processes, one version is as good as the other. We
observed however that the relevant data from the ``classical'' fusing rules,
which correspond to two equations in section 3.3., may be merged together
into one single equation by the q-deformation. This is the key feature which
can be exploited in the quantum field theory and which appears to be
absolutely necessary for the construction of generic expressions for the
scattering matrices.

\section{Block Representation}

The scattering matrices for affine Toda field theories have been the subject
of numerous investigations \cite{TodaS,FO,nons,Donons,Khast,Oota}.
Restricting the attention to the case when the coupling constant is real,
the two-particle scattering matrix for all simple Lie algebras, involving
particles of the species $i$ and $j$ as a function of the relative rapidity $%
\theta $, may be cast into the universal expression 
\begin{equation}
S_{ij}(\theta )=\prod\limits_{x=1}^{h}\prod\limits_{y=1}^{H}\left\{
x,y\right\} _{\theta }^{2\mu _{ij}(x,y)}\,\,\,\,.  \label{SPP}
\end{equation}
Here $\left\{ x,y\right\} _{\theta }$ are certain combinations of hyperbolic
functions and the $\mu _{ij}(x,y)$ are positive semi-integers for the given
range in (\ref{SPP}).

\subsection{The Building Blocks}

Before explaining how the powers $\mu _{ij}(x,y)$ may be computed, we
present several representations of the general building blocks, which will
serve different purposes. As a crucial step in the process of formulating
generic expressions for scattering matrices one should view the observation
of P. Dorey \cite{Donons} who noticed that the building blocks may all be
expressed in a very elegant form. We slightly modify them to simplify
certain computations and define\footnote{%
In \cite{nons,FKS} a different type of blocks was used. They may be
translated into each other by simple replacements, e.g. for $G_{2}$ and $%
F_{4}$ one sets $H^{-1}=\theta _{h}+\theta _{H}$.} 
\begin{equation}
\left\{ x,y\right\} _{\theta }:=\frac{\left[ x,y\right] _{\theta }}{\left[
x,y\right] _{-\theta }}  \label{blockhyp}
\end{equation}
and

\begin{equation}
\lbrack x,y]_{\theta }:=\frac{\left\langle x-1,y-1\right\rangle _{\theta
}\left\langle x+1,y+1\right\rangle _{\theta }}{\left\langle
x-1,y+1\right\rangle _{\theta }\left\langle x+1,y-1\right\rangle _{\theta }}%
\quad \left\langle x,y\right\rangle _{\theta }:=\sinh \frac{1}{2}\left(
\theta +x\theta _{h}+y\theta _{H}\right) \,.
\end{equation}
We used the angles $\theta _{h}$ and $\theta _{H}$ as introduced in section
3.4. Notice that the strong-weak duality transformation $\beta \rightarrow
4\pi /\beta $ ($B\rightarrow 2-B$), $h\leftrightarrow H$ leaves the
scattering matrix invariant. One should stress that besides the strong-weak
interchange the invariance also demands the interchange of the Coxeter
numbers.

Alternatively, each block (\ref{blockhyp}) admits an integral representation
in the form 
\begin{equation}
\left\{ x,y\right\} _{\theta }=\exp \int\limits_{0}^{\infty }\frac{dt}{%
t\sinh t}\,\,f_{x,y}^{h,H}(t)\sinh \left( \frac{\theta t}{i\pi }\right)
\label{block}
\end{equation}
with 
\begin{equation}
\,f_{x,y}^{h,H}(t)=8\sinh \left( \vartheta _{h}t\right) \sinh \left(
\vartheta _{H}t\right) \sinh \left( t-x\vartheta _{h}t-y\vartheta
_{H}t\right) \,\,.\quad
\end{equation}
This may be verified for instance by the explicit computation of the
integral in (\ref{block}). We abbreviated here $\vartheta _{h}:=(2-B)/2h$
and $\vartheta _{H}:=B/2H$. Particular attention has to be paid to the
convergence of the integral representation (\ref{block}), especially when we
analytically continue. Shifting $\theta \rightarrow \theta +x^{\prime
}\theta _{h}+y^{\prime }\theta _{H}$, convergence requires that 
\begin{equation}
0\leq (x-x^{\prime }-1)\vartheta _{h}+(y-y^{\prime }-1)\vartheta _{H}\leq
2(1-(1+x^{\prime })\vartheta _{h}-(1+y^{\prime })\vartheta _{H})\,\,.\,
\label{converg}
\end{equation}
In particular for real rapidity $\theta $ the convergence is guaranteed if $%
0<x<h$ and $0<y<H$.

With regard to several applications, the values of the scattering matrices
at $\theta =0$ are of special interest and we therefore comment on it for
definiteness. In general we have $\left\{ x,y\right\} _{\theta =0}=1$, apart
from the case $\left\{ 1,1\right\} _{\theta =0}=-1$. This means we have to
pay attention to the ordering of certain limits. When writing the blocks in
form of hyperbolic functions (\ref{blockhyp}), we have to set first $x=y=1$
and then take the limit $\theta \rightarrow 0$, whereas in the integral
representation (\ref{block}) we have to set $x=y=1$, integrate thereafter
and finally take the limit $\theta \rightarrow 0$.

The following obvious identities will turn out to be useful in the course of
our argumentation 
\begin{eqnarray}
\left\{ x,y\right\} _{\theta } &=&\left\{ x+2h,y+2H\right\} _{\theta
}=\left\{ -x,-y\right\} _{\theta }^{-1}\quad  \label{shif1} \\
\left\{ x,y\right\} _{\theta +x^{\prime }\theta _{h}+y^{\prime }\theta _{H}}
&=&\frac{\left[ x+x^{\prime },y+y^{\prime }\right] _{\theta }}{\left[
x-x^{\prime },y-y^{\prime }\right] _{-\theta }}  \label{shif2} \\
\left\{ x,y\right\} _{\theta +p\theta _{h}+q\theta _{H}}\left\{ x,y\right\}
_{\theta -p\theta _{h}-q\theta _{H}} &=&\left\{ x+p,y+q\right\} _{\theta
}\left\{ x-p,y-q\right\} _{\theta }\,\,.  \label{shif}
\end{eqnarray}
Furthermore, it will be convenient to adopt the slightly more compact
notation for the product of several blocks 
\begin{equation}
\left\{ x_{1},y_{1}\right\} _{\theta }^{\mu _{1}}\left\{ x_{2},y_{2}\right\}
_{\theta }^{\mu _{2}}\ldots \left\{ x_{n},y_{n}\right\} _{\theta }^{\mu
_{n}}=:\left\{ x_{1},y_{1}^{\mu _{1}};x_{2},y_{2}^{\mu _{2}};\cdots
;x_{n},y_{n}^{\mu _{n}}\right\} _{\theta }
\end{equation}
from time to time.

We shall now come to the characterization of the powers $\mu _{ij}(x,y)$ of
particular blocks $\left\{ x,y\right\} _{\theta }$, which may be computed
either by using the properties of $X_{r}^{(1)}$ or $\hat{X}_{n}^{(l)}$.

\subsection{The Powers from $X_{r}^{(1)}$}

The powers in (\ref{SPP}) can be evaluated from the matrix-valued generating
function 
\begin{equation}
\sum\limits_{y}\mu _{ij}\left( 2x-\frac{c_{i}+c_{j}}{2},y\right) q^{y}=-%
\frac{[t_{j}]_{q}}{2}q^{\frac{(1-c_{j})t_{j}-(1+c_{i})t_{i}}{2}}(\,\tilde{%
\lambda}_{j}\cdot \sigma _{q}^{x}\gamma _{i})\,\,,  \label{detpower}
\end{equation}
for fixed $x$. Taking $x$ in the range $(3-c_{i})/2\leq x\leq h+(1-c_{i})/2$
ensures that the first argument of $\mu $ is between $1$ and $2h$. This
formula is a natural generalization of the one for the simply laced case (%
\ref{possim}), where now the q-deformation incorporates the information of
both dual algebras. At this point we have only stated (\ref{detpower}) and
we shall now convince ourselves that it is indeed satisfying all the
requirements we need.

When applying formula (\ref{SPP}), we have to guarantee that the properties
of the combinations of hyperbolic functions in the building blocks $\left\{
x,y\right\} _{\theta }$ are reflected in the correct way by the Lie
algebraic quantities. This means, that according to the identities in (\ref
{shif1}) we should have 
\begin{equation}
\mu _{ij}(x,y)=\mu _{ij}(x+2h,y+2H)\quad \text{and\quad }\mu _{ij}(x,y)=-\mu
_{ij}(2h-x,2H-y)\,\,.  \label{po}
\end{equation}
Considering (\ref{detpower}), the first relation in (\ref{po}) follows
trivially from (\ref{quasi}). Together with the r.h.s. of (\ref{detpower})
the second relation in (\ref{po}) may be proven directly with the help of (%
\ref{kro}). The second relation is important, since it ensures that we can
always find two blocks which combine in such a way that the total power of
each building block becomes an integer. Therefore it guarantees that the
scattering matrix is a meromorphic function, even if we choose (this is
sometimes very convenient) the ranges in (\ref{SPP}) to be $1\leq x\leq h$
and $1\leq y\leq H$.

Having established the formal legitimacy of (\ref{SPP}), it is clear that
properties of the $\mu $'s may be carried over into properties of the
scattering matrix. We will therefore prove several identities which we
exploit below when discussing the scattering matrix.

First we note that 
\begin{equation}
\mu _{ij}(x,y)=\mu _{ji}(x,y)=-\mu _{\bar{\imath}j}(x+h,y+H)\,\,.
\label{antimu}
\end{equation}
The symmetry in the subscripts follows directly from the defining relation
for the $\mu $'s (\ref{detpower}) and the symmetry property of the inner
product (\ref{Pari}). The second equation follows in view of the definition
of the anti-particle (\ref{anti}) and (\ref{detpower}). The latter identity
relates the powers involving the particle on one hand and the anti-particle
on the other and will therefore turn out to be useful to show the crossing
relation.

From the fusing rule in $\Omega ^{q}$ follows by similar manipulations as we
have just performed 
\begin{equation}
\sum\limits_{l=i,j,k}\mu _{lp}\left( x\pm \eta _{l},y\pm \bar{\eta}%
_{l}\right) =0\,\,,  \label{23}
\end{equation}
where the lower sign relates to the first (\ref{FU}) and the upper sign to
the second solution (\ref{FU2}). The integers $\eta _{l}$ and $\bar{\eta}%
_{l} $ are related to the two solutions of the fusing rules by (\ref{eta11}%
). It still needs to be established that they are indeed the same as the
ones occurring in the equations involving the conserved quantities, (\ref
{fumat}) and (\ref{fumat2}). It will turn out that both relations in (\ref
{23}) will be crucial to prove the bootstrap equations for the scattering
matrices.

The final relation in this section follows from (\ref{bootp}) and (\ref
{detpower}) 
\begin{equation}
\mu _{ij}(x+1,y+t_{i})+\mu
_{ij}(x-1,y-t_{i})=\sum\limits_{n=1}^{I_{il}}\sum\limits_{l\in \Delta }\mu
_{lj}(x,y+2n-1-I_{il})\,\,,  \label{idd}
\end{equation}
where we understand that the sum $\sum_{n=1}^{I_{il}}$ yields zero when $%
I_{il}=0$. We can view (\ref{idd}) as a particular solution of the recursive
equations (2.4) quoted in \cite{Oota}. One may take these equations as a
starting point and use them to construct the powers $\mu _{ij}$ recursively.
However, it remains unclear how to obtain the equations (\ref{idd}) from
first principles. In fact (\ref{idd}) should be regarded as a consequence of
(\ref{23}) and we therefore view the latter equations as more fundamental.
We demonstrate this fact only for the equivalent equations of the scattering
matrices, since in that setting they correspond to a simple physical
property, see section 7.

\subsection{The Powers from $\hat{X}_{n}^{(l)}$}

Alternatively we can use the data of the twisted algebra $\hat{X}_{n}^{(l)}$
in order to compute the powers of the building blocks. In this case the role
of two arguments $x$ and $y$ in the generating function is reversed, that is
now we fix a particular $y$ and read off the possible values for $x$ from
the generating functions 
\begin{equation}
\sum\limits_{x}\nu _{ij}\left( x,2y-c_{i}+\frac{c_{i}-1}{2}l_{i}-\frac{%
c_{j}-1}{2}l_{j}\right) q^{x}=-\frac{q^{-\frac{c_{i}+c_{j}}{2}}}{2}(\,\hat{%
\lambda}_{j}\cdot \hat{\sigma}_{q}^{y}\hat{\gamma}_{i}^{+})\,\,.
\label{detpower2}
\end{equation}
Since the two descriptions, i.e. in terms of the data of $X_{r}^{(1)}$ or in
terms of the data of $\hat{X}_{n}^{(l)}$ are supposed to be the same, we
expect similar relations as we obtained in the previous section for the $\mu 
$'s also to hold for the $\nu $'s. Now property (\ref{shif1}) of the blocks
demands that

\begin{equation}
\nu _{ij}(x,y)=\nu _{ij}(x+2h,y+2H)\quad \text{and}\quad \nu _{ij}(x,y)=-\nu
_{ij}(2h-x,2H-y)\,\,.  \label{po22}
\end{equation}
The first relation in (\ref{po22}) follows trivially from (\ref{quasi2}).
Once again we may guarantee that the scattering matrix is a meromorphic
function by means of the second relation in (\ref{po22}), which follows from
(\ref{mm}). We also have the identities which imply parity and crossing 
\begin{equation}
\nu _{ij}(x,y)=\nu _{ji}(x,y)=-\nu _{\bar{\imath}j}(x+h,y+H)\,\,.
\label{paricr}
\end{equation}
The first equation follows now from (\ref{paritt}) and the second from (\ref
{antit})\footnote{%
We should keep in mind here that we did not yet prove (\ref{mm}) and (\ref
{paritt}). In fact we reverse the logic and prove first the properties for
the $\nu $'s in section 5.2. and deduce from them the inner product
identities in section 2.2.3.}. The relation which implies the bootstrap
identity

\begin{equation}
\sum\limits_{l=i,j,k}\nu _{lp}\left( x\pm \eta _{l},y\pm \bar{\eta}%
_{l}\right) =0\,\,,  \label{boott}
\end{equation}
follows from the version of the fusing rules related to the q-deformed
twisted Coxeter element in $\hat{\Omega}^{q}$ (section 3.2). As the
counterpart of (\ref{idd}) we derive from the defining relations of the $\nu 
$'s and (\ref{tw}) 
\begin{equation}
\nu _{ij}(x+c_{i},y)+\nu _{\omega
^{-c_{i}}(i)j}(x-c_{i},y-2c_{i})=\!\!\!\sum\limits_{\alpha _{l}\in \hat{%
\Delta}_{-c_{i}}}\!\!\!I_{il}\nu _{lj}\left( x,y+\frac{1-c_{i}}{2}l_{i}-%
\frac{1+c_{i}}{2}l_{l}\right)  \label{788}
\end{equation}

Having finally assembled the main properties of all the ingredients from
which we construct the scattering matrices, we are now in the position to
utilize them in order to study the properties of $S$.

\subsection{Bootstrap Properties}

The exact expressions for two-particle scattering matrices of integrable
quantum field theories may be obtained by solving certain consistency
equations, the so-called bootstrap equations. We will now demonstrate that (%
\ref{SPP}) fulfills indeed all the requirements and take this as a proof for
the conjectured formulae stated in the previous subsection.

\subsubsection{Unitarity, Crossing and Parity Invariance}

The unitarity-analyticity equation $S_{ij}(\theta )S_{ij}(-\theta )=1$
follows trivially from the property $\left\{ x,y\right\} _{\theta }\left\{
x,y\right\} _{-\theta }=1$ of each individual building blocks. The crossing
relation $S_{ij}(\theta )=S_{\bar{\imath}j}(i\pi -\theta )$ requires in
general a little bit more effort, e.g. \cite{KO}. Using (\ref{shif1}) and (%
\ref{shif2}) we obtain 
\begin{equation}
S_{ij}(i\pi -\theta )=S_{ij}(h\theta _{h}+H\theta _{H}-\theta
)=\prod\limits_{x=1}^{h}\prod\limits_{y=1}^{H}\left\{ x+h,y+H\right\}
_{\theta }^{-\mu _{ij}(x+h,y+H)}\,\,.  \label{133}
\end{equation}
Employing now the second identity in (\ref{antimu}), the r.h.s. of (\ref{133}%
) equals $S_{\bar{\imath}j}(\theta )$, which establishes the crossing
relation. The parity invariance of the scattering matrix, i.e. $%
S_{ij}(\theta )=S_{ji}(\theta )$, is guaranteed by the symmetry property of
the $\mu $'s in the lower indices, i.e. the first equation in (\ref{antimu}).

Alternatively we can use the data of the q-deformed twisted Coxeter element
and repeat the argumentation once more, using now the relations (\ref{paricr}%
) instead of (\ref{antimu}).

\subsubsection{Bootstrap Identities}

We will now come to the key equations, whose names are sometimes associated
with this whole approach, the bootstrap equations. The claim is that once
the fusing rules in section 3 hold, the following identity is true for the
scattering matrices 
\begin{equation}
\prod\limits_{l=i,j,k}S_{pl}(\theta +\eta _{l}\theta _{h}+\bar{\eta}%
_{l}\theta _{H})=1\,.  \label{boot}
\end{equation}
The integers $\eta _{l}$ and $\bar{\eta}_{l}$ may be expressed by using the
data from the various versions of the fusing rules (\ref{eta11}). The proofs
of the relations (\ref{boot}) are straightforward. We obtain with the help
of (\ref{shif2}) 
\begin{equation}
\prod\limits\Sb x,y  \\ l=i,j,k  \endSb \left\{ x,y\right\} _{\theta +\eta
_{l}\theta _{h}+\bar{\eta}_{l}\theta _{H}}^{\mu _{pl}(x,y)}=\prod\limits\Sb %
x,y  \\ l=i,j,k  \endSb \frac{\left[ x+\eta _{l},y+\bar{\eta}_{l}\right]
_{\theta }^{\mu _{pl}(x,y)}}{\left[ x-\eta _{l},y-\bar{\eta}_{l}\right]
_{-\theta }^{\mu _{pl}(x,y)}}=1\,\,.
\end{equation}
The last step follows by shifting $x\rightarrow x-\eta _{l}$ and $%
x\rightarrow x+\bar{\eta}_{l}$ in the numerator and denominator,
respectively, such that we can employ the two equations in (\ref{23}). We
note that it is crucial to have both solutions at hand. Alternatively we can
derive the bootstrap equation (\ref{boot}) by exploiting the property (\ref
{boott}) of the $\nu $'s and repeating the arguments once more.

With the help of (\ref{shif}) we translate (\ref{idd}) into what we refer to
as the ``combined bootstrap'' identity for the scattering matrix 
\begin{equation}
S_{ij}\left( \theta +\theta _{h}+t_{i}\theta _{H}\right) S_{ij}\left( \theta
-\theta _{h}-t_{i}\theta _{H}\right)
=\prod\limits_{l=1}^{r}\prod\limits_{n=1}^{I_{il}}S_{jl}\left( \theta
+(2n-1-I_{il})\theta _{H}\right) \,\quad .  \label{Ravqq}
\end{equation}
Here we understand that the product $\prod_{n=1}^{I_{il}}$ contributes $1$
when $I_{il}=0$. Sometimes this identity is identical to some bootstrap
equation, but in general it has to be constructed by combining several
identities of the type (\ref{boot}) in a very particular way. Its
significance is, that it may be employed in order to derive the matrix
representation for the scattering matrix (see section 4.3). Reducing (\ref
{Ravqq}) to the simply laced case, i.e. $[I_{il}]_{q}\rightarrow I_{il}$, $%
H\rightarrow h,t_{i}\rightarrow 1$, we recover an identity quoted in \cite
{Rava}, see section 7.

\subsubsection{Occurrence of certain special Blocks}

For various purposes it is important to exhibit explicitly the occurrence of
particular blocks $\{x,y\}$ in the general formula (\ref{SPP}). It is
possible to extract the blocks of the form $\left\{ 1,y\right\} _{\theta
}\,\,\left\{ 2,y^{\prime }\right\} _{\theta }$ from the general product and
re-write the scattering matrix as 
\begin{equation}
S_{ij}(\theta )=\left\{ 1,1_{t_{i}}\right\} _{\theta }^{\delta
_{ij}}\,\,\left\{ 2,2_{I_{ij}\,t_{j}}\right\} _{\theta }\prod\limits_{x\neq
1,2}\prod\limits_{y}\left\{ x,y\right\} _{\theta }\,\,\,.  \label{SP}
\end{equation}
For the proof of (\ref{SP}) we exploit the properties of the q-deformed
Coxeter element $\sigma _{q}$. Considering the identity (\ref{detpower}), we
notice that for $i=j$ a block of the form $\{1,y\}$ may only occur for $%
x=0,c_{i}=-1$ or $x=1,c_{i}=1$. From (\ref{bootp}) and the orthogonality of
simple roots and co-fundamental weights, we obtain $\,\tilde{\lambda}%
_{i}\cdot \sigma _{q}^{\frac{c_{i}+1}{2}}\gamma _{i}=-q^{(1+c_{i})t_{i}}$
and therefore we get 
\begin{equation}
\sum\limits_{y}\mu _{ii}\left( 1,y\right) q^{y}=\frac{[t_{i}]_{q}}{2}%
q^{t_{i}}=\frac{1}{2}\sum\limits_{n=1}^{t_{i}}q^{2n-1}\,\,,
\end{equation}
which establishes the first factor in (\ref{SP}). In order to prove the
occurrence of the second factor, we observe that a block of the form $%
\{2,y\} $ may only be generated if $c_{i}\neq c_{j}$. Due to the parity
property of the $\mu $'s (\ref{po}), we may choose $c_{i}=1$ and $c_{j}=-1$
w.l.g., such that we obtain from (\ref{bootp}) $\,\tilde{\lambda}_{j}\cdot
\sigma _{q}\gamma _{i}=-q^{2t_{i}}[I_{ij}]_{q}$. Hence we obtain 
\begin{equation}
\sum\limits_{y}\mu _{ij}\left( 2,y\right) q^{y}=\frac{q^{t_{i}+t_{j}}}{2}[%
I_{ij}]_{q}[t_{j}]_{q}=\frac{1}{2}\sum\limits_{n=1}^{t_{j}}q^{2n-1}\,\sum%
\limits_{n=1}^{I_{ij}}q^{2n-1}=\frac{1}{2}\sum%
\limits_{n=1}^{I_{ij}t_{j}}q^{2n}\,.
\end{equation}
In the last equality we have used the fact that either $t_{j}$ or $I_{ij}$
has to be one. This establishes (\ref{SP}).

There are several consequences we may draw from (\ref{SP}). An immediate
conclusion concerns the value of the scattering matrix at vanishing
rapidities. With the remark made in section 4.1 we deduce from (\ref{SP})
that 
\begin{equation}
S_{ij}(0)=(-1)^{\delta _{ij}}\,\,.
\end{equation}
The knowledge of this value is for instance important in the context of the
thermodynamic Bethe ansatz \cite{TBAZam}.

\subsubsection{Singularities and the generalized Bootstrap}

As we have seen the blocks of the form (\ref{blockhyp}) are extremely useful
to exhibit the Lie algebraic structure of the scattering matrix. However,
they are quite misleading with regard to the singularity structure due to
the possible cancellation of zeros and poles. This may happen whenever we
have a product of two blocks $\left\{ x,y\right\} \left\{ x^{\prime
},y^{\prime }\right\} $ and $x,x^{\prime }$ or $y,y^{\prime }$ differ by $2$%
. It suffices to consider the latter case, since it will cover all examples
we shall be constructing. Motivated by this observation we introduce the
quantity 
\begin{eqnarray}
\left\{ x,y_{n}\right\} _{\theta } &=&\prod\limits_{l=0}^{n-1}\left\{
x,y+2l\right\} _{\theta } \\
&=&\frac{\left\langle x-1,y-1\right\rangle _{\theta }\left\langle
x+1,y-1+2n\right\rangle _{\theta }}{\left\langle x+1,y-1\right\rangle
_{\theta }\left\langle x-1,y-1+2n\right\rangle _{\theta }}\times (\theta
\rightarrow -\theta )^{-1}\,\,,
\end{eqnarray}
and also define the angles 
\begin{equation}
\theta _{x,y,n}^{\pm }=(x\pm 1)\theta _{h}+(2n+y-1)\theta _{H}\text{\quad }%
\,\,.
\end{equation}
which serve to characterize the precise location of the singularities of the
blocks $\left\{ x,y_{n}\right\} _{\theta }$. Obviously the four zeros are
situated at $\pm \theta _{x,y,0}^{\pm },\mp \theta _{x,y,n}^{\pm }$ and the
four poles at $\pm \theta _{x,y,n}^{\pm },\pm \theta _{x,y,0}^{\mp }$
respectively. In order to interpret these singularities from the physical
point of view we should know when they are situated on the physical sheet,
i.e. $0\leq \func{Im}\theta \leq i\pi $. Recalling that the range for the
possible arguments of the blocks $0<x<h$, $0<y<H$ and the range in which the
effective coupling takes its value, i.e. $0\leq B\leq 2$ we evaluate 
\begin{equation}
0\leq \func{Im}(\theta _{x,y,n}^{\pm })\leq \pi \qquad \text{for }B\leq 
\frac{2H(h-x\mp 1)}{\left| h(2n+y-1)-H(x\pm 1)\right| }\,.  \label{64}
\end{equation}
The relevant residues are computed to 
\begin{eqnarray}
&& 
\begin{array}{c}
\limfunc{Res}\limits_{\theta =\theta _{x,y,0}^{-}}\left\{ x,y_{n}\right\} =\,%
\frac{-2\sinh \theta _{h}\sinh (n\theta _{H})\sinh \left( x\theta
_{h}+(n+y-1)\theta _{H}\right) \sinh \left( \theta _{x,y,0}^{-}\right) }{%
\sinh \left( \theta _{h}+n\theta _{H}\right) \sinh \left( x\theta
_{h}+(y-1)\theta _{H}\right) \sinh \left( (x-1)\theta _{h}+(y+n-1)\theta
_{H}\right) }
\end{array}
\\
&& 
\begin{array}{c}
\limfunc{Res}\limits_{\theta =\theta _{x,y,n}^{+}}\!\left\{ x,y_{n}\right\} =%
\frac{2\sinh \theta _{h}\sinh (n\theta _{H})\sinh \left( x\theta
_{h}+(n-1+y)\theta _{H}\right) \sinh \left( \theta _{x,y,n}^{+}\right) }{%
\sinh \left( \theta _{h}+n\theta _{H}\right) \sinh \left( (1+x)\theta
_{h}+(n+y-1)\theta _{H}\right) \sinh \left( x\theta _{h}+(2n+y-1)\theta
_{H}\right) }\,\,.\qquad
\end{array}
\end{eqnarray}
It is easy to convince oneself that with the stated range for $x,y,B,n$
together with (\ref{64}) we have 
\begin{equation}
\func{Im}\left( \limfunc{Res}_{\theta =\theta _{x,y,0}^{-}}\!\left\{
x,y_{n}\right\} _{\theta }\right) <0\quad \quad \text{and \qquad }\func{Im}%
\left( \limfunc{Res}_{\theta =\theta _{x,y,n}^{+}}\!\left\{ x,y_{n}\right\}
_{\theta }\right) >0\,\,,
\end{equation}
such that the $\theta _{x,y,n}^{+}$ could correspond to the direct channel
poles. In the simply laced case this knowledge is enough to judge about the
sign of the residue of the whole S-matrix, e.g. \cite{FO}. For the case at
hand matters are more involved since the remaining blocks in the scattering
matrix do in general not possess a definite sign. It is this feature which
lead the authors of \cite{nons} to the formulation of the generalized
bootstrap. According to this prescription only odd order poles, whose
imaginary part of the residue is positive in the \emph{whole }range of the
effective coupling $B$, participate in the bootstrap.

So let us have a closer look at the behaviour of a block $\left\{ x^{\prime
},y_{n^{\prime }}^{\prime }\right\} _{\theta _{x,y,n}^{+}}$. We obtain a
first criterion for a possible sign change by considering the extreme limits
in the coupling constant. In general we have $\lim_{\beta \rightarrow
0,\infty }\left\{ x^{\prime },y_{n^{\prime }}^{\prime }\right\} _{\theta
_{x,y,n}^{+}}$ $=1$. However, if $x^{\prime }=x$ we have 
\begin{eqnarray}
\lim_{\beta \rightarrow 0}\left\{ x^{\prime },y_{n^{\prime }}^{\prime
}\right\} _{\theta _{x,y,n}^{\pm }} &=&\left( \frac{y^{\prime }-y-2n}{%
y^{\prime }-y+2n^{\prime }-2n}\right) ^{\pm 1}  \label{res1} \\
\lim_{\beta \rightarrow \infty }\left\{ x^{\prime },y_{n^{\prime }}^{\prime
}\right\} _{\theta _{x,y,n}^{\pm }} &=&1\,\,.
\end{eqnarray}
This means if the block responsible for the pole is $\left\{ x,y_{n}\right\}
_{\theta }$ and the right hand side of (\ref{res1}) is negative the
imaginary parts of the possible additional blocks 
\begin{equation}
\left\{ x,y_{n^{\prime }}^{\prime }\right\} _{\theta _{x,y,n}^{+}}\text{%
\qquad and\qquad }\left\{ x+2,y_{n^{\prime \prime }}^{\prime \prime
}\right\} _{\theta _{x,y,n}^{+}}  \label{kii}
\end{equation}
both change their sign while $\beta $ runs from zero to infinity. This means
the pole $\theta _{x,y,n}^{+}$ does not participate in the bootstrap if in
the scattering matrix also the blocks (\ref{kii}) occur to an odd power and
if they do not cross the real axis at the same position. This means having
the scattering matrix given explicitly in blockform the condition on $%
y,y^{\prime },n,n^{\prime }$ by which the l.h.s. of (\ref{res1}) becomes
negative, together with the occurrence of blocks like (\ref{kii}) provides a
simple criterion which allows to judge whether a pole resulting from a
certain block should be excluded from the generalized bootstrap or not.

Exploiting the fusing rules and reading off the relative rapidities from (%
\ref{fumat}) we obtain the precise location, say $\phi $, of a pole in the
scattering matrix which participates in the generalized bootstrap 
\begin{equation}
\phi =\pm (\eta _{i}-\eta _{j})\theta _{h}\pm (\bar{\eta}_{i}-\bar{\eta}%
_{j})\theta _{H}\,\,.
\end{equation}
The two signs result from the two non-equivalent solutions of the fusing
rule.

\section{Matrix-Integral Representation}

Alternatively to the universal form for the scattering matrix in form of
blocks there exists a remarkable integral representation. This version of
the scattering matrix is particularly useful when applied in the context of
the thermodynamic Bethe ansatz \cite{TBAZam,FKS} or off-shell when computing
form factors \cite{FF}. We can express the scattering matrix as\footnote{%
A very similar formula was first obtained by Oota ((5.2) in \cite{Oota}) on
the base of a case-by-case study. In comparision, the formula (\ref{SPPi})
differs only by a factor $(-1)^{\delta _{ij}}$ $\exp \left( 2\delta
_{ij}\int_{0}^{\infty }\frac{dt}{t}\sinh \left( \frac{\theta t}{i\pi }%
\right) \right) $, which is $1$ for $\theta $ real, but different from one
if the rapidity becomes complex. Similar expressions also appear in \cite{FR}%
.}

\begin{equation}
S_{ij}(\theta )=\,\exp \int\limits_{0}^{\infty }\frac{dt}{t}\,\,\Phi
_{ij}(t)\sinh \left( \frac{\theta t}{i\pi }\right) \,\,,  \label{SPPi}
\end{equation}
with 
\begin{equation}
\Phi _{ij}\left( t\right) =8\sinh (\vartheta _{h}t)\sinh (t_{j}\vartheta
_{H}t)\left( [K]_{q(t)\bar{q}(t)}\right) _{ij}^{-1}\,\,\,.  \label{Oota}
\end{equation}
We introduced here the particular deformation parameters $q(t)=\exp
(\vartheta _{h}t)$ and $\bar{q}(t)=\exp (\vartheta _{H}t)$ and the matrix 
\begin{equation}
\lbrack K_{ij}]_{q\bar{q}}=(q\bar{q}^{t_{i}}+q^{-1}\bar{q}^{-t_{i}})\delta
_{ij}-[I_{ij}]_{\bar{q}}\,\,.  \label{qqK}
\end{equation}
In the limit $q\rightarrow 1$ and $\bar{q}\rightarrow 1$ the matrix $%
[K_{ij}]_{q\bar{q}}$ obviously reduces to the ordinary Cartan matrix $K$,
such that one is tempted to view this matrix as a doubly q-deformed Cartan
matrix. However, this viewpoint is slightly misleading as we now argue. For
the simply laced cases it was proven \cite{FLO}, that the conserved
quantities may be organized as right eigenvectors of the Cartan matrix $%
\sum_{j}K_{ij}y_{j}(n)=4\sin ^{2}(s_{n}\pi /h)$ $y_{i}(n)$ with $s_{n}$
labeling the exponents of the algebra as already introduced. In particular
we have that $y_{i}(1)\sim m_{i}$. It is then easy to see that this may also
be re-written as 
\begin{equation}
\sum_{j=1}^{r}[K_{ij}]_{q(i\pi s_{n})\bar{q}(i\pi s_{n})}y_{j}(n)=0\,\,.
\label{null}
\end{equation}
Hence, we can alternatively organize the conserved charges as nullvectors of
the matrix $[K_{ij}]_{q(i\pi t)\bar{q}(i\pi t)}$ evaluated at exponents of
the Lie algebra, i.e. $t=is_{n}$. Based on a case-by-case investigation,
Oota pointed out \cite{Oota} that equation (\ref{null}) also holds for the
non-simply laced case. A general proof of this statement is still
outstanding. There is, however, one important difference in comparison with
the simply laced case. In general we can not reverse the interpretation
anymore, such that we are not able to recover a genuine eigenvalue equation.
In particular for $s_{1}=1$ this leads to 
\begin{equation}
\sum_{j}[I_{ij}]_{\bar{q}(i\pi )}m_{j}=2\cosh \left( \theta _{h}+t_{i}\theta
_{H}\right) m_{i}\,\,.
\end{equation}
We observe that the eigenvalue depends now through the symmetrizer $t_{i}$
on the component of the ``eigenvector''. In the limit $\beta \rightarrow 0$
we restore the old picture and recover the equation $%
\sum_{j}I_{ij}y_{j}(n)=2\cos (\pi s_{n}/h)y_{i}(n)$ valid for all simple Lie
algebras. With the help of (\ref{deft}) we also obtain the equation for the
left nullvector $x_{i}(n)$ related to the right as $y_{i}(n)=[t_{i}]_{\bar{q}%
}\,x_{i}(n)$.

The determinant of the matrix (\ref{qqK}) may be computed \cite{Oota} to 
\begin{equation}
\det [K]_{q(i\pi t)\bar{q}(i\pi t)}=\prod\limits_{n=1}^{r}4\sin
((t+s_{n})\pi /h)\sin ((s_{n}-t)\pi /h)\,\,.  \label{deter}
\end{equation}
We do not have a general proof of this formula, but we have verified it
case-by-case. Two important features which we exploit below should be
noticed here, first the determinant becomes independent of the coupling
constant $\beta $ and second it vanishes for $t$ being an exponent.

Before we provide the proof for the representation (\ref{SPPi}), we will
introduce two further auxiliary matrices.

\subsection{The M-Matrix}

We restrict now the sum of the generating function for the powers of the
building blocks (\ref{detpower}) to a finite range and also include an
additional deformation parameter $\bar{q}$ into our consideration. We define
the matrix 
\begin{equation}
M_{ij}(q,\bar{q})=\sum_{x=1}^{2h}\sum_{y=1}^{2H}\mu _{ij}(x,y)q^{x}\bar{q}%
^{y}\,\,\,\,,  \label{M}
\end{equation}
where initially we keep both deformation parameters completely generic. From
the properties for the $\mu $'s, which we deduced in section 4.2., we can
immediately derive several features for the matrix $M$ 
\begin{equation}
M_{ij}(q,\bar{q})=-q^{2h}\bar{q}^{2H}M_{ij}(q^{-1},\bar{q}^{-1})=M_{ji}(q,%
\bar{q})\,.  \label{Mid}
\end{equation}

The first identity in (\ref{Mid}) is a consequence of the two relations in (%
\ref{po}) together with the fact that $\mu _{ij}(0,y)=\mu _{ij}(2h,y)=0$ for
all $y$. The second follows trivially from the symmetry properties of the $%
\mu $'s from the first relation in (\ref{antimu}).

Most crucial is once more the combined bootstrap equation, which on the Lie
algebraic side corresponds to the property (\ref{idd}). In fact, this
identity will enable us to compute the matrix $M$ explicitly. By some
straightforward manipulations of this relation we deduce with (\ref{idd})
that $M(q,\bar{q})$ has to satisfy

\begin{equation}
(q^{-1}\bar{q}^{\,-t_{i}}+q\bar{q}^{t_{i}})M_{ij}(q,\bar{q}%
)-\sum_{k=1}^{r}[I_{ik}]_{\bar{q}}M_{kj}(q,\bar{q})=\frac{1-q^{2h}\bar{q}%
^{2H}}{2}\,[t_{i}]_{\bar{q}}\delta _{ij}\,.  \label{M1}
\end{equation}
Solving this equation for $M(q,\bar{q})$ yields 
\begin{equation}
M_{ij}(q,\bar{q})=\frac{1-q^{2h}\bar{q}^{2H}}{2}\left( [K]_{q\bar{q}}\right)
_{ij}^{-1}\,[t_{j}]_{\bar{q}}\,\,.  \label{M2}
\end{equation}
At first sight (\ref{M2}) does not seem to be a finite polynomial of the
form (\ref{M}). However, the doubly q-deformed Cartan matrix becomes
singular at certain values and the pre-factor $(1-q^{2h}\bar{q}^{2H})$
ensures the whole expression to remain finite. In other words this term may
always be factorized into the determinant of $[K_{ij}]_{q\bar{q}}$ and some
rest, such that the r.h.s. of (\ref{M2}) will indeed be a polynomial as
defined in (\ref{M}). In \cite{FR} a similar matrix as (\ref{M2}) also
occurs. However, apart from the ordering of $[t],[K]$, the pre-factor $%
(1-q^{2h}\bar{q}^{2H})/2$, which is crucial for the polynomial aspect we
discuss below, is not mentioned in there.

We will deviate now from our generic consideration and specify the
deformation parameters to be $q(t)$ and $\bar{q}(t)$ as introduced after
equation (\ref{Oota}). Noting first of all that $q(t)^{2h}\bar{q}%
(t)^{2H}=e^{2t}$, we observe that for $t=i\pi m$ the r.h.s. of (\ref{M1})
always vanishes. Furthermore it follows from (\ref{M2}) that $M(q(i\pi m),%
\bar{q}(i\pi m))$ is also always zero unless $m$ is an exponent by (\ref
{deter}). From this we deduce that $M(q(i\pi s_{n}),\bar{q}(i\pi s_{n}))$ is
proportional to the right nullvector $y(n)$ as specified in (\ref{null}). In
view of the symmetry property (\ref{Mid}), we conclude that 
\begin{equation}
M_{ij}(q(i\pi s_{n}),\bar{q}(i\pi s_{n}))\sim y_{i}(n)y_{j}(n)  \label{propo}
\end{equation}
where the factor of proportionality does neither depend on the particle
index $i$ nor on $j$. Most importantly we derive from (\ref{23}) a matrix
version of the fusing rule (\ref{fumat}) and (\ref{fumat2}) 
\begin{equation}
\sum\limits_{l=i,j,k}q(i\pi s_{n})^{\eta _{l}}\,\bar{q}(i\pi s_{n})^{\bar{%
\eta}_{l}}\,M_{lp}(q(i\pi s_{n}),\bar{q}(i\pi s_{n}))=0\,\,\quad \quad \text{%
for }1\leq p\leq r\,\,.  \label{fusmat}
\end{equation}
By means of (\ref{propo}) we may divide out $y_{p}(n)$ and the factor of
proportionality from (\ref{fusmat}), such that we have at last established
the relation (\ref{fumat}) involving the conserved quantities.

We may specify the deformation parameters further and take $q$ and $\bar{q}$
to be roots of unity of order $2h$ and $2H$, respectively. This may be done
safely after we have cancelled the determinant against the pre-factor. As a
consequence this means in particular that together with the periodicity
property of the $\mu $'s (the first property in (\ref{po})), we may
simultaneously shift the upper and lower limit in the sum (\ref{M})
arbitrarily. The properties of the blocks are now also reflected by the
polynomial (\ref{M}), such that we can not only carry out a one-to-one
identification between $\{x,y\}_{\theta }$ and $q^{x}\bar{q}^{y}$, but in
addition we can also manipulate them in an identical way. If in analogy to $%
\{-x,-y\}_{\theta }=\{x,y\}_{\theta }^{-1}$, we further define $q^{-x}\bar{q}%
^{-y}=-q^{x}\bar{q}^{y}$ we can even guarantee that the range of $x$ and $y$
is $1\leq x\leq h$, $1\leq y\leq H$. With these assumptions in mind we
derive 
\begin{equation}
M_{ij}(q,\bar{q})=q^{h}\bar{q}^{H}M_{\bar{\imath}j}(q^{-1},\bar{q}%
^{-1})=\,-q^{h}\bar{q}^{H}M_{\bar{\imath}j}(q,\bar{q})\,\,\,  \label{Mcr}
\end{equation}
from the last relation in (\ref{antimu}).

As a final remark of this section, we note that at roots of unity the
defining relation for the $M$-matrix (\ref{M}) may be viewed as the discrete
Fourier transformation of $\mu _{ij}(x,y)$, the inverse of which reads 
\begin{equation}
\mu _{ij}(x,y)=\frac{1}{4\,h\,H}\sum_{m=1}^{2h}\sum_{n=1}^{2H}M_{ij}\left(
\omega ^{\,m},\hat{\omega}^{\,\,n}\right) \omega ^{-\,\,mx}\hat{\omega}%
^{-ny},  \label{invers}
\end{equation}
with $\omega $ and $\hat{\omega}$ being the $2h$'th and $2H$'th primitive
roots of unity, respectively. This allows us to compute the powers of the
blocks, i.e. the $\mu $'s, in an alternative way from the explicit
expression of $M(q,\bar{q})$ in matrix form (\ref{M2}). We may also utilize (%
\ref{invers}) to verify the properties of the $\mu $'s by exploiting now
explicitly matrix representation of $M(q,\bar{q})$, instead of the orbits of
the q-deformed Coxeter element as in of section 4.2. In addition the
computing rules, which we stated in the previous paragraph for generic $q$
and $\bar{q}$ are automatically satisfied for $q$ and $\bar{q}$ being roots
of unity.

\subsection{The N-Matrix}

As to be expected, we may also express the scattering matrix in terms of the
data of the twisted algebra $\hat{X}_{n}^{(l)}$. In analogy to the M-matrix (%
\ref{M}) we define the $n\times n$-matrix 
\begin{equation}
N_{ij}(q,\bar{q})=\sum_{x=1}^{2h}\sum_{y=1}^{2H}\nu _{ij}(x,y)q^{x}\bar{q}%
^{y}\,\,\,\,,
\end{equation}
where once again we keep both deformation parameters completely generic for
the time being. It should be clear that our notation in (\ref{detpower2}) is
slighly abused here at the cost of avoiding the introduction of new symbols.
From the Lie algebraic analogue to the combined bootstrap equation (\ref{788}%
) we derive 
\begin{eqnarray}
&&(-1)^{\hat{t}_{i}+1}(q^{\hat{t}_{i}}\bar{q})^{-2c_{i}}N_{ij}+N_{\omega
^{-c_{i}}(i)j}-\sum_{\alpha _{l}\in \hat{\Delta}_{-c_{i}}}q^{-c_{i}}\bar{q}%
^{-2c_{i}+\frac{c_{i}-1}{2}l_{i}+\frac{c_{i}+1}{2}l_{l}}I_{il}N_{lj} 
\nonumber \\
&=&(q\bar{q})^{-c_{i}}\frac{(1-q^{2h}\bar{q}^{2H})}{2}\delta _{i\omega ^{%
\frac{1+c_{i}}{2}}(j)}\,\,.  \label{N1}
\end{eqnarray}
Unlike the corresponding equation for the non-twisted case (\ref{M1}), we
can not solve (\ref{N1}) directly due to the occurrence of indices
transformed by the automorphism $\omega $. However, we may consider equation
(\ref{N1}) for $i\rightarrow \omega ^{-c_{i}}(i)$ and iterate the resulting
equations as long as we obtain $N_{\omega ^{-lc_{i}}(i)j}=N_{ij}$.
Thereafter we can safely solve the equation for the $r\times r$-submatrix,
say $\hat{N},$ and obtain 
\begin{equation}
\hat{N}_{ij}(q,\bar{q})=\frac{1-q^{2h}\bar{q}^{2H}}{2}\left( [\hat{K}]_{q%
\bar{q}}\right) _{ij}^{-1}\,[l_{j}]_{\bar{q}}\,\,.  \label{N2}
\end{equation}
Here we have introduced the doubly q-deformed twisted Cartan matrix 
\begin{equation}
\lbrack \hat{K}]_{q\bar{q}}=(q\bar{q}^{l_{i}}+q^{-1}\bar{q}^{-l_{i}})\delta
_{ij}-\left[ \sum_{k=1}^{l_{i}}\hat{I}_{\omega ^{k}(i)j}\right] _{\bar{q}%
}\,\,\,.  \label{twisK}
\end{equation}
Note that in the classical limit $q,\bar{q}\rightarrow 1$ we recover the
transpose of the usual twisted Cartan matrix. The transposition results from
our convention that particles in both dual algebras are denoted by the same
particle index. Similarly as in the nontwisted case the determinant of the
matrix (\ref{twisK}) acquires a very neat form 
\begin{equation}
\det [\hat{K}]_{q(i\pi t)\bar{q}(i\pi t)}=\prod\limits_{k=1}^{n}4\sin ((t+%
\hat{s}_{k})\pi /H)\sin ((\hat{s}_{k}-t)\pi /H)\,\,,
\end{equation}
where the $\hat{s}_{k}$ denote the $l$-th exponents of $\hat{X}_{n}^{(l)}$ 
\cite{Kac}. We also do not have a general proof of this formula, but we have
verified it once again case-by-case.

By direct computation, we may now derive several identities for the matrix $%
\hat{N}$, namely 
\begin{equation}
\hat{N}_{ij}(q,\bar{q})=\hat{N}_{ji}(q,\bar{q})=-q^{2h}\bar{q}^{2H}\hat{N}%
_{ij}(q^{-1},\bar{q}^{-1})\,\,\,.  \label{555}
\end{equation}
The first and second relation in (\ref{555}) imply the first property for
the $\nu $'s in (\ref{paricr}) and the second relation in (\ref{po22}),
respectively, which on the other hand finally prove the inner product
identities of section 2.2.3. Comparing (\ref{M2}) and (\ref{N2}) we see
immediately that $M=\hat{N}$ and therefore $\nu (x,y)=\mu (x,y)$. A direct
Lie algebraic proof of the latter equality would be desirable since it
allows to express quantities of the twisted algebra in terms of the
non-twisted algebra and vice versa. Having established several features of
the matrix $M(q,\bar{q})$ and $\hat{N}(q,\bar{q})$ we will now supply the
context in which they naturally originate.

\subsection{From Block- to Integral Representation}

Concerning the representation of the scattering matrix in blockform (\ref
{SPP}), an obvious question which arises is, whether it is possible to
compute explicitly the product over $x$ and $y$. Taking the explicit
integral representations of the blocks (\ref{block}) into account , this
problem amounts to the evaluation of 
\begin{eqnarray}
\Phi _{ij}(t) &=&\frac{1}{\sinh t}\sum_{x=1}^{2h}\sum_{y=1}^{2H}\mu
_{ij}(x,y)\,f_{x,y}^{h,H}(t,B),  \label{phi} \\
&=&-\frac{8\sinh (\vartheta _{h}t)\sinh (\vartheta _{H}t)}{\sinh t}%
e^{-t}M(q(t),\bar{q}(t)) \\
&=&-\frac{8\sinh (\vartheta _{h}t)\sinh (\vartheta _{H}t)}{\sinh t}e^{-t}%
\hat{N}(q(t),\bar{q}(t))
\end{eqnarray}
if we want to transform the scattering matrix into the form (\ref{SPPi}).
From the first identity in (\ref{Mid}), noting that $q(t)^{2h}\bar{q}%
(t)^{2H}=e^{2t}$, together with the explicit form of the M-matrix (\ref{M2}%
), we deduce the integral representation (\ref{SPPi}) with (\ref{Oota}).

Some comments are due, since it appears that the convergence condition (\ref
{converg}) is violated by the range we chose for $x$ and $y$ in the defining
relation for $M$. However, for each individual block $\{x,y\}_{\theta }$ we
can exploit the properties (\ref{shif1}) and bring the arguments $x$ and $y$
into a range for which the integral representation (\ref{block}) is
convergent. These features are reflected in the $M$-matrix if it is taken at
roots of unity together with the already mentioned rule $q^{-x}\bar{q}%
^{-y}=-q^{x}\bar{q}^{y}$.

As an alternative proof we may proceed similar as in \cite{FKS} for the
simply laced case. This method turns out to be instructive with regard to
particular applications as the thermodynamic Bethe ansatz and it will
illustrate the origin of the slight difference between (\ref{Oota}) and the
formula in \cite{Oota}. First we notice that the scattering matrix may also
be written as 
\begin{equation}
S_{ij}(\theta )=\mathcal{N}\,\exp \left( \int\limits_{0}^{\infty }\frac{dt}{%
\pi t}\widetilde{\varphi }_{ij}(t/\pi )\sinh \left( \frac{it\theta }{\pi }%
\right) \right) \,\,\,\,,  \label{Stba}
\end{equation}
when we introduce the quantities 
\begin{equation}
\varphi _{ij}(\theta )=-i\frac{d}{d\theta }\ln S_{ij}(\theta )\qquad \text{%
and \qquad }\widetilde{\varphi }_{ij}(k)=\int\limits_{-\infty }^{\infty
}d\theta \varphi _{ij}(\theta )e^{ik\theta }\,.  \label{TBAK}
\end{equation}
Due to the differentiation in (\ref{TBAK}), we have the freedom of a
normalization constant $\mathcal{N}$ in (\ref{Stba}), which may be fixed by
some asymptotic condition. Acting now with $-i$ times the logarithmic
derivative on the combined bootstrap identity (we concentrate here on the
case $I_{il}=1$) (\ref{Ravqq}), multiplying with $\exp (ik\theta )$ and
integrating thereafter with respect to $\theta $ we obtain 
\begin{equation}
\mathcal{P}\int d\theta \left( \varphi _{ij}\left( \theta +\theta
_{h}+t_{i}\theta _{H}\right) +\varphi _{ij}\left( \theta -\theta
_{h}-t_{i}\theta _{H}\right) \right) e^{ik\theta
}=\sum\limits_{l=1}^{r}I_{il}\,\,\widetilde{\varphi }_{lj}(k)\,\,.
\label{h22}
\end{equation}
Here $\mathcal{P}$ denotes the Cauchy principal value. Alternatively we may
compute $\widetilde{\varphi }_{ij}(k)$ directly. For this purpose we shift
the Fourier integral into the complex plane and integrate along the contours 
$\mathcal{C}_{\varrho }^{\pm }$ as depicted in figure 2.

\noindent Due to (\ref{SP}) we know explicitly the occurrence of the
relevant blocks which will give a contribution when we integrate along $%
\mathcal{C}_{\varrho }^{\pm }$. 
\begin{eqnarray}
\lim\limits_{\varrho \rightarrow \infty }\oint\limits_{\mathcal{C}_{\varrho
}^{\pm }}d\theta \varphi _{ij}(\theta )e &=&\widetilde{\varphi }_{ij}(k)-%
\mathcal{P}\int d\theta \varphi _{ij}\left( \theta \pm \theta _{h}\pm
t_{i}\theta _{H}\right) e^{ik(\theta \pm \theta _{h}\pm t_{i}\theta _{H})}
\label{h2} \\
&=&2\pi \delta _{ij}e^{\mp 2\pi \vartheta _{h/H}}\,-\pi I_{ij}e^{\mp k\pi
(\vartheta _{h}+\vartheta _{H})}\,.  \label{h3}
\end{eqnarray}
On the other hand the l.h.s. of (\ref{h22}) may be computed alternatively
from the right hand sides of (\ref{h2}) and (\ref{h3}), such that we obtain 
\begin{equation}
\widetilde{\varphi }_{ij}(k/\pi )=2\pi \left( \delta _{ij}-4\sinh k\vartheta
_{h}\sinh k\vartheta _{H}\left( 2\cosh t(\vartheta _{h}+\vartheta
_{H})-I\right) _{ij}^{-1}\right) \,\,,  \label{h33}
\end{equation}
and therefore (\ref{Oota}) by means of (\ref{Stba}). The other cases when $%
I_{il}=2,3$ may be obtained similarly with the singularity structure as
indicated in figure 2.

\[
\put(-150,0){\vector(1,0){300}} \put(0,-100){\vector(0,1){200}}
\put(-100,-50){\line(0,1){100}} \put(100,-50){\line(0,1){100}}
\put(-100,50){\line(1,0){200}} \put(-100,-50){\line(1,0){200}}
\put(100,0){\vector(0,1){25}} \put(100,0){\vector(0,-1){25}}
\put(-100,50){\vector(0,-1){25}} \put(-100,-50){\vector(0,1){25}}
\put(0,50){\vector(-1,0){50}} \put(0,-50){\vector(-1,0){50}}
\put(100,50){\vector(-1,0){50}} \put(100,-50){\vector(-1,0){50}}
\put(0,0){\vector(1,0){50}} \put(-100,0){\vector(1,0){50}}
\put(-117,-7){{\small $-\varrho$}} \put(102,-7){{\small $\varrho$}}
\put(2,55){{\small $ 2 \pi (\vartheta_h + t_i \vartheta_H)$ }}
\put(2,-59){{\small -$ 2 \pi (\vartheta_h + t_i \vartheta_H) $ }}
\put(130,-10){ {\small Re($\theta$)}} \put(2,95){{\small Im($\theta$)} }
\put(-3.3,48){$\bullet$ } \put(-3.3,-53){$\bullet$ } \put(-3.3,20){$\circ$ }
\put(-3.3,75){$\circ$ } \put(-3.3,-75){$\circ$ } \put(2,75){{\small $ 2 \pi
t_i \vartheta_H$ }} \put(2,-78){{\small -$ 2 \pi t_i \vartheta_H$ }}
\put(-3.3,-20){$\circ$ } \put(2,20){ {\small $ 2 \pi \vartheta_h $ }}
\put(2,-20){ {\small -$ 2 \pi \vartheta_h $ } } \put(80,35){ ${\cal
C}_{\varrho }^{+}$ } \put(80,-43){ ${\cal C}_{\varrho }^{-}$ } 
\]

\noindent {\small Figure 2: The contours $\mathcal{C}_{\varrho }^{\pm }$ in
the complex $\theta $-plane. The bullets $\bullet $ belong to poles
resulting from $-id/d\theta \ln \{1,1\}_{\theta }$ and the open circles $%
\circ $ to poles of $-id/d\theta \ln \{2,2\}_{\theta }$, for the situation $%
B>2H/(H+t_{i}h)$. When $B<2H/(H+t_{i}h)$ the poles at $\pm 2\pi
t_{i}\vartheta _{H}$ and $\pm 2\pi \vartheta _{h}$ reverse their roles. }

\section{Reduction to the simply laced Case}

It is instructive to investigate how the general formulae valid for all
simple Lie algebras behave when we specialize to simply laced Lie algebras.
Considering the data of $X_{r}^{(1)}$, we notice first of all that there is
no distinction anymore between $H$ and $h$. The length of all roots is the
same in the simply laced case, such that $t_{i}=1$ for all $i$ and the
incidence matrix becomes therefore symmetric. The q-deformed incidence
matrix reduces now to the usual incidence matrix, i.e. $[I_{ij}]_{q}%
\rightarrow I_{ij}$, since it does not have entries different from 1. As a
consequence the q-deformed Weyl reflections in (\ref{Weyl}) become the
ordinary Weyl reflections, such that $\sigma _{c(i)}^{q}\rightarrow \sigma
_{c(i)}$. The map $\tau $ commutes now with the $\sigma _{c(i)}$ and
therefore the q-deformed Coxeter element becomes 
\begin{equation}
\sigma _{q}\rightarrow q^{2}\sigma _{-}\sigma _{+\,\,}=q^{2}\sigma \,\,,
\end{equation}
with $\sigma $ being the ordinary non-deformed Coxeter element of $%
X_{n}^{(1)}$. Noting further that co-weights become identical to weights,
i.e. $\hat{\lambda}_{i}\rightarrow \lambda _{i}$, the generating function (%
\ref{detpower}) acquires the form 
\begin{equation}
\sum\limits_{y}\mu _{ij}\left( 2x-\frac{c_{i}+c_{j}}{2},y\right) q^{y}=-%
\frac{1}{2}q^{2x-\frac{c_{i}+c_{j}}{2}}(\,\lambda _{j}\cdot \sigma
^{x}\gamma _{i})\,\,.  \label{possim}
\end{equation}
Hence we always have $y=2x-\frac{c_{i}+c_{j}}{2}$ and the only type of
blocks which emerges is $\left\{ x,x\right\} _{\theta }$\footnote{%
The block $\left\{ x,x\right\} _{\theta }$ corresponds to the block $\left\{
x\right\} _{\theta }$ as defined in \cite{FO} or \cite{FKS}.}. Therefore the
block form of the scattering matrix reads 
\begin{equation}
S_{ij}\left( \theta \right) =\prod\limits_{q=1}^{h}\left\{ 2q-\frac{%
c_{i}+c_{j}}{2},2q-\frac{c_{i}+c_{j}}{2}\right\} _{\theta }^{-\frac{1}{2}%
\lambda _{i}\cdot \sigma ^{q}\gamma _{j}},\quad X_{r}^{(1)}\equiv ADE.
\label{ssimply}
\end{equation}
This means, that also conceptually the simply laced case admits a slightly
different formulation. In the generic case we compute the powers of the
building blocks indirectly via a generating function, whilst in the simply
laced case we may compute them directly.

We can also consider the data of $\hat{X}_{n}^{(l)}$ and undo the twist,
which means that $\omega \rightarrow 1$, $l_{i}\rightarrow 1$ and $\hat{t}%
_{i}\rightarrow 1$ for all $i$, such that the twisted q-deformed Coxeter
element becomes 
\begin{equation}
\hat{\sigma}_{q}\rightarrow q^{2}\sigma _{-}\sigma _{+\,\,}=q^{2}\sigma \,\,.
\end{equation}
Therefore the generating function (\ref{detpower2}) becomes 
\begin{equation}
\sum\limits_{x}\nu _{ij}\left( x,2y-\frac{c_{i}+c_{j}}{2}\right) q^{x}=-%
\frac{q^{2y-\frac{c_{i}+c_{j}}{2}}}{2}(\,\lambda _{j}\cdot \sigma ^{y}\gamma
_{i})\,\,\,\,,
\end{equation}
which means that $x=2y-\frac{c_{i}+c_{j}}{2}$ and the only type of blocks
which emerge are once again $\left\{ x,x\right\} _{\theta }$. Hence, the
scattering matrix reduces also in this analysis to the form (\ref{ssimply}).

The matrix inside the integral representation (\ref{Oota}) for the simply
laced case follows likewise and acquires the form 
\begin{equation}
\Phi _{ij}\left( t\right) =8\sinh \left( \frac{Bt}{2h}\right) \sinh \left( 
\frac{(2-B)t}{2h}\right) \left( 2\cosh t/h-I\right) _{ij}^{-1}\,\,\,.
\end{equation}
Hence we have recovered the formulae of \cite{FO} or \cite{FKS}.

\section{Case-by-Case}

In order to illustrate the working of our general formulae it is useful to
work them out explicitly for some concrete examples. We concentrate here on
the non-simply laced case, since the simply laced case is covered
extensively in the literature \cite{TodaS}. We will be most detailed for the
($G_{2}^{(1)},D_{4}^{(3)}$)-case. Our conventions with regard to numbering
and colouring may be read off from the Dynkin diagrams. As usual the arrow
points towards the short roots. A black and white vertex corresponds to the
colour value $c_{i}=-1$ and $c_{i}=1$, respectively.

\subsection{$(G_{2}^{(1)},D_{4}^{(3)})$}

\unitlength=0.780000pt 
\begin{picture}(223.00,105.00)(-120.00,200.00)
\put(130.00,0.00){\line(0,1){0.00}}
\qbezier(160.00,223.00)(176.00,213.00)(191.00,223.00)
\put(223.00,218.00){\makebox(0.00,0.00){$\alpha_3$}}
\put(175.00,316.01){\makebox(0.00,0.00){$\alpha_4$}}
\put(127.00,218.00){\makebox(0.00,0.00){$\hat{\alpha}_2$}}
\put(193.00,266.00){\makebox(0.00,0.00){$\hat{\alpha}_1$}}
\put(46.67,274.17){\makebox(0.00,0.00){$\alpha_2$}}
\put(6.00,274.01){\makebox(0.00,0.00){$\alpha_1$}}
\qbezier(189.33,298.34)(227.67,279.33)(213.33,242.00)
\qbezier(160.67,297.67)(126.00,281.01)(137.67,242.33)
\put(178.67,256.33){\line(1,-1){23.33}}
\put(171.67,256.33){\line(-1,-1){23.00}}
\put(175.00,294.68){\line(0,-1){31.33}}
\put(175.00,300.00){\circle{10.00}}
\put(205.00,230.00){\circle{10.00}}
\put(145.00,230.00){\circle{10.00}}
\put(175.00,260.00){\circle*{10.00}}
\put(20.00,260.00){\line(1,-2){7.00}}
\put(20.00,260.00){\line(1,2){7.00}}
\put(5.00,255.00){\line(1,0){40.00}}
\put(5.33,265.00){\line(1,0){39.67}}
\put(10.00,260.00){\line(1,0){30.00}}
\put(5.00,260.00){\circle*{10.00}}
\put(45.00,260.00){\circle{10.00}}
\end{picture}

\noindent The S-matrices of the theory read \cite{G2} 
\begin{eqnarray}
S_{11}(\theta ) &=&\{\stackrel{2}{\overbrace{1,1}};\stackrel{1}{\overbrace{%
3,5_{2}}};5,11\}_{\theta }  \label{11} \\
S_{12}(\theta ) &=&\{2,2_{3};\stackrel{1}{\overbrace{4,6_{3}}\}}_{\theta }
\label{22} \\
S_{22}(\theta ) &=&\{1,1_{3};\stackrel{2}{\overbrace{3,3_{3}}}%
;3,5_{3};5,7_{3}\}_{\theta }\,\,.  \label{33}
\end{eqnarray}
Here we indicated which block is responsible for which type of fusing
process. We have $h=6$ and $H=12$ for the Coxeter numbers. With the help of (%
\ref{shif}), we easily verify that for (\ref{11}), (\ref{22}) and (\ref{33})
the following bootstrap identities hold 
\begin{eqnarray}
S_{1l}\left( \theta +\theta _{h}+\theta _{H}\right) S_{1l}\left( \theta
-\theta _{h}-\theta _{H}\right) &=&S_{2l}\left( \theta \right) \quad \quad
l=1,2  \label{b2} \\
S_{1l}\left( \theta +2\theta _{h}+4\theta _{H}\right) S_{1l}\left( \theta
-2\theta _{h}-4\theta _{H}\right) &=&S_{1l}\left( \theta \right) \,\quad
\quad l=1,2\,  \label{b4} \\
S_{2l}\left( \theta +2\theta _{h}+4\theta _{H}\right) S_{2l}\left( \theta
-2\theta _{h}-4\theta _{H}\right) &=&S_{2l}\left( \theta \right) \,\quad
\quad l=1,2\,\,\,.  \label{b5}
\end{eqnarray}
As an example for the working of the generalized bootstrap and our criterion
provided in section 4.4.4., we plotted the imaginary part of the residues of 
$S_{22}(\theta )$ in figure 3 for several poles. We observe that the sign
changes throughout the range for poles resulting from $\{1,1_{3}\}$ and $%
\{3,5_{3}\}$. Only the poles responsible for the self-coupling of particle $%
2 $ has a positive imaginary part of the residue throughout the range of the
coupling constant $\beta $. Except at $B=4/3$ where it is zero, such that
this fusing process decouples.

\vspace*{-1.0cm}

\begin{center}
\includegraphics[width=10cm,height=16cm,angle=-90]{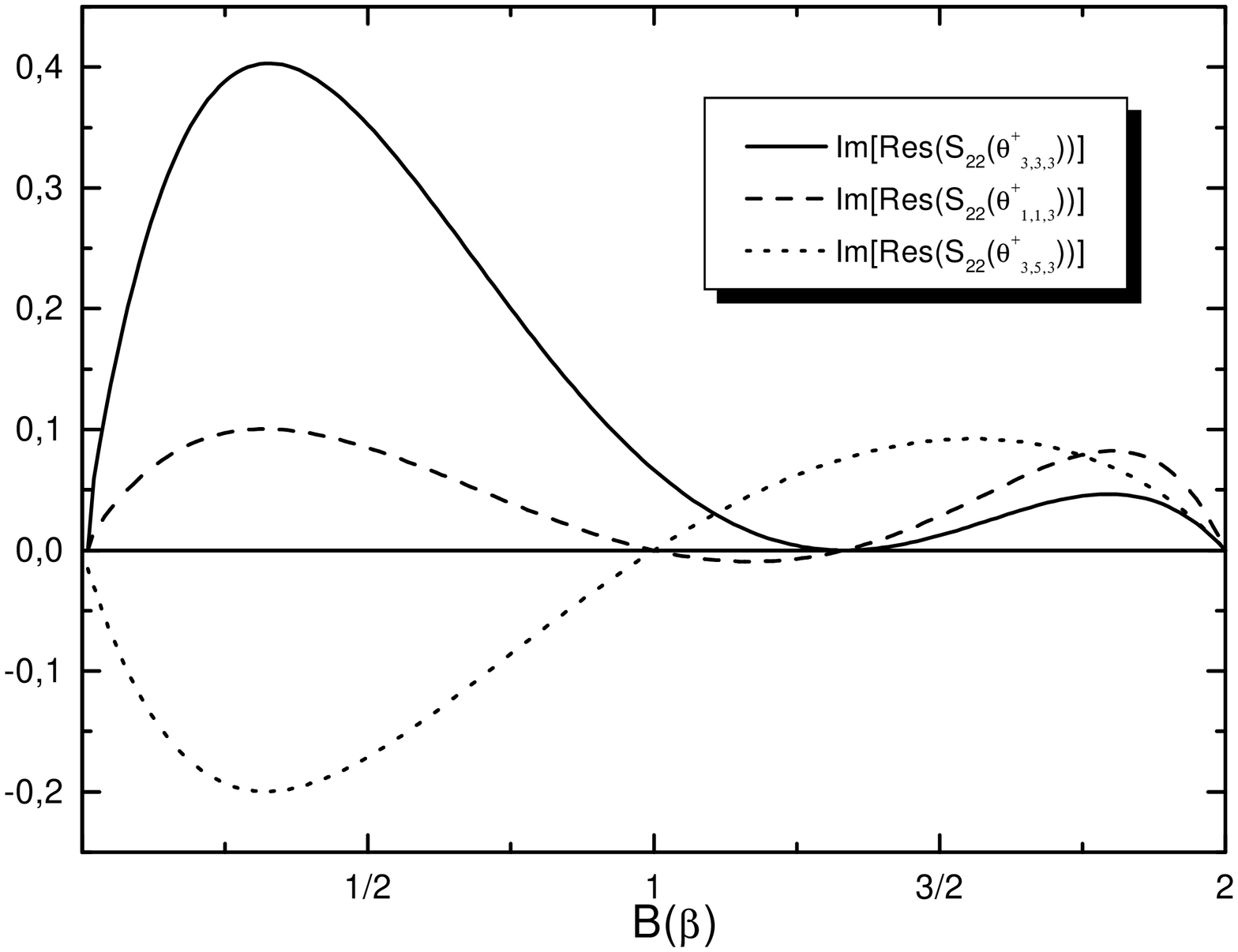}
\end{center}

\vspace*{0.5cm} \noindent {\small Figure 3: The imaginary part of several
residues of $S_{22}(\theta )$ as a function of the effective coupling
constant.} \vspace*{1.2mm}

\noindent Besides (\ref{b2}) the combined bootstrap identities (\ref{Ravqq})
also yield 
\begin{equation}
S_{l2}\left( \theta +\theta _{h}+3\theta _{H}\right) S_{l2}\left( \theta
-\theta _{h}-3\theta _{H}\right) \!\!=\!\!S_{l1}\left( \theta \right)
S_{l1}\left( \theta +2\theta _{H}\right) S_{l1}\left( \theta -2\theta
_{H}\right) ,  \label{g2boot}
\end{equation}
for $l=1,2$. These equations may be derived from (\ref{b2}) and (\ref{b4})
or verified directly for (\ref{11}), (\ref{22}) and (\ref{33}), with the
help of (\ref{shif}). The process corresponding to the combined bootstrap
identity (\ref{g2boot}) is depicted in figure 4.

\medskip

\medskip 
\begin{picture}(299.67,150.00)(-57.00,0.00)
\put(281.33,95.00){\makebox(0.00,0.00){$1$}}
\put(245.67,75.00){\makebox(0.00,0.00){$1$}}
\put(234.67,75.00){\makebox(0.00,0.00){$1$}}
\put(76.33,74.67){\makebox(0.00,0.00){$1$}}
\put(248.67,130.00){\makebox(0.00,0.00){$1$}}
\put(197.67,94.67){\makebox(0.00,0.00){$1$}}
\put(259.67,0.00){\makebox(0.00,0.00){$2$}}
\put(219.67,0.00){\makebox(0.00,0.00){$2$}}
\put(170.00,109.33){\makebox(0.00,0.00){$l$}}
\put(125.00,130.00){\makebox(0.00,0.00){$1$}}
\put(79.00,129.67){\makebox(0.00,0.00){$1$}}
\put(35.33,129.67){\makebox(0.00,0.00){$1$}}
\put(63.33,74.67){\makebox(0.00,0.00){$1$}}
\put(0.00,29.67){\makebox(0.00,0.00){$l$}}
\put(89.67,0.00){\makebox(0.00,0.00){$2$}}
\put(49.67,0.00){\makebox(0.00,0.00){$2$}}
\put(139.67,86.00){\line(1,0){21.00}}
\put(140.67,90.00){\line(1,0){19.00}}
\put(179.67,110.00){\line(6,1){120.00}}
\put(259.67,60.00){\line(0,-1){50.00}}
\put(239.67,110.00){\line(0,1){40.00}}
\put(259.67,60.00){\line(1,3){30.00}}
\put(239.67,110.00){\line(2,-5){20.00}}
\put(219.67,60.00){\line(2,5){20.00}}
\put(219.67,60.00){\line(0,-1){50.00}}
\put(189.67,150.00){\line(1,-3){30.00}}
\put(9.67,30.00){\line(6,1){120.00}}
\put(89.67,60.00){\line(0,-1){50.00}}
\put(49.67,60.00){\line(0,-1){50.00}}
\put(69.67,150.00){\line(0,-1){40.00}}
\put(89.67,60.00){\line(1,3){30.00}}
\put(69.67,110.00){\line(2,-5){20.00}}
\put(49.67,60.00){\line(2,5){20.00}}
\put(19.67,150.00){\line(1,-3){30.00}}
\end{picture}
\medskip\medskip

\noindent {\small Figure 4: }$(G_{2}^{(1)},D_{4}^{(3)})${\small -combined
bootstrap identities (\ref{g2boot}). }

\medskip \medskip

\noindent Reading off the fusing angles from the bootstrap equations we
obtain the mass ratios according to (\ref{massr}) 
\begin{equation}
\frac{m_{1}}{m_{2}}=\frac{\sinh \left( \theta _{h}+\theta _{H}\right) }{%
\sinh \left( 2\theta _{h}+2\theta _{H}\right) }\,\,.
\end{equation}

We may construct all these formulae from the Lie algebraic data in two
alternative ways.

\subsubsection{$S_{ij}\left( \theta \right) $ from $G_{2}^{(1)}$}

We start by exploiting the properties of $G_{2}^{(1)}$. The non-vanishing
entries of the incidence matrix are $I_{12}=1$ and $I_{21}=3$. Consequently
equation (\ref{deft}) yields $t_{1}=1$ and $t_{2}=3$. As indicated in the
Dynkin diagram we choose $c_{1}=-1$ and $c_{2}=1$, such that the q-deformed
Coxeter element reads $\sigma _{q}=\sigma _{1}^{q}\tau \sigma _{2}^{q}\tau $%
. The result of successive actions of this element on the simple roots is
reported in table 1. Here and in all further tables we choose the following
conventions: To each $\gamma _{i}$ we associate a column in which we report
the powers of the $q$ of the coefficients of the simple roots. We abbreviate 
\begin{equation}
\pm (q^{\mu _{1}^{1}}+\ldots +q^{\mu _{1}^{l_{1}}})\alpha _{1}\pm \ldots \pm
(q^{\mu _{r}^{1}}+\ldots +q^{\mu _{r}^{l_{r}}})\alpha _{r}\rightarrow \pm
\mu _{1}^{1},\ldots ,\mu _{l_{1}}^{1};\ldots ;\mu _{r}^{1},\ldots ,\mu
_{l_{r}}^{1}\,\,,  \label{notat}
\end{equation}
with $r=$ rank $\mathbf{g}$. When $q^{\mu }$ occurs $x$-times we denote this
by $\mu ^{x}$. Like in the undeformed case the overall sign of any element
in $\Omega _{q}$ is definite. Therefore it suffices to report the sign only
once as stated in (\ref{notat}). In the complete orbit we always have an
equal number of plus and minus signs. When we do not report any signs in the
column at all, the signs of the column to the left are adopted. In case the
coefficient of the root is zero, we indicate this by a $*$. For instance
from table 1 we read off : $\sigma _{q}\gamma _{1}=-(q^{4}+q^{6})\alpha
_{1}-q^{4}\alpha _{2}$.

\begin{center}
\begin{tabular}{|c||c|c|}
\hline\hline
$\sigma_q ^{x}$ & $\alpha _{1}=-\gamma _{1}$ & $\alpha _{2}=\gamma _{2}$ \\ 
\hline\hline
$1$ & $4,6;4$ & $-4,6,8;6$ \\ \hline
$2$ & $10;8$ & $-8,10,12;8,10$ \\ \hline
$3$ & $-12;*$ & $-*;12$ \\ \hline
$4$ & $-16,18;16$ & $16,18,20;18$ \\ \hline
$5$ & $-22;20$ & $20,22,24;20,22$ \\ \hline
$6$ & $24;*$ & $*;24$ \\ \hline\hline
\end{tabular}
\end{center}

\noindent {\small Table 1: The orbits $\Omega _{i}^{q}$ created by the
action of $\sigma _{q}^{x}$ on $\gamma _{i}$} \medskip

\noindent For the conventions chosen the generating functions (\ref{detpower}%
) for the powers of the building blocks are obtainable from the generating
functions 
\begin{eqnarray}
\sum\limits_{y}\mu _{11}\left( 2x+1,y\right) q^{y} &=&-q^{1}(\,\tilde{\lambda%
}_{1}\cdot (\sigma _{q})^{x}\gamma _{1})/2\,\, \\
\sum\limits_{y}\mu _{21}\left( 2x,y\right) q^{y} &=&-q^{-2}(\,\tilde{\lambda}%
_{1}\cdot (\sigma _{q})^{x}\gamma _{2})/2 \\
\sum\limits_{y}\mu _{22}\left( 2x-1,y\right) q^{y} &=&-q^{-3}\left[ 3\right]
_{q}(\,\tilde{\lambda}_{2}\cdot (\sigma _{q})^{x}\gamma _{2})/2\,\,.
\end{eqnarray}
We may now read off the Lie algebraic data from the table 1 and we can
construct the scattering matrices (\ref{11}), (\ref{22}) and (\ref{33})
according to formula (\ref{SPP}).

The two non-equivalent solutions to (\ref{FU}) corresponding to the S-matrix
bootstrap equations (\ref{b2}), (\ref{b4}) and (\ref{b5}) read 
\begin{equation}
q\sigma _{q}^{-1}\gamma _{1}+q^{-1}\gamma _{1}=q^{-3}\gamma _{2},\text{%
\qquad }q^{-1}\gamma _{1}+q\sigma _{q}^{-1}\gamma _{1}=q^{-3}\gamma _{2}\,,
\label{67}
\end{equation}
\begin{equation}
q^{3}\sigma _{q}^{-1}\gamma _{1}+q^{-5}\sigma _{q}\gamma _{1}=q^{-1}\gamma
_{1},\text{\qquad }q^{-3}\gamma _{1}+q^{5}\sigma _{q}^{-2}\gamma
_{1}=q\sigma _{q}^{-1}\gamma _{1}\,\,,  \label{68}
\end{equation}
\begin{equation}
q^{16}\sigma _{q}\gamma _{2}+\sigma _{q}^{5}\gamma _{2}=q^{20}\gamma _{2},%
\text{\qquad }q^{4}\sigma _{q}^{4}\gamma _{2}+q^{20}\gamma _{2}=\sigma
_{q}^{5}\gamma _{2}\,\,,  \label{69}
\end{equation}
respectively. These relations may be obtained either from (\ref{b2}), (\ref
{b4}) and (\ref{b5}) together with the formulae which relate the fusing
angles to the solution of the fusing rules in terms of the q-deformed
Coxeter element (\ref{eta11}) or alternatively they may be read off directly
from table 1. For a direct comparison with (\ref{eta11}) one should cross
all term to one side of the equation by means of (\ref{anti}).

It is also instructive to consider explicitly the matrix representation and
verify the general formulae of section 5. The doubly q--deformed Cartan
matrix for generic $q$ and $\bar{q}$ reads 
\begin{equation}
\lbrack K]_{q\bar{q}}=\left( 
\begin{array}{ll}
q\bar{q}+q^{-1}\bar{q}^{-1} & -1 \\ 
-(1+\bar{q}^{2}+\bar{q}^{-2}) & q\bar{q}^{3}+q^{-1}\bar{q}^{-3}
\end{array}
\right)  \label{KG}
\end{equation}
with determinant $\det [K]_{q\bar{q}}=q^{2}\bar{q}^{4}+q^{-2}\bar{q}^{-4}-1$%
. The right nullvectors are evaluated to 
\begin{eqnarray}
y(1) &=&(\sinh (\theta _{h}+\theta _{H}),\sinh (2\theta _{h}+2\theta _{H}))
\\
y(2) &=&(\sinh (5\theta _{h}+5\theta _{H}),\sinh (10\theta _{h}+10\theta
_{H}))\,\,.
\end{eqnarray}
From (\ref{KG}) we compute the $M$-matrix according to (\ref{M2}) 
\begin{equation}
M(q,\bar{q})=\frac{1-q^{12}\bar{q}^{24}}{2}\left( 
\begin{array}{ll}
\frac{q\bar{q}+q^{3}\bar{q}^{7}}{1-q^{2}\bar{q}^{4}+q^{4}\bar{q}^{8}} & 
\frac{1+\bar{q}^{2}+\bar{q}^{-2}}{q^{2}\bar{q}^{4}+q^{-2}\bar{q}^{-4}-1} \\ 
\frac{1+\bar{q}^{2}+\bar{q}^{-2}}{q^{2}\bar{q}^{4}+q^{-2}\bar{q}^{-4}-1} & 
\frac{(q\bar{q}+q^{3}\bar{q}^{3})(1+\bar{q}^{2}+\bar{q}^{4})}{1+q^{2}\bar{q}%
^{4}+q^{4}\bar{q}^{8}}
\end{array}
\right)  \label{MAA}
\end{equation}
\[
=\left( 
\begin{array}{ll}
{\tiny \allowbreak }\frac{{\footnotesize (1+q}^{2}{\footnotesize \bar{q}}^{4}%
{\footnotesize -q}^{6}{\footnotesize \bar{q}}^{12}{\footnotesize -q}^{8}%
{\footnotesize \bar{q}}^{16}{\footnotesize )(q\bar{q}+q}^{3}{\footnotesize 
\bar{q}}^{7}{\footnotesize )}}{2} & \frac{{\tiny (1+\bar{q}}^{2}{\tiny +\bar{%
q}}^{-2}{\tiny )(q}^{2}{\tiny \bar{q}}^{4}{\tiny +q}^{4}{\tiny \bar{q}}^{8}%
{\tiny -q}^{10}{\tiny \bar{q}}^{20}{\tiny -q}^{8}{\tiny \bar{q}}^{16}{\tiny )%
}}{2} \\ 
\frac{{\small (1+\bar{q}}^{2}{\small +\bar{q}}^{-2}{\small )(q}^{2}{\small 
\bar{q}}^{4}{\small +q}^{4}{\small \bar{q}}^{8}{\small -q}^{10}{\small \bar{q%
}}^{20}{\small -q}^{8}{\small \bar{q}}^{16}{\small )}}{2} & \frac{{\tiny (1+%
\bar{q}}^{2}{\tiny +\bar{q}}^{4}{\tiny )(q\bar{q}+q^{3}\bar{q}^{3}%
{\scriptsize )}(1+q}^{6}{\tiny \bar{q}}^{12}{\tiny -q}^{2}{\tiny \bar{q}}^{4}%
{\tiny -q}^{8}{\tiny \bar{q}}^{16}{\tiny )}}{2}
\end{array}
\right) \,. 
\]
Evaluating the $M$-matrix at $M(q(i\pi s_{n}),\bar{q}(i\pi s_{n}))$ leads to 
\begin{eqnarray}
M_{ij}(q(i\pi ),\bar{q}(i\pi ))\! &=&\!\!\frac{2i\sqrt{3}(1+2\cosh \theta
_{H})}{\sinh (\theta _{h}+\theta _{H})\sinh (2\theta _{h}+2\theta _{H})}%
\,\,y_{i}(1)y_{j}(1)\,\, \\
M_{ij}(q(5i\pi ),\bar{q}(5i\pi ))\! &=&\!\!\!\frac{-2i\sqrt{3}(1+2\cosh
(5\theta _{H}))}{\sinh (5\theta _{h}+5\theta _{H})\sinh (10\theta
_{h}+10\theta _{H})}\,\,y_{i}(2)y_{j}(2)\,,\,\,\,\,\,\,\,\,
\end{eqnarray}
which confirms equation (\ref{propo}) including also the precise factor of
proportionality.

\subsubsection{$S_{ij}\left( \theta \right) $ from $D_{4}^{(3)}$}

Instead of using the data from $G_{2}^{(1)}$, we can also employ the
properties of $D_{4}^{(3)}$. As indicated in the Dynkin diagram, we choose
the values of the bi-colouration to be $c_{1}=-1$ and $c_{2}=c_{3}=c_{4}=1$.
Our conventions for the incidence matrix $I$, the action of $\hat{\tau}$ on
the simple roots and the action of the automorphism $\omega $ on the simple
roots are 
\begin{equation}
I=\left( 
\begin{array}{llll}
0 & 1 & 1 & 1 \\ 
1 & 0 & 0 & 0 \\ 
1 & 0 & 0 & 0 \\ 
1 & 0 & 0 & 0
\end{array}
\right) ,\quad \quad \hat{\tau}(\vec{\alpha})=\left( 
\begin{array}{l}
q^{2}\alpha _{1} \\ 
q^{2}\alpha _{2} \\ 
\alpha _{3} \\ 
\alpha _{4}
\end{array}
\right) ,\quad \quad \omega (\vec{\alpha})=\left( 
\begin{array}{l}
\alpha _{1} \\ 
\alpha _{4} \\ 
\alpha _{2} \\ 
\alpha _{3}
\end{array}
\right) \,.  \label{inc}
\end{equation}
The lengths of the orbits are $l_{1}=1$, $l_{2}=l_{3}=l_{4}=3$ and the
q-deformed twisted Coxeter element reads therefore $\hat{\sigma}_{q}=\omega
^{-1}\hat{\sigma}_{1}^{q}\hat{\tau}\hat{\sigma}_{2}^{q}$. Successive actions
of this element on the representatives of $\Omega _{i}^{\omega }$ are
reported in table 2.\bigskip

\begin{center}
\begin{tabular}{|c||c|c|}
\hline\hline
$\hat{\sigma}_{q}^{x}$ & $\hat{\alpha}_{1}=-\hat{\gamma}_{1}^{+}$ & $\hat{%
\alpha}_{2}=\hat{\gamma}_{2}^{+}$ \\ \hline\hline
$1$ & $*;*;2;*$ & $-2;*;2;*$ \\ \hline
$2$ & $2;*;*;2$ & $-2;*;4;2$ \\ \hline
$3$ & $2;2;4;*$ & $-2,4;2;4;4$ \\ \hline
$4$ & $*;*;*;4$ & $-4;4;6;4$ \\ \hline
$5$ & $4;4;*;*$ & $-4;4;*;6$ \\ \hline
$6$ & $-6;*;*;*$ & $-*;6;*;*$ \\ \hline
$7$ & $-*;*;8;*$ & $8;*;8;*$ \\ \hline
$8$ & $-8;*;*;8$ & $8;*;10;8$ \\ \hline
$9$ & $-8;8;10;*$ & $8,10;8;10;10$ \\ \hline
$10$ & $-*;*;*;10$ & $10;10;12;10$ \\ \hline
$11$ & $-10;10;*;*$ & $10;10;*;12$ \\ \hline
$12$ & $12;*;*;*$ & $*;12;*;*$ \\ \hline\hline
\end{tabular}
\end{center}

\noindent {\small Table 2: The orbits $\hat{\Omega}_{i}^{q}$ created by the
action of $\hat{\sigma}_{q}^{x}$ on $\gamma _{i}$} \medskip

\noindent For the generating functions (\ref{detpower2}) we obtain 
\begin{eqnarray}
\sum\limits_{x}\nu _{11}\left( x,2y+1\right) q^{x} &=&-q(\,\hat{\lambda}%
_{1}\cdot (\hat{\sigma}_{q})^{y}\hat{\gamma}_{1})/2 \\
\,\,\sum\limits_{x}\nu _{12}\left( x,2y\right) q^{x} &=&-(\,\hat{\lambda}%
_{2}\cdot (\hat{\sigma}_{q})^{y}\hat{\gamma}_{1})/2 \\
\sum\limits_{x}\nu _{22}\left( x,2y-1\right) q^{x} &=&-q^{-1}(\,\hat{\lambda}%
_{2}\cdot (\hat{\sigma}_{q})^{y}\hat{\gamma}_{2})/2
\end{eqnarray}
which yield the scattering matrices (\ref{11}), (\ref{22}) and (\ref{33})
with the help of table 2.

The two non-equivalent solutions to (\ref{FUt}) corresponding to (\ref{b2}),
(\ref{b4}) and (\ref{b5}) read 
\begin{equation}
q^{2}\hat{\gamma}_{1}^{+}+\hat{\sigma}_{q}\hat{\gamma}_{1}^{+}=\hat{\sigma}%
_{q}\hat{\gamma}_{2}^{+},\qquad \hat{\sigma}_{q}\hat{\gamma}_{1}^{+}+q^{2}%
\hat{\gamma}_{1}^{+}=\hat{\sigma}_{q}\hat{\gamma}_{2}^{+},  \label{77}
\end{equation}
\begin{equation}
q\hat{\sigma}_{q}^{-1}\hat{\gamma}_{1}^{+}+q^{-3}\hat{\sigma}_{q}^{3}\hat{%
\gamma}_{1}^{+}=q^{-1}\hat{\sigma}_{q}\hat{\gamma}_{1}^{+},\qquad q^{-3}\hat{%
\sigma}_{q}^{3}\hat{\gamma}_{1}^{+}+\,q\hat{\sigma}_{q}^{-1}\hat{\gamma}%
_{1}^{+}=q^{-1}\hat{\sigma}_{q}\hat{\gamma}_{1}^{+},  \label{78}
\end{equation}
\begin{equation}
q^{-2}\hat{\sigma}_{q}^{6}\hat{\gamma}_{2}^{+}+q^{2}\hat{\sigma}_{q}^{2}\hat{%
\gamma}_{2}^{+}=\hat{\sigma}_{q}^{4}\hat{\gamma}_{2}^{+},\qquad q^{2}\hat{%
\sigma}_{q}^{2}\hat{\gamma}_{2}^{+}+q^{-2}\hat{\sigma}_{q}^{6}\hat{\gamma}%
_{2}^{+}=\hat{\sigma}_{q}^{4}\hat{\gamma}_{2}^{+},  \label{79}
\end{equation}
respectively. These relations may be obtained either from (\ref{b2}), (\ref
{b4}) and (\ref{b5}) together with the relation which relates the fusing
angles to the solution of the fusing rules in terms of the q-deformed
twisted Coxeter element (\ref{eta11}) or alternatively they may be read off
directly from table 2. Exploiting the relationship between the different
versions of the fusing rules (\ref{eta11}), we may also obtain (\ref{77}), (%
\ref{78}) and (\ref{79}) from (\ref{67}), (\ref{68}) and (\ref{69}).

\subsection{$(F_{4}^{(1)},E_{6}^{(2)})$}

\unitlength=0.780000pt 
\begin{picture}(370.00,107.58)(-100.00,0.00)
\put(70.00,38.00){\line(-1,2){7.00}}
\put(70.00,38.00){\line(-1,-2){7.00}}
\qbezier(211.00,28.00)(285.00,-28.00)(360.00,28.00)
\qbezier(251.00,28.00)(285.00,7.01)(320.00,28.00)
\put(365.00,52.00){\makebox(0.00,0.00){$\alpha_6$}}
\put(325.00,52.00){\makebox(0.00,0.00){$\alpha_5$}}
\put(296.25,52.17){\makebox(0.00,0.00){$\hat{\alpha}_3$}}
\put(285.00,93.00){\makebox(0.00,0.00){$\hat{\alpha}_4$}}
\put(245.00,52.00){\makebox(0.00,0.00){$\hat{\alpha}_2$}}
\put(206.00,52.00){\makebox(0.00,0.00){$\hat{\alpha}_1$}}
\put(125.00,53.00){\makebox(0.00,0.00){$\alpha_4$}}
\put(85.83,53.00){\makebox(0.00,0.00){$\alpha_3$}}
\put(45.00,53.00){\makebox(0.00,0.00){$\alpha_2$}}
\put(5.00,53.00){\makebox(0.00,0.00){$\alpha_1$}}
\put(330.00,38.00){\line(1,0){30.00}}
\put(285.33,72.67){\line(0,-1){31.00}}
\put(290.00,38.00){\line(1,0){30.00}}
\put(250.00,38.00){\line(1,0){30.00}}
\put(210.00,38.00){\line(1,0){30.00}}
\put(45.33,33.00){\line(1,0){40.33}}
\put(45.33,43.00){\line(1,0){40.33}}
\put(90.00,38.00){\line(1,0){30.00}}
\put(10.00,38.00){\line(1,0){30.00}}
\put(285.00,78.00){\circle{10.00}}
\put(245.00,38.00){\circle{10.00}}
\put(325.00,38.00){\circle{10.00}}
\put(365.00,38.00){\circle*{10.00}}
\put(285.00,38.00){\circle*{10.00}}
\put(45.00,38.00){\circle{10.00}}
\put(125.00,38.00){\circle{10.00}}
\put(5.00,38.00){\circle*{10.00}}
\put(85.00,38.00){\circle*{10.00}}
\put(205.00,38.00){\circle*{10.00}}
\end{picture}

\noindent The S-matrices of the theory read \cite{PD} 
\begin{eqnarray*}
S_{11}(\theta ) &=&\left\{ 1,1_{2};5,7_{2};7,9_{2};11,15_{2}\right\}
_{\theta } \\
S_{12}(\theta ) &=&\left\{
2,3_{2};4,5_{2};6,7_{2};6,9_{2};8,11_{2};10,13_{2}\right\} _{\theta } \\
S_{13}(\theta ) &=&\left\{ 3,4_{2};5,6_{2};7,10_{2};9,12_{2}\right\}
_{\theta } \\
S_{14}(\theta ) &=&\left\{ 4,5_{2};8,11_{2}\right\} _{\theta } \\
S_{22}(\theta )
&=&%
\{1,1_{2};3,3_{2};3,5_{2};5,5_{2};5,7_{2}^{2};7,9_{2}^{2};7,11_{2};9,11_{2};9,13_{2};11,15_{2}\}_{\theta }
\\
S_{23}(\theta )
&=&\{2,2_{2};4,4_{2};4,6_{2};6,8_{2}^{2};8,10_{2};8,12_{2};10,14_{2}\}_{%
\theta } \\
S_{24}(\theta ) &=&\left\{ 3,3_{2};5,7_{2};7,9_{2};9,13_{2}\right\} _{\theta
} \\
S_{33}(\theta ) &=&\left\{
1,1;3,3_{2};5,7;5,7_{2};7,9_{2};7,11;9,13_{2};11,17\right\} _{\theta } \\
S_{34}(\theta ) &=&\left\{ 2,2;4,6;6,8_{2};8,12;10,16\right\} _{\theta } \\
S_{44}(\theta ) &=&\{1,1;5,7;7,11;11,17\}_{\theta }\,\,.
\end{eqnarray*}
We have $h=12$ and $H=18$ for the Coxeter numbers. We will not report here
all boostrap identities, but we state the combined bootstrap identities (\ref
{Ravqq}) 
\begin{eqnarray}
S_{1l}(\theta +\theta _{h}+2\theta _{H})S_{1l}(\theta -\theta _{h}-2\theta
_{H}) &=&S_{l2}(\theta ) \\
S_{2l}(\theta +\theta _{h}+2\theta _{H})S_{2l}(\theta -\theta _{h}-2\theta
_{H}) &=&S_{l1}(\theta )S_{l3}(\theta -\theta _{H})S_{l3}(\theta +\theta
_{H})\,\,\,\,\,\,\,\,\,\,  \label{f4boot} \\
S_{3l}(\theta +\theta _{h}+\theta _{H})S_{3l}(\theta -\theta _{h}-\theta
_{H}) &=&S_{l2}(\theta )S_{l4}(\theta ) \\
S_{4l}(\theta +\theta _{h}+\theta _{H})S_{4l}(\theta -\theta _{h}-\theta
_{H}) &=&S_{l3}(\theta )
\end{eqnarray}
for $l=1,2,3,4$. Once again there occurs one equation which is more involved
than the usual bootstrap which we depict in figure 5. \bigskip

\unitlength=0.780000pt 
\begin{picture}(299.67,150.00)(-58.00,0.00)
\put(281.33,95.00){\makebox(0.00,0.00){$3$}}
\put(245.67,75.00){\makebox(0.00,0.00){$4$}}
\put(234.67,75.00){\makebox(0.00,0.00){$1$}}
\put(76.33,74.67){\makebox(0.00,0.00){$4$}}
\put(248.67,130.00){\makebox(0.00,0.00){$3$}}
\put(197.67,94.67){\makebox(0.00,0.00){$1$}}
\put(259.67,0.00){\makebox(0.00,0.00){$2$}}
\put(219.67,0.00){\makebox(0.00,0.00){$2$}}
\put(170.00,109.33){\makebox(0.00,0.00){$l$}}
\put(125.00,130.00){\makebox(0.00,0.00){$3$}}
\put(79.00,129.67){\makebox(0.00,0.00){$3$}}
\put(35.33,129.67){\makebox(0.00,0.00){$1$}}
\put(63.33,74.67){\makebox(0.00,0.00){$1$}}
\put(0.00,29.67){\makebox(0.00,0.00){$l$}}
\put(89.67,0.00){\makebox(0.00,0.00){$2$}}
\put(49.67,0.00){\makebox(0.00,0.00){$2$}}
\put(140.67,86.00){\line(1,0){20.00}}
\put(140.67,90.00){\line(1,0){20.00}}
\put(179.67,110.00){\line(6,1){120.00}}
\put(259.67,60.00){\line(0,-1){50.00}}
\put(239.67,110.00){\line(0,1){40.00}}
\put(259.67,60.00){\line(1,3){30.00}}
\put(239.67,110.00){\line(2,-5){20.00}}
\put(219.67,60.00){\line(2,5){20.00}}
\put(219.67,60.00){\line(0,-1){50.00}}
\put(189.67,150.00){\line(1,-3){30.00}}
\put(9.67,30.00){\line(6,1){120.00}}
\put(89.67,60.00){\line(0,-1){50.00}}
\put(49.67,60.00){\line(0,-1){50.00}}
\put(69.67,150.00){\line(0,-1){40.00}}
\put(89.67,60.00){\line(1,3){30.00}}
\put(69.67,110.00){\line(2,-5){20.00}}
\put(49.67,60.00){\line(2,5){20.00}}
\put(19.67,150.00){\line(1,-3){30.00}}
\end{picture}

\bigskip

\noindent {\small Figure 5: $(F_{4}^{(1)},E_{6}^{(2)})$-combined bootstrap
identities \ (\ref{f4boot}).}

\noindent Reading off the fusing angles from the bootstrap equations we
obtain the mass ratios from (\ref{massr})

\begin{eqnarray}
\frac{m_{1}}{m_{2}} &=&\frac{\sinh (\theta _{h}+2\theta _{H})}{\sinh
(10\theta _{h}+14\theta _{H})}\qquad \qquad \frac{m_{1}}{m_{3}}=\frac{\sinh
(3\theta _{h}+5\theta _{H})}{\sinh (7\theta _{h}+10\theta _{H})} \\
\frac{m_{1}}{m_{4}} &=&\frac{\sinh (3\theta _{h}+5\theta _{H})}{\sinh
(2\theta _{h}+3\theta _{H})}\qquad \quad \qquad \frac{m_{2}}{m_{3}}=\frac{%
\sinh (9\theta _{h}+15\theta _{H})}{\sinh (2\theta _{h}+2\theta _{H})} \\
\frac{m_{2}}{m_{4}} &=&\frac{\sinh (9\theta _{h}+15\theta _{H})}{\sinh
(\theta _{h}+\theta _{H})}\qquad \qquad \,\,\,\frac{m_{3}}{m_{4}}=\frac{%
\sinh (2\theta _{h}+2\theta _{H})}{\sinh (\theta _{h}+\theta _{H})}.
\label{mrf4_end}
\end{eqnarray}
As in the previous case these formulae can be re-constructed from the
twisted as well as the untwisted Lie algebra.

\subsubsection{$S_{ij}\left( \theta \right) $ from $F_{4}^{(1)}$}

According to our conventions the q-deformed Coxeter element reads $\sigma
_{q}=\sigma _{1}^{q}\sigma _{3}^{q}\tau \sigma _{2}^{q}\sigma _{4}^{q}\tau $%
. The result of successive actions of this element on the simple roots is
reported in table 3.

{\footnotesize 
\begin{tabular}{|c||c|c|c|c|}
\hline\hline
$\sigma _{q}^{x}$ & $\alpha _{1}=-\gamma _{1}$ & $\alpha _{3}=-\gamma _{3}$
& $\alpha _{2}=\gamma _{2}$ & $\alpha _{4}=\gamma _{4}$ \\ \hline\hline
$1$ & $*;4;3,5;*$ & $3;3;2,4;2$ & $-4;4;3,5;*$ & $*;*;2;2$ \\ \hline
$2$ & $6;6;5,7;5,7$ & $5;5,7;6^{2},8;6$ & $-6;6,8;5,7^{2},9;5,7$ & $5;5;6;*$
\\ \hline
$3$ & $8;8,10;9,11;*$ & $9;9^{2};8,10^{2};8,10$ & $%
-8,10;8,10^{2};9^{2},11^{2};9,11$ & $*;9;8,10;8$ \\ \hline
$4$ & $*;12;11,13;11,13$ & $11;11,13;12,14;12$ & $%
-12;12^{2},14;11,13^{2},15;11,13$ & $11;11;12;12$ \\ \hline
$5$ & $14;14;*;*$ & $*;15;16;16$ & $-14;14,16;15,17;15,17$ & $*;15;16;*$ \\ 
\hline
$6$ & $-18;*;*;*$ & $*;*;18;*$ & $-*;18;*;*$ & $*;*;*;18$ \\ \hline
$7$ & $-*;22;21,23;*$ & $21;21;20,22;20$ & $22;22;21,23;*$ & $*;*;20;20$ \\ 
\hline
$8$ & $-24;24;23,25;23,25$ & $23;23,25;24^{2},26;24$ & $%
24;24,26;23,25^{2},27;23,25$ & $23;23;24;*$ \\ \hline
$9$ & $-26;26,28;27,29;*$ & $27;27^{2};26,28^{2};26,28$ & $%
26,28;26,28^{2};27^{2},29^{2};27,29$ & $*;27;26,28;26$ \\ \hline
$10$ & $-*;30;29,31;29,31$ & $29;29,31;30,32;30$ & $%
30;30^{2},32;29,31^{2},33;29,31$ & $29;29;30;30$ \\ \hline
$11$ & $-32;32;*;*$ & $*;33;34;34$ & $32;32,34;33,35;33,35$ & $*;33;34;*$ \\ 
\hline
$12$ & $36;*;*;*$ & $*;*;36;*$ & $*;36;*;*$ & $*;*;*;36$ \\ \hline\hline
\end{tabular}
}

{\small \noindent Table 3: The orbits $\Omega _{i}^{q}$ created by the
action of $\sigma _{q}^{x}$ on $\gamma _{i}.$} \medskip

\noindent By using table 3 we may recover the {\small $%
(F_{4}^{(1)},E_{6}^{(2)})-$}S-matrices with the help of generating functions
(\ref{detpower}). The two non-equivalent solutions of the fusing rule in $%
\Omega _{q}$ are 
\begin{eqnarray*}
\gamma _{l}+q^{-12}\sigma _{q}^{4}\gamma _{l} &=&q^{-6}\sigma _{q}^{2}\gamma
_{l},\;\;\;\sigma _{q}^{-1}\gamma _{l}+q^{12}\sigma _{q}^{-5}\gamma
_{l}=q^{6}\sigma _{q}^{-3}\gamma _{l},\;\;l=1,2,3,4 \\
\gamma _{1}+q^{-4}\sigma _{q}\gamma _{1} &=&q^{-4}\sigma _{q}\gamma
_{2},\;\;\;\sigma _{q}^{-1}\gamma _{1}+q^{4}\sigma _{q}^{-2}\gamma
_{1}=\sigma _{q}^{-1}\gamma _{2}, \\
\gamma _{2}+q^{-14}\sigma _{q}^{5}\gamma _{1} &=&\gamma
_{1},\;\;\;q^{-4}\gamma _{2}+q^{14}\sigma _{q}^{-6}\gamma _{1}=\sigma
_{q}^{-1}\gamma _{1}, \\
\gamma _{4}+q^{-2}\sigma _{q}\gamma _{4} &=&\gamma _{3},\;\;\;q^{-2}\gamma
_{4}+\sigma _{q}^{-1}\gamma _{4}=\sigma _{q}^{-1}\gamma _{3}, \\
\gamma _{4}+q^{-16}\sigma _{q}^{5}\gamma _{3} &=&q^{-16}\sigma
_{q}^{5}\gamma _{4},\;\;\;q^{-2}\gamma _{4}+q^{16}\sigma _{q}^{-6}\gamma
_{3}=q^{14}\sigma _{q}^{-5}\gamma _{4}, \\
\gamma _{1}+q^{-15}\sigma _{q}^{5}\gamma _{3} &=&q^{-11}\sigma
_{q}^{4}\gamma _{4},\;\;\;\sigma _{q}^{-1}\gamma _{1}+q^{15}\sigma
_{q}^{-6}\gamma _{3}=q^{9}\sigma _{q}^{-4}\gamma _{4}, \\
\gamma _{1}+q^{-9}\sigma _{q}^{3}\gamma _{4} &=&q^{-3}\sigma _{q}\gamma
_{3},\;\;\;\sigma _{q}^{-1}\gamma _{1}+q^{7}\sigma _{q}^{-3}\gamma
_{4}=q^{3}\sigma _{q}^{-2}\gamma _{3}, \\
\gamma _{3}+q^{-14}\sigma _{q}^{5}\gamma _{4} &=&q^{-3}\sigma _{q}\gamma
_{1},\;\;\;\sigma _{q}^{-1}\gamma _{3}+q^{12}\sigma _{q}^{-5}\gamma
_{4}=q^{3}\sigma _{q}^{-2}\gamma _{1}, \\
\gamma _{2}+q^{-15}\sigma _{q}^{5}\gamma _{3} &=&q^{-1}\sigma _{q}\gamma
_{4},\;\;\;q^{-4}\gamma _{2}+q^{15}\sigma _{q}^{-6}\gamma _{3}=q^{-1}\sigma
_{q}^{-1}\gamma _{4}, \\
\gamma _{2}+q^{-15}\sigma _{q}^{5}\gamma _{4} &=&q\gamma
_{3},\;\;\;q^{-4}\gamma _{2}+q^{13}\sigma _{q}^{-5}\gamma _{4}=q^{-1}\sigma
_{q}^{-1}\gamma _{3}, \\
\gamma _{3}+q^{-4}\sigma _{q}^{2}\gamma _{4} &=&q^{-3}\sigma _{q}\gamma
_{2},\;\;\;\sigma _{q}^{-1}\gamma _{3}+q^{2}\sigma _{q}^{-2}\gamma
_{4}=q^{-1}\sigma _{q}^{-1}\gamma _{2}, \\
\gamma _{4}+q^{-8}\sigma _{q}^{3}\gamma _{4} &=&q^{-3}\sigma _{q}\gamma
_{1},\;\;\;q^{-2}\gamma _{4}+q^{6}\sigma _{q}^{-3}\gamma _{4}=q^{3}\sigma
_{q}^{-2}\gamma _{1}, \\
\gamma _{4}+q^{-13}\sigma _{q}^{4}\gamma _{1} &=&q^{-10}\sigma
_{q}^{3}\gamma _{4},\;\;\;q^{-2}\gamma _{4}+q^{13}\sigma _{q}^{-5}\gamma
_{1}=q^{8}\sigma _{q}^{-3}\gamma _{4}.
\end{eqnarray*}
Once again we can confirm from these solution the equivalence of the
bootstrap equations and the fusing rules by means of (\ref{eta11}) and also
verify the relation for the mass ratios (\ref{massr}).

\subsubsection{$S_{ij}\left( \theta \right) $ from $\hat{E}_{6}^{(2)}$}

The q-deformed twisted Coxeter element in our conventions reads $\hat{\sigma}%
_{q}=\omega ^{-1}\hat{\sigma}_{1}^{q}\hat{\sigma}_{3}^{q}\hat{\tau}\hat{%
\sigma}_{2}^{q}\hat{\sigma}_{4}^{q}$. We report successive actions of this
element on the representatives of $\Omega _{i}^{\omega }$ in table 4.\bigskip

{\footnotesize 
\begin{tabular}{|c||c|c|c|c|}
\hline\hline
$\hat{\sigma}_{q}^{x}$ & $\alpha _{6}=-\hat{\gamma}_{1}^{+}$ & $\hat{\alpha}%
_{3}=-\hat{\gamma}_{3}^{+}$ & $\hat{\alpha}_{2}=\hat{\gamma}_{2}^{+}$ & $%
\hat{\alpha}_{4}=\hat{\gamma}_{4}^{+}$ \\ \hline\hline
$1$ & $0;*;*;*;*;*$ & $*;*;2;2;2;2$ & $-*;*;2;*;2;2$ & $*;*;2;2;*;*$ \\ 
\hline
$2$ & $*;*;2;*;2;*$ & $2;2;2;*;4;4$ & $-2;2;2,4;4;4;4$ & $*;*;*;*;4;4$ \\ 
\hline
$3$ & $*;2;2,4;4;4;4$ & $4;4;4^{2};4;4;*$ & $-4;4;4^{2};4;4,6;6$ & $%
4;4;4;*;*;*$ \\ \hline
$4$ & $4;4;4;4;6;6$ & $*;4;4,6;6;6^{2};6$ & $-6;4,6;4,6^{2};6;6^{2};6$ & $%
*;*;6;6;6;*$ \\ \hline
$5$ & $6;6;6;*;6;*$ & $6;6^{2};6^{2};6;8;8$ & $-6;6^{2};6^{2},8;6,8;8^{2};8$
& $*;6;6;*;8;8$ \\ \hline
$6$ & $*;6;6,8;8;8;*$ & $8;8;8;8;8;*$ & $-8;8^{2};8^{2};8;8,10;10$ & $%
8;8;8;8;*;*$ \\ \hline
$7$ & $*;8;8;8;10;10$ & $*;8;8;*;10;*$ & $-10;8,10;8,10;10;10;*$ & $%
*;*;*;*;10;*$ \\ \hline
$8$ & $10;10;*;*;*;*$ & $*;10;10;10;*;*$ & $-*;10;10;10;12;*$ & $%
*;10;10;*;*;*$ \\ \hline
$9$ & $-*;*;*;*;*;12$ & $*;*;12;*;*;*$ & $-*;12;*;*;*;*$ & $*;*;*;12;*;*$ \\ 
\hline
$10$ & $-12;*;*;*;*;*$ & $*;*;14;14;14;14$ & $*;*;14;*;14;14$ & $%
*;*;14;14*;* $ \\ \hline
$11$ & $-*;*;14;*;14;*$ & $14;14;14;*;16;16$ & $14;14;14,16;16;16;16$ & $%
*;*;*;*;16;16$ \\ \hline
$12$ & $-*;14;14,16;16;16;16$ & $16;16;16^{2};16;16;*$ & $%
16;16;16^{2};16;16,18;18$ & $16;16;16;*;*;*$ \\ \hline
$13$ & $-16;16;16;16;18;18$ & $*;16;16,18;18;18^{2};18$ & $%
18;16,18;16,18^{2};18;18^{2};18$ & $*;*;18;18;18;*$ \\ \hline
$14$ & $-18;18;18;*;18;*$ & $18;18^{2};18^{2};18;20;20$ & $%
18;18^{2};18^{2},20;18,20;20^{2};20$ & $*;18;18;*;20;20$ \\ \hline
$15$ & $-*;18;18,20;20;20;*$ & $20;20;20;20;20;*$ & $%
20;20^{2};20^{2};20;20,22;22$ & $20;20;20;20;*;*$ \\ \hline
$16$ & $-*;20;20;20;22;22$ & $*;20;20;*;22;*$ & $22;20,22;20,22;22;22;*$ & $%
*;*;*;*;22;*$ \\ \hline
$17$ & $-22;22;*;*;*;*$ & $*;22;22;22;*;*$ & $*;22;22;22;24;*$ & $%
*;22;22;*;*;*$ \\ \hline
$18$ & $*;*;*;*;*;24$ & $*;*;24;*;*;*$ & $*;24;*;*;*;*$ & $*;*;*;24;*;*$ \\ 
\hline\hline
\end{tabular}
}

\noindent \noindent {\small Table 4: The orbits $\hat{\Omega}_{i}^{q}$
created by the action of $\hat{\sigma}_{q}^{x}$ on $\gamma _{i}$.} \medskip

\noindent Using the orbits $\hat{\Omega}_{i}^{q}$ listed in table 4 we
recover with help of the generating functions (\ref{detpower2}) the {\small $%
(F_{4}^{(1)},E_{6}^{(2)})$-}S-matrices. The two non-equivalent solutions to
the fusing rule in $\hat{\Omega}_{q}$ read 
\begin{eqnarray*}
\hat{\gamma}_{l}^{+}+q^{-8}\hat{\sigma}_{q}^{6}\hat{\gamma}_{l}^{+} &=&q^{-4}%
\hat{\sigma}_{q}^{3}\hat{\gamma}_{l}^{+},\;\;\;q^{2}\hat{\sigma}_{q}^{2}\hat{%
\gamma}_{l}^{+}+q^{10}\hat{\sigma}_{q}^{-4}\hat{\gamma}_{l}^{+}=q^{6}\hat{%
\sigma}_{q}^{-1}\hat{\gamma}_{l}^{+},\;\;l=1,2,3,4 \\
\hat{\gamma}_{1}^{+}+q^{-2}\hat{\sigma}_{q}^{2}\hat{\gamma}_{1}^{+} &=&q^{-2}%
\hat{\sigma}_{q}\hat{\gamma}_{2}^{+},\;\;\;q^{2}\hat{\sigma}_{q}^{2}\hat{%
\gamma}_{1}^{+}+q^{4}\hat{\gamma}_{1}^{+}=q^{2}\hat{\sigma}_{q}\hat{\gamma}%
_{2}^{+}, \\
\hat{\gamma}_{2}^{+}+q^{-10}\hat{\sigma}_{q}^{8}\hat{\gamma}_{1}^{+} &=&\hat{%
\sigma}_{q}\hat{\gamma}_{1}^{+},\;\;\;\hat{\sigma}_{q}^{2}\hat{\gamma}%
_{2}^{+}+q^{12}\hat{\sigma}_{q}^{-6}\hat{\gamma}_{1}^{+}=q^{2}\hat{\sigma}%
_{q}\hat{\gamma}_{1}^{+}, \\
\hat{\gamma}_{4}^{+}+q^{-2}\hat{\sigma}_{q}\hat{\gamma}_{4}^{+} &=&\hat{%
\gamma}_{3}^{+},\;\;\;\hat{\sigma}_{q}^{2}\hat{\gamma}_{4}^{+}+q^{2}\hat{%
\sigma}_{q}\hat{\gamma}_{4}^{+}=q^{2}\hat{\sigma}_{q}\hat{\gamma}_{3}^{+}, \\
\hat{\gamma}_{4}^{+}+q^{-10}\hat{\sigma}_{q}^{8}\hat{\gamma}_{3}^{+}
&=&q^{-10}\hat{\sigma}_{q}^{8}\hat{\gamma}_{4}^{+},\;\;\;\hat{\sigma}_{q}^{2}%
\hat{\gamma}_{4}^{+}+q^{12}\hat{\sigma}_{q}^{-7}\hat{\gamma}_{3}^{+}=q^{10}%
\hat{\sigma}_{q}^{-6}\hat{\gamma}_{4}^{+}, \\
\hat{\gamma}_{1}^{+}+q^{-10}\hat{\sigma}_{q}^{7}\hat{\gamma}_{3}^{+}
&=&q^{-8}\hat{\sigma}_{q}^{5}\hat{\gamma}_{4}^{+},\;\;\;q^{2}\hat{\sigma}%
_{q}^{2}\hat{\gamma}_{1}^{+}+q^{12}\hat{\sigma}_{q}^{-6}\hat{\gamma}%
_{3}^{+}=q^{8}\hat{\sigma}_{q}^{-3}\hat{\gamma}_{4}^{+}, \\
\hat{\gamma}_{1}^{+}+q^{-6}\hat{\sigma}_{q}^{4}\hat{\gamma}_{4}^{+} &=&q^{-2}%
\hat{\sigma}_{q}\hat{\gamma}_{3}^{+},\;\;\;q^{2}\hat{\sigma}_{q}^{2}\hat{%
\gamma}_{1}^{+}+q^{6}\hat{\sigma}_{q}^{-2}\hat{\gamma}_{4}^{+}=q^{4}\hat{%
\gamma}_{3}^{+}, \\
\hat{\gamma}_{3}^{+}+q^{-10}\hat{\sigma}_{q}^{7}\hat{\gamma}_{4}^{+}
&=&q^{-2}\hat{\sigma}_{q}^{2}\hat{\gamma}_{1}^{+},\;\;\;q^{2}\hat{\sigma}_{q}%
\hat{\gamma}_{3}^{+}+q^{10}\hat{\sigma}_{q}^{-5}\hat{\gamma}_{4}^{+}=q^{4}%
\hat{\gamma}_{1}^{+}, \\
\hat{\gamma}_{2}^{+}+q^{-10}\hat{\sigma}_{q}^{8}\hat{\gamma}_{3}^{+}
&=&q^{-2}\hat{\sigma}_{q}\hat{\gamma}_{4}^{+},\;\;\;\hat{\sigma}_{q}^{2}\hat{%
\gamma}_{2}^{+}+q^{12}\hat{\sigma}_{q}^{-7}\hat{\gamma}_{3}^{+}=q^{2}\hat{%
\sigma}_{q}\hat{\gamma}_{4}^{+}, \\
\hat{\gamma}_{2}^{+}+q^{-10}\hat{\sigma}_{q}^{8}\hat{\gamma}_{4}^{+} &=&\hat{%
\gamma}_{3}^{+},\;\;\;\hat{\sigma}_{q}^{2}\hat{\gamma}_{2}^{+}+q^{10}\hat{%
\sigma}_{q}^{-6}\hat{\gamma}_{4}^{+}=q^{2}\hat{\sigma}_{q}\hat{\gamma}%
_{3}^{+}, \\
\hat{\gamma}_{3}^{+}+q^{-4}\hat{\sigma}_{q}^{2}\hat{\gamma}_{4}^{+} &=&q^{-2}%
\hat{\sigma}_{q}\hat{\gamma}_{2}^{+},\;\;\;q^{2}\hat{\sigma}_{q}\hat{\gamma}%
_{3}^{+}+q^{4}\hat{\gamma}_{4}^{+}=q^{2}\hat{\sigma}_{q}\hat{\gamma}_{2}^{+},
\\
\hat{\gamma}_{4}^{+}+q^{-6}\hat{\sigma}_{q}^{4}\hat{\gamma}_{4}^{+} &=&q^{-2}%
\hat{\sigma}_{q}^{2}\hat{\gamma}_{1}^{+},\;\;\;\hat{\sigma}_{q}^{2}\hat{%
\gamma}_{4}^{+}+q^{6}\hat{\sigma}_{q}^{-2}\hat{\gamma}_{4}^{+}=q^{4}\hat{%
\gamma}_{1}^{+}, \\
\hat{\gamma}_{4}^{+}+q^{-8}\hat{\sigma}_{q}^{7}\hat{\gamma}_{1}^{+} &=&q^{-6}%
\hat{\sigma}_{q}^{5}\hat{\gamma}_{4}^{+},\;\;\;\hat{\sigma}_{q}^{2}\hat{%
\gamma}_{4}^{+}+q^{10}\hat{\sigma}_{q}^{-5}\hat{\gamma}_{1}^{+}=q^{6}\hat{%
\sigma}_{q}^{-3}\hat{\gamma}_{4}^{+}.
\end{eqnarray*}

Again we confirm from these solution the equivalence between the bootstrap
equations and the fusing rules by means of (\ref{eta11}) and also verify the
relation for the mass ratios (\ref{massr}).

\subsection{$(C_{2}^{(1)},D_{3}^{(2)})$}

\unitlength=0.780000pt 
\begin{picture}(437.92,149.59)(-40.00,50.00)
\qbezier(420.00,170.00)(440.00,150.00)(420.00,130.00)
\put(57.00,125.00){\makebox(0.00,0.00){$c_1=-1$ if $N$ even}}
\put(437.50,104.17){\makebox(0.00,0.00){$\alpha_{N+1}$}}
\put(437.92,184.17){\makebox(0.00,0.00){$\hat{\alpha}_N$}}
\put(411.17,149.59){\makebox(0.00,0.00){$\hat{\alpha}_{N-1}$}}
\put(347.50,165.00){\makebox(0.00,0.00){$\hat{\alpha}_{N-2}$}}
\put(296.00,165.00){\makebox(0.00,0.00){$\hat{\alpha}_2$}}
\put(255.00,165.00){\makebox(0.00,0.00){$\hat{\alpha}_1$}}
\put(177.92,165.42){\makebox(0.00,0.00){$\alpha_N$}}
\put(137.92,165.42){\makebox(0.00,0.00){$\alpha_{N-1}$}}
\put(97.50,165.00){\makebox(0.00,0.00){$\alpha_{N-2}$}}
\put(45.00,165.00){\makebox(0.00,0.00){$\alpha_2$}}
\put(5.00,165.00){\makebox(0.00,0.00){$\alpha_1$}}
\put(389.00,146.33){\line(1,-1){22.67}}
\put(411.67,176.67){\line(-1,-1){23.00}}
\put(350.00,150.00){\line(1,0){30.00}}
\put(330.00,150.00){\line(1,0){10.00}}
\put(300.00,150.00){\line(1,0){10.00}}
\put(260.00,150.00){\line(1,0){30.00}}
\put(150.00,150.00){\line(1,-2){7.00}}
\put(150.00,150.00){\line(1,2){7.00}}
\put(135.33,145.00){\line(1,0){39.67}}
\put(135.33,155.00){\line(1,0){39.67}}
\put(100.00,150.00){\line(1,0){30.00}}
\put(80.00,150.00){\line(1,0){10.00}}
\put(50.00,150.00){\line(1,0){10.00}}
\put(10.00,150.00){\line(1,0){30.00}}
\put(415.00,120.00){\circle{10.00}}
\put(415.00,180.00){\circle{10.00}}
\put(385.00,150.00){\circle*{10.00}}
\put(345.00,150.00){\circle{10.00}}
\put(135.00,150.00){\circle*{10.00}}
\put(175.00,150.00){\circle{10.00}}
\put(255.00,150.00){\circle{10.00}}
\put(295.00,150.00){\circle{10.00}}
\put(95.00,150.00){\circle{10.00}}
\put(45.00,150.00){\circle{10.00}}
\put(5.00,150.00){\circle{10.00}}
\end{picture}

\noindent The S-matrices are given as 
\[
S_{11}(\theta )=\{1,1;3,5\}_{\theta }\quad S_{12}(\theta
)=\{2,2_{2}\}_{\theta }\quad S_{22}(\theta )=\{1,1_{2};3,3_{2}\}_{\theta }. 
\]
We have $h=4$ and $H=6$ for the Coxeter numbers. The combined bootstrap
equations (\ref{Ravqq}) yield 
\begin{eqnarray}
S_{1l}(\theta +\theta _{h}+\theta _{H})S_{1l}(\theta -\theta _{h}-\theta
_{H}) &=&S_{l2}(\theta ) \\
S_{2l}(\theta +\theta _{h}+2\theta _{H})S_{2l}(\theta -\theta _{h}-2\theta
_{H}) &=&S_{l1}(\theta -\theta _{H})S_{l1}(\theta +\theta _{H})  \label{xxx}
\end{eqnarray}
for $l=1,2$.

\unitlength=0.780000pt 
\begin{picture}(260.00,100.00)(-65.00,0.00)
\put(210.00,40.00){\makebox(0.00,0.00){$1$}}
\put(60.00,60.00){\makebox(0.00,0.00){$1$}}
\put(250.00,100.00){\makebox(0.00,0.00){$1$}}
\put(170.00,100.00){\makebox(0.00,0.00){$1$}}
\put(100.00,100.00){\makebox(0.00,0.00){$1$}}
\put(20.00,100.00){\makebox(0.00,0.00){$1$}}
\put(250.00,0.00){\makebox(0.00,0.00){$2$}}
\put(170.00,0.00){\makebox(0.00,0.00){$2$}}
\put(100.00,0.00){\makebox(0.00,0.00){$2$}}
\put(20.00,0.00){\makebox(0.00,0.00){$2$}}
\put(160.00,60.00){\makebox(0.00,0.00){$l$}}
\put(0.00,20.00){\makebox(0.00,0.00){$l$}}
\put(170.00,60.00){\line(5,1){90.00}}
\put(230.00,50.00){\line(1,-2){20.00}}
\put(230.00,50.00){\line(1,2){20.00}}
\put(190.00,50.00){\line(1,0){40.00}}
\put(190.00,50.00){\line(-1,-2){20.00}}
\put(170.00,90.00){\line(1,-2){20.00}}
\put(126.00,50.00){\line(1,0){20.00}}
\put(126.00,54.00){\line(1,0){20.00}}
\put(10.00,20.00){\line(5,1){100.00}}
\put(80.00,50.00){\line(1,-2){20.00}}
\put(80.00,50.00){\line(1,2){20.00}}
\put(40.00,50.00){\line(1,0){40.00}}
\put(40.00,50.00){\line(-1,2){20.00}}
\put(40.00,50.00){\line(-1,-2){20.00}}
\end{picture}

\noindent {\small Figure 6: $(C_{2}^{(1)},D_{3}^{(2)})$-combined bootstrap
identities (\ref{xxx}).}

\noindent The mass ratio according to (\ref{massr}) are

\begin{equation}
\frac{m_{1}}{m_{2}}=\frac{\sinh (\theta _{h}+\theta _{H})}{\sinh (2\theta
_{h}+4\theta _{H})}.
\end{equation}

\subsubsection{$S_{ij}(\theta )$ from $C_{2}^{(1)}$:}

The result of successive actions of the q-deformed Coxeter element on the
simple roots is reported in table 5.

\begin{center}
\begin{tabular}{|c||c|c|}
\hline\hline
$\sigma _{q}^{x}$ & $\alpha _{1}=-\gamma _{1}$ & $\alpha _{2}=\gamma _{2}$
\\ \hline\hline
$1$ & $4;3$ & $-3,5;4$ \\ \hline
$2$ & $-6;*$ & $-*;6$ \\ \hline
$3$ & $-10;9$ & $9,11;10$ \\ \hline
$4$ & $12;*$ & $*;12$ \\ \hline\hline
\end{tabular}
\end{center}

\noindent {\small Table 5: The orbits }$\ \Omega _{i}^{q}${\small \ created
by the action of }$\sigma _{q}^{x}$ {\small on }$\gamma _{i}$

\noindent The two non-equivalent solutions to the fusing rule in $\Omega
_{q} $ read 
\begin{eqnarray*}
\gamma _{1}+q^{-2}\sigma _{q}\gamma _{1} &=&q^{-3}\sigma _{q}\gamma
_{2},\;\;\;\sigma _{q}^{-1}\gamma _{1}+q^{2}\sigma _{q}^{-2}\gamma
_{1}=q^{-1}\sigma _{q}^{-1}\gamma _{2}, \\
\gamma _{1}+q^{-7}\sigma _{q}^{2}\gamma _{2} &=&q^{-4}\sigma _{q}\gamma
_{1},\;\;\;\sigma _{q}^{-1}\gamma _{1}+q^{3}\sigma _{q}^{-2}\gamma
_{2}=q^{4}\sigma _{q}^{-2}\gamma _{1}.
\end{eqnarray*}

\subsubsection{$S_{ij}(\theta )$ from $\hat{D}_{3}^{(2)}$:}

The result of successive actions of the q-deformed twisted Coxeter element
on the simple roots is reported in table 6.

\begin{center}
\begin{tabular}{|c||c|c|}
\hline\hline
$\hat{\sigma}_{q}^{x}$ & $\hat{\alpha}_{1}=-\hat{\gamma}_{1}^{+}$ & $\hat{%
\alpha}_{2}=\hat{\gamma}_{2}^{+}$ \\ \hline\hline
$1$ & $*;*;2$ & $-2;*;2$ \\ \hline
$2$ & $2;2;*$ & $-2;2;4$ \\ \hline
$3$ & $-4;*;*$ & $-*;4;*$ \\ \hline
$4$ & $-*;*;6$ & $6;*;6$ \\ \hline
$5$ & $-6;6;*$ & $6;6;8$ \\ \hline
$6$ & $8;*;*$ & $*;8;*$ \\ \hline\hline
\end{tabular}
\end{center}

\noindent {\small Table 6: The orbits }$\ \hat{\Omega}_{i}^{q}${\small \
created by the action of }$\hat{\sigma}_{q}^{x}$ {\small on }$\hat{\gamma}%
_{i}^{+}$

\noindent The two non-equivalent solutions to the fusing rule in $\hat{\Omega%
}_{q}$ read 
\begin{eqnarray*}
\hat{\gamma}_{1}^{+}+q^{-2}\hat{\sigma}_{q}\hat{\gamma}_{1}^{+} &=&q^{-2}%
\hat{\sigma}_{q}\hat{\gamma}_{2}^{+},\;\;\;q^{2}\hat{\sigma}_{q}\hat{\gamma}%
_{1}^{+}+q^{4}\hat{\gamma}_{1}^{+}=q^{2}\hat{\sigma}_{q}\hat{\gamma}_{2}^{+},
\\
\hat{\gamma}_{1}^{+}+q^{-4}\hat{\sigma}_{q}^{3}\hat{\gamma}_{2}^{+} &=&q^{-2}%
\hat{\sigma}_{q}^{2}\hat{\gamma}_{1},\;\;\;q^{2}\hat{\sigma}_{q}\hat{\gamma}%
_{1}^{+}+q^{4}\hat{\sigma}_{q}^{-1}\hat{\gamma}_{2}^{+}=q^{4}\hat{\sigma}%
_{q}^{-1}\hat{\gamma}_{1}^{+}.
\end{eqnarray*}

\subsection{$(C_{3}^{(1)},D_{4}^{(2)})$}

The S-matrices are 
\begin{eqnarray*}
S_{11}(\theta ) &=&\{1,1;5,7\}_{\theta }\quad S_{12}(\theta
)=\{2,2;4,6\}_{\theta }\quad S_{33}(\theta
)=\{1,1_{2};3,3_{2};5,5_{2}\}_{\theta } \\
S_{22}(\theta ) &=&\{1,1;3,3_{2};5,7\}_{\theta }\quad S_{23}(\theta
)=\{2,2_{2};4,4_{2}\}_{\theta }\quad S_{13}(\theta )=\{3,3_{2}\}_{\theta }.
\end{eqnarray*}
We have $h=6$ and $H=8$ for the Coxeter numbers. The combined bootstrap
identities read 
\begin{eqnarray}
S_{1l}(\theta +\theta _{h}+\theta _{H})S_{1l}(\theta -\theta _{h}-\theta
_{H}) &=&S_{l2}(\theta ) \\
S_{2l}(\theta +\theta _{h}+\theta _{H})S_{2l}(\theta -\theta _{h}-\theta
_{H}) &=&S_{l1}(\theta )S_{l3}(\theta ) \\
S_{3l}(\theta +\theta _{h}+2\theta _{H})S_{3l}(\theta -\theta _{h}-2\theta
_{H}) &=&S_{l2}(\theta -\theta _{H})S_{l2}(\theta +\theta _{H}).
\end{eqnarray}

\noindent The mass ratios turn out to be 
\begin{equation}
\frac{m_{1}}{m_{2}}=\frac{\sinh (\theta _{h}+\theta _{H})}{\sinh (4\theta
_{h}+6\theta _{H})}\quad \frac{m_{1}}{m_{3}}=\frac{\sinh (\theta _{h}+\theta
_{H})}{\sinh (3\theta _{h}+5\theta _{H})}\quad \frac{m_{2}}{m_{3}}=\frac{%
\sinh (2\theta _{h}+2\theta _{H})}{\sinh (3\theta _{h}+5\theta _{H})}.
\end{equation}

\subsubsection{$S_{ij}(\theta )$ from $C_{3}^{(1)}$}

The result of successive actions of the q-deformed Coxeter element on the
simple roots is reported in table 7.

\begin{center}
\begin{tabular}{|c||c|c|c|}
\hline\hline
$\sigma _{q}^{x}$ & $\alpha _{1}=\gamma _{1}$ & $\alpha _{3}=\gamma _{3}$ & $%
\alpha _{2}=-\gamma _{2}$ \\ \hline\hline
$1$ & $-2;2;*$ & $*;3,5;4$ & $2;2,4;3$ \\ \hline
$2$ & $-*;6;5$ & $5,7;5,7;6$ & $6;6;5$ \\ \hline
$3$ & $-8;*;*$ & $*;*;8$ & $-*;8;*$ \\ \hline
$4$ & $10;10;*$ & $*;11,13;12$ & $-10;10,12;11$ \\ \hline
$5$ & $*;14;13$ & $13,15;13,15;14$ & $-14;14;13$ \\ \hline
$6$ & $16;*;*$ & $*;*;16$ & $*;16;*$ \\ \hline\hline
\end{tabular}
\end{center}

\noindent {\small Table 7 : The orbits }$\ \Omega _{i}^{q}${\small \ created
by the action of }$\sigma _{q}^{x}$ {\small on }$\gamma _{i}$

\noindent The solutions of the fusing rule in $\Omega ^{q}$ are{\small \ } 
\begin{eqnarray*}
\gamma _{1}+q^{-2}\sigma _{q}\gamma _{1} &=&\gamma _{2},\;\;\;q^{-2}\gamma
_{1}+\sigma _{q}^{-1}\gamma _{1}=\sigma _{q}^{-1}\gamma _{2}, \\
\gamma _{1}+q^{-6}\sigma _{q}^{2}\gamma _{2} &=&q^{-6}\sigma _{q}^{2}\gamma
_{1},\;\;\;q^{-2}\gamma _{1}+q^{6}\sigma _{q}^{-3}\gamma _{2}=q^{4}\sigma
_{q}^{-2}\gamma _{1}, \\
\gamma _{1}+q^{-2}\sigma _{q}\gamma _{2} &=&q^{-3}\sigma _{q}\gamma
_{3},\;\;\;q^{-2}\gamma _{1}+q^{2}\sigma _{q}^{-2}\gamma _{2}=q^{-1}\sigma
_{q}^{-1}\gamma _{3}, \\
\gamma _{1}+q^{-7}\sigma _{q}^{2}\gamma _{3} &=&q^{-4}\sigma _{q}\gamma
_{2},\;\;\;q^{-2}\gamma _{1}+q^{3}\sigma _{q}^{-2}\gamma _{3}=q^{4}\sigma
_{q}^{-2}\gamma _{2}, \\
\gamma _{2}+q^{-9}\sigma _{q}^{3}\gamma _{3} &=&q^{-6}\sigma _{q}^{2}\gamma
_{1},\;\;\;\sigma _{q}^{-1}\gamma _{2}+q^{5}\sigma _{q}^{-3}\gamma
_{3}=q^{4}\sigma _{q}^{-2}\gamma _{1}.
\end{eqnarray*}

\subsubsection{$S_{ij}(\theta )$ from $\hat{D}_{4}^{(2)}$}

The result of successive actions of the q-deformed twisted Coxeter element
on the simple roots is reported in table 8.

\begin{center}
\begin{tabular}{|c||c|c|c|}
\hline\hline
$\hat{\sigma}_{q}^{x}$ & $\hat{\alpha}_{1}=\hat{\gamma}_{1}^{+}$ & $\hat{%
\alpha}_{3}=\hat{\gamma}_{3}^{+}$ & $\hat{\alpha}_{2}=-\hat{\gamma}_{2}^{+}$
\\ \hline\hline
$1$ & $-2;2;*;*$ & $*;2;*;2$ & $2;2;*;2$ \\ \hline
$2$ & $-*;*;*;4$ & $4;2,4;2;4$ & $*;2;2;4$ \\ \hline
$3$ & $-*;4;4;*$ & $4;4;4;6$ & $4;4;4;*$ \\ \hline
$4$ & $-6;*;*;*$ & $*;*;6;*$ & $-*;6;*;*$ \\ \hline
$5$ & $8;8;*;*$ & $*;8;*;8$ & $-8;8;*;8$ \\ \hline
$6$ & $*;*;*;10$ & $10;8,10;8;10$ & $-*;8;8;10$ \\ \hline
$7$ & $*;10;10;*$ & $10;10;10;12$ & $-10;10;10;*$ \\ \hline
$8$ & $12;*;*;*$ & $*;*;12;*$ & $*;12;*;*$ \\ \hline\hline
\end{tabular}
\end{center}

\noindent {\small Table 8 : The orbits }$\ \hat{\Omega}_{i}^{q}${\small \
created by the action of }$\hat{\sigma}_{q}^{x}$ {\small on }$\hat{\gamma}%
_{i}^{+}$

\noindent The solutions of the fusing rule in $\hat{\Omega}^{q}$ are{\small %
\ } 
\begin{eqnarray*}
\hat{\gamma}_{1}^{+}+q^{-2}\hat{\sigma}_{q}\hat{\gamma}_{1}^{+} &=&\hat{%
\gamma}_{2}^{+},\;\;\;\hat{\sigma}_{q}^{2}\hat{\gamma}_{1}^{+}+q^{2}\hat{%
\sigma}_{q}\hat{\gamma}_{1}^{+}=q^{2}\hat{\sigma}_{q}\hat{\gamma}_{2}^{+}, \\
\hat{\gamma}_{1}^{+}+q^{-4}\hat{\sigma}_{q}^{3}\hat{\gamma}_{2}^{+} &=&q^{-4}%
\hat{\sigma}_{q}^{3}\hat{\gamma}_{1}^{+},\;\;\;\hat{\sigma}_{q}^{2}\hat{%
\gamma}_{1}^{+}+q^{6}\hat{\sigma}_{q}^{-2}\hat{\gamma}_{2}^{+}=q^{4}\hat{%
\sigma}_{q}^{-1}\hat{\gamma}_{1}^{+}, \\
\hat{\gamma}_{1}^{+}+q^{-2}\hat{\sigma}_{q}\hat{\gamma}_{2}^{+} &=&q^{-2}%
\hat{\sigma}_{q}\hat{\gamma}_{3}^{+},\;\;\;\hat{\sigma}_{q}^{2}\hat{\gamma}%
_{1}^{+}+q^{4}\hat{\gamma}_{2}^{+}=q^{2}\hat{\sigma}_{q}\hat{\gamma}_{3}^{+},
\\
\hat{\gamma}_{1}^{+}+q^{-4}\hat{\sigma}_{q}^{3}\hat{\gamma}_{3}^{+} &=&q^{-2}%
\hat{\sigma}_{q}^{2}\hat{\gamma}_{2}^{+},\;\;\;\hat{\sigma}_{q}^{2}\hat{%
\gamma}_{1}^{+}+q^{4}\hat{\sigma}_{q}^{-1}\hat{\gamma}_{3}^{+}=q^{4}\hat{%
\sigma}_{q}^{-1}\hat{\gamma}_{2}^{+}, \\
\hat{\gamma}_{2}^{+}+q^{-6}\hat{\sigma}_{q}^{4}\hat{\gamma}_{3}^{+} &=&q^{-4}%
\hat{\sigma}_{q}^{3}\hat{\gamma}_{1}^{+},\;\;\;q^{2}\hat{\sigma}_{q}\hat{%
\gamma}_{2}^{+}+q^{6}\hat{\sigma}_{q}^{-2}\hat{\gamma}_{3}^{+}=q^{4}\hat{%
\sigma}_{q}^{-1}\hat{\gamma}_{1}^{+}.
\end{eqnarray*}

\subsection{$(B_{2}^{(1)},A_{3}^{(2)})$}

{
\unitlength=0.780000pt 
\begin{picture}(430.00,78.12)(0.00,0.00)
\qbezier(270.00,38.12)(330.00,18.13)(380.00,38.12)
\qbezier(230.00,38.12)(326.25,-13.12)(420.00,38.12)
\put(390.00,48.12){\line(1,0){30.00}}
\put(270.00,48.12){\line(1,0){10.00}}
\put(230.00,48.12){\line(1,0){30.00}}
\put(430.25,62.71){\makebox(0.00,0.00){$\alpha_{2N-1}$}}
\put(386.67,62.71){\makebox(0.00,0.00){$\alpha_{2N-2}$}}
\put(326.00,63.12){\makebox(0.00,0.00){$\hat{\alpha}_N$}}
\put(266.25,62.29){\makebox(0.00,0.00){$\hat{\alpha}_2$}}
\put(226.00,62.12){\makebox(0.00,0.00){$\hat{\alpha}_1$}}
\put(175.42,63.54){\makebox(0.00,0.00){$\alpha_N$}}
\put(136.00,63.12){\makebox(0.00,0.00){$\alpha_{N-1}$}}
\put(95.00,63.12){\makebox(0.00,0.00){$\alpha_{N-2}$}}
\put(46.00,63.12){\makebox(0.00,0.00){$\alpha_2$}}
\put(5.00,64.12){\makebox(0.00,0.00){$\alpha_1$}}
\put(278.75,0.00){\makebox(0.00,0.00){$c_1=-1$ if $N$ odd}}
\put(57.50,26.88){\makebox(0.00,0.00){$c_1=-1$ if $N$ odd}}
\put(160.00,48.12){\line(-1,-2){7.00}}
\put(160.00,48.12){\line(-1,2){7.00}}
\put(135.00,42.79){\line(1,0){40.33}}
\put(135.00,53.12){\line(1,0){39.67}}
\put(330.00,48.12){\line(1,0){10.00}}
\put(310.00,48.12){\line(1,0){10.00}}
\put(100.00,48.12){\line(1,0){30.00}}
\put(80.00,48.12){\line(1,0){10.00}}
\put(50.00,48.12){\line(1,0){10.00}}
\put(10.00,48.12){\line(1,0){30.00}}
\put(425.00,48.12){\circle{10.00}}
\put(385.00,48.12){\circle{10.00}}
\put(325.00,48.12){\circle*{10.00}}
\put(265.00,48.12){\circle{10.00}}
\put(225.00,48.12){\circle{10.00}}
\put(175.00,48.12){\circle*{10.00}}
\put(135.00,48.12){\circle{10.00}}
\put(95.00,48.12){\circle*{10.00}}
\put(45.00,48.12){\circle{10.00}}
\put(5.00,48.12){\circle{10.00}}
\end{picture}}

\noindent The S-matrices read 
\[
S_{11}(\theta )=\{1,1_{2};3,3_{2}\}_{\theta }\quad S_{12}(\theta
)=\{2,2_{2}\}_{\theta }\quad S_{22}(\theta )=\{1,1;3,5\}_{\theta }. 
\]
We have $h=4$ and $H=6$ for the Coxeter numbers. The combined bootstrap
identities are

\begin{eqnarray}
S_{1l}(\theta +\theta _{h}+2\theta _{H})S_{1l}(\theta -\theta _{h}-2\theta
_{H}) &=&S_{l2}(\theta -\theta _{H})S_{l2}(\theta +\theta _{H}) \\
S_{2l}(\theta +\theta _{h}+\theta _{H})S_{2l}(\theta -\theta _{h}-\theta
_{H}) &=&S_{l1}(\theta ).
\end{eqnarray}
The mass ratio is 
\begin{equation}
\frac{m_{1}}{m_{2}}=\frac{\sinh (2\theta _{h}+4\theta _{H})}{\sinh (\theta
_{h}+\theta _{H})}.
\end{equation}

\subsubsection{$S_{ij}(\theta )$ from $B_{2}^{(1)}$}

The result of successive actions of the q-deformed Coxeter element on the
simple roots is reported in table 9.

\begin{center}
$
\begin{tabular}{|c||c|c|}
\hline\hline
$\sigma _{q}^{x}$ & $\alpha _{1}=\gamma _{1}$ & $\alpha _{2}=-\gamma _{2}$
\\ \hline\hline
$1$ & $-4;3,5$ & $3;4$ \\ \hline
$2$ & $-6;*$ & $-*;6$ \\ \hline
$3$ & $10;9,11$ & $-9;10$ \\ \hline
$4$ & $12;*$ & $*;12$ \\ \hline\hline
\end{tabular}
$
\end{center}

\noindent {\small Table 9: The orbits }$\ \Omega _{i}^{q}${\small \ created
by the action of }$\sigma _{q}^{x}$ {\small on }$\gamma _{i}$

\noindent Solutions of the fusing rule in $\Omega ^{q}${\small \ } 
\begin{eqnarray*}
\gamma _{1}+q^{-3}\sigma _{q}\gamma _{2} &=&q\gamma _{2},\;\;\;q^{-4}\gamma
_{1}+q^{3}\sigma _{q}^{-2}\gamma _{2}=q^{-1}\sigma _{q}^{-1}\gamma _{2}, \\
\gamma _{2}+q^{-2}\sigma _{q}\gamma _{2} &=&q^{-3}\sigma _{q}\gamma
_{1},\;\;\;\sigma _{q}^{-1}\gamma _{2}+q^{2}\sigma _{q}^{-2}\gamma
_{2}=q^{-1}\sigma _{q}^{-1}\gamma _{1}.
\end{eqnarray*}

\subsubsection{$S_{ij}(\theta )$ from $\hat{A}_{3}^{(2)}$}

The result of successive actions of the q-deformed twisted Coxeter element
on the simple roots is reported in table 10.

\begin{center}
\begin{tabular}{|c||c|c|}
\hline\hline
$\hat{\sigma}_{q}^{x}$ & $\hat{\alpha}_{1}=\hat{\gamma}_{1}^{+}$ & $\hat{%
\alpha}_{2}=-\hat{\gamma}_{2}^{+}$ \\ \hline\hline
$1$ & $-*;2;2$ & $*;*;2$ \\ \hline
$2$ & $-2;2;4$ & $2;2;*$ \\ \hline
$3$ & $-4;*;*$ & $-*;4;*$ \\ \hline
$4$ & $*;6;6$ & $-*;*;6$ \\ \hline
$5$ & $6;6;8$ & $-6;6;*$ \\ \hline
$6$ & $8;*;*$ & $*;8;*$ \\ \hline\hline
\end{tabular}
\end{center}

\noindent {\small Table 10 : The orbits }$\ \hat{\Omega}_{i}^{q}${\small \
created by the action of }$\hat{\sigma}_{q}^{x}$ {\small on }$\hat{\gamma}%
_{i}^{+}$

\noindent The solutions to the fusing rule in $\hat{\Omega}^{q}${\small \ } 
\begin{eqnarray*}
\hat{\gamma}_{1}^{+}+q^{-2}\hat{\sigma}_{q}^{2}\hat{\gamma}_{2}^{+} &=&\hat{%
\gamma}_{2}^{+},\;\;\;\hat{\sigma}_{q}^{2}\hat{\gamma}_{1}^{+}+q^{4}\hat{%
\sigma}_{q}^{-1}\hat{\gamma}_{2}^{+}=q^{2}\hat{\sigma}_{q}\hat{\gamma}%
_{2}^{+}, \\
\hat{\gamma}_{2}^{+}+q^{-2}\hat{\sigma}_{q}\hat{\gamma}_{2}^{+} &=&q^{-2}%
\hat{\sigma}_{q}\hat{\gamma}_{1},\;\;\;q^{2}\hat{\sigma}_{q}\hat{\gamma}%
_{2}^{+}+q^{4}\hat{\gamma}_{2}^{+}=q^{2}\hat{\sigma}_{q}\hat{\gamma}_{1}^{+}.
\end{eqnarray*}

\subsection{$(B_{3}^{(1)},A_{5}^{(2)})$}

The S-matrices read 
\begin{eqnarray*}
S_{11}(\theta ) &=&\{1,1_{2};5,7_{2}\}_{\theta }\quad S_{12}(\theta
)=\{2,3_{2};4,5_{2}\}_{\theta }\quad S_{33}(\theta )=\{1,1;3,5;5,9\}_{\theta
} \\
S_{22}(\theta ) &=&\{1,1_{2};3,3_{2};3,5_{2};5,7_{2}\}_{\theta }\quad
S_{23}(\theta )=\{2,2_{2};4,6_{2}\}_{\theta }\quad S_{13}(\theta
)=\{3,4_{2}\}_{\theta }.
\end{eqnarray*}
We have $h=6$ and $H=10$ for the Coxeter numbers. The combined bootstrap
identities read 
\begin{eqnarray}
S_{1l}(\theta +\theta _{h}+2\theta _{H})S_{1l}(\theta -\theta _{h}-2\theta
_{H}) &=&S_{l2}(\theta ) \\
S_{2l}(\theta +\theta _{h}+2\theta _{H})S_{2l}(\theta -\theta _{h}-2\theta
_{H}) &=&S_{l1}(\theta )S_{l3}(\theta -\theta _{H})S_{l3}(\theta +\theta
_{H})\,\,\,\,\,\,\,\,\, \\
S_{3l}(\theta +\theta _{h}+\theta _{H})S_{3l}(\theta -\theta _{h}-\theta
_{H}) &=&S_{l2}(\theta ).
\end{eqnarray}
The mass ratios are 
\begin{equation}
\frac{m_{1}}{m_{2}}=\frac{\sinh (\theta _{h}+2\theta _{H})}{\sinh (4\theta
_{h}+6\theta _{H})}\quad \frac{m_{1}}{m_{3}}=\frac{\sinh (2\theta
_{h}+4\theta _{H})}{\sinh (2\theta _{h}+3\theta _{H})}\quad \frac{m_{2}}{%
m_{3}}=\frac{\sinh (4\theta _{h}+8\theta _{H})}{\sinh (\theta _{h}+\theta
_{H})}.
\end{equation}

\subsubsection{$S_{ij}(\theta )$ from $B_{3}^{(1)}$}

The result of successive actions of the q-deformed Coxeter element on the
simple roots is reported in table 11.

\begin{center}
\begin{tabular}{|c||c|c|c|}
\hline\hline
$\sigma _{q}^{x}$ & $\alpha _{1}=-\gamma _{1}$ & $\alpha _{3}=-\gamma _{3}$
& $\alpha _{2}=\gamma _{2}$ \\ \hline\hline
$1$ & $*;4;3,5$ & $3;3;4$ & $-4;4;3,5$ \\ \hline
$2$ & $6;6;*$ & $*;7;8$ & $-6;6,8;7,9$ \\ \hline
$3$ & $-10;*;*$ & $*;*;10$ & $-*;10;*$ \\ \hline
$4$ & $-*;14;13,15$ & $13;13;14$ & $14;14;13,15$ \\ \hline
$5$ & $-16;16;*$ & $*;17;18$ & $16;16,18;17,19$ \\ \hline
$6$ & $20;*;*$ & $*;*;20$ & $*;20;*$ \\ \hline\hline
\end{tabular}
\end{center}

\noindent {\small Table 11: The orbits }$\ \Omega _{i}^{q}${\small \ created
by the action of }$\sigma _{q}^{x}$ {\small on }$\gamma _{i}$

\noindent The solutions of the fusing rule in $\Omega ^{q}$ are 
\begin{eqnarray*}
\gamma _{1}+q^{-4}\sigma _{q}\gamma _{1} &=&q^{-4}\sigma _{q}\gamma
_{2},\;\;\;\sigma _{q}^{-1}\gamma _{1}+q^{4}\sigma _{q}^{-2}\gamma
_{1}=\sigma _{q}^{-1}\gamma _{2}, \\
\gamma _{1}+q^{-10}\sigma _{q}^{3}\gamma _{2} &=&q^{-6}\sigma _{q}^{2}\gamma
_{1},\;\;\;\sigma _{q}^{-1}\gamma _{1}+q^{6}\sigma _{q}^{-3}\gamma
_{2}=q^{6}\sigma _{q}^{-3}\gamma _{1}, \\
\gamma _{1}+q^{-7}\sigma _{q}^{2}\gamma _{3} &=&q^{-3}\sigma _{q}\gamma
_{3},\;\;\;\sigma _{q}^{-1}\gamma _{1}+q^{7}\sigma _{q}^{-3}\gamma
_{3}=q^{3}\sigma _{q}^{-2}\gamma _{3}, \\
\gamma _{2}+q^{-7}\sigma _{q}^{2}\gamma _{3} &=&q\gamma
_{3},\;\;\;q^{-4}\gamma _{2}+q^{7}\sigma _{q}^{-3}\gamma _{3}=q^{-1}\sigma
_{q}^{-1}\gamma _{3}, \\
\gamma _{3}+q^{-6}\sigma _{q}^{2}\gamma _{3} &=&q^{-3}\sigma _{q}\gamma
_{1},\;\;\;\sigma _{q}^{-1}\gamma _{3}+q^{6}\sigma _{q}^{-3}\gamma
_{3}=q^{3}\sigma _{q}^{-2}\gamma _{1}, \\
\gamma _{3}+q^{-2}\sigma _{q}\gamma _{3} &=&q^{-3}\sigma _{q}\gamma
_{2},\;\;\;\sigma _{q}^{-1}\gamma _{3}+q^{2}\sigma _{q}^{-2}\gamma
_{3}=q^{-1}\sigma _{q}^{-1}\gamma _{2}.
\end{eqnarray*}

\subsubsection{$S_{ij}(\theta )$ from $\hat{A}_{5}^{(2)}$:}

The result of successive actions of the q-deformed twisted Coxeter element
on the simple roots is reported in table 12.

\begin{center}
\begin{tabular}{|c||c|c|c|}
\hline\hline
$\hat{\sigma}_{q}^{x}$ & $\alpha _{5}=-\hat{\gamma}_{1}^{+}$ & $\hat{\alpha}%
_{3}=-\hat{\gamma}_{3}^{+}$ & $\hat{\alpha}_{2}=\hat{\gamma}_{2}^{+}$ \\ 
\hline\hline
$1$ & $0;*;*;*;*$ & $*;*;*;2;2\ $ & $-*;*;2;2;2$ \\ \hline
$2$ & $\ *;*;2;2;*$ & $\ 2;2;2;*;*$ & $-2;2;2;4;4$ \\ \hline
$3$ & $\ *;2;2;4;4$ & $\ *;*;*;4;*$ & $\ -4;4;4;4;*$ \\ \hline
$4$ & $\ 4;4;*;*;*$ & $\ *;4;4;*;*$ & $\ -*;4;4;6;*$ \\ \hline
$5$ & $\ \ -*;*;*;*;6$ & $\ *;*;6;*;*$ & $\ -*;6;*;*;*$ \\ \hline
$6$ & $-6;*;*;*;*$ & $\ *;*;*;8;8$ & $\ *;*;8;8;8$ \\ \hline
$7$ & $\ -*;*;8;8;*$ & $\ 8;8;8;*;*$ & $\ 8;8;8;10;10$ \\ \hline
$8$ & $\ -*;8;8;10;10$ & $*;*;*;10;*$ & $\ 10;10;10;10;*$ \\ \hline
$9$ & $-10;10;*;*;*$ & $*;10;10;*;*$ & $*;10;10;12;*$ \\ \hline
$10$ & $*;*;*;*;12$ & $*;*;12;*;*$ & $*;12;*;*,*$ \\ \hline\hline
\end{tabular}
\end{center}

\noindent {\small Table 12 : The orbits }$\ \hat{\Omega}_{i}^{q}${\small \
created by the action of }$\hat{\sigma}_{q}^{x}$ {\small on }$\hat{\gamma}%
_{i}^{+}$

\noindent The solutions of the fusing rule in $\ \hat{\Omega}^{q}$%
\begin{eqnarray*}
\hat{\gamma}_{1}^{+}+q^{-2}\hat{\sigma}_{q}^{2}\hat{\gamma}_{1}^{+} &=&q^{-2}%
\hat{\sigma}_{q}\hat{\gamma}_{2}^{+},\;\;\;q^{2}\hat{\sigma}_{q}^{2}\hat{%
\gamma}_{1}^{+}+q^{4}\hat{\gamma}_{1}^{+}=q^{2}\hat{\sigma}_{q}\hat{\gamma}%
_{2}^{+}, \\
\hat{\gamma}_{1}^{+}+q^{-6}\hat{\sigma}_{q}^{4}\hat{\gamma}_{2}^{+} &=&q^{-4}%
\hat{\sigma}_{q}^{3}\hat{\gamma}_{1}^{+},\;\;\;q^{2}\hat{\sigma}_{q}^{2}\hat{%
\gamma}_{1}^{+}+q^{6}\hat{\sigma}_{q}^{-2}\hat{\gamma}_{2}^{+}=q^{6}\hat{%
\sigma}_{q}^{-1}\hat{\gamma}_{1}^{+}, \\
\hat{\gamma}_{1}^{+}+q^{-4}\hat{\sigma}_{q}^{3}\hat{\gamma}_{3}^{+} &=&q^{-2}%
\hat{\sigma}_{q}\hat{\gamma}_{3}^{+},\;\;\;q^{2}\hat{\sigma}_{q}^{2}\hat{%
\gamma}_{1}^{+}+q^{6}\hat{\sigma}_{q}^{-2}\hat{\gamma}_{3}^{+}=q^{4}\hat{%
\gamma}_{3}^{+}, \\
\hat{\gamma}_{2}^{+}+q^{-4}\hat{\sigma}_{q}^{4}\hat{\gamma}_{3}^{+} &=&\hat{%
\gamma}_{3}^{+},\;\;\;\hat{\sigma}_{q}^{2}\hat{\gamma}_{2}^{+}+q^{6}\hat{%
\sigma}_{q}^{-3}\hat{\gamma}_{3}^{+}=q^{2}\hat{\sigma}_{q}\hat{\gamma}%
_{3}^{+}, \\
\hat{\gamma}_{3}^{+}+q^{-4}\hat{\sigma}_{q}^{3}\hat{\gamma}_{3}^{+} &=&q^{-2}%
\hat{\sigma}_{q}^{2}\hat{\gamma}_{1}^{+},\;\;\;q^{2}\hat{\sigma}_{q}\hat{%
\gamma}_{3}^{+}+q^{6}\hat{\sigma}_{q}^{-2}\hat{\gamma}_{3}^{+}=q^{4}\hat{%
\gamma}_{1}^{+}, \\
\hat{\gamma}_{3}^{+}+q^{-2}\hat{\sigma}_{q}\hat{\gamma}_{3}^{+} &=&q^{-2}%
\hat{\sigma}_{q}\hat{\gamma}_{2}^{+},\;\;\;q^{2}\hat{\sigma}_{q}\hat{\gamma}%
_{3}^{+}+q^{4}\hat{\gamma}_{3}^{+}=q^{2}\hat{\sigma}_{q}\hat{\gamma}_{2}^{+}.
\end{eqnarray*}

\section{Conclusion}

We have systematically developed the properties of the q-deformed Coxeter
element and its twisted counterpart. The vanishing of the
three-point-coupling is governed by the so-called fusing rules. They rules
may be formulated either in the orbits $\Omega _{q}$, $\hat{\Omega}_{q}$ or $%
\Omega $ and $\hat{\Omega}$. The precise relation between these alternative
rules is worked out (\ref{eta11}). All of these identities may be proven by
appealing to physical arguments. The scattering matrices of affine Toda
field theories with real coupling constant related to any dual pair of
simple Lie algebras may be expressed in a completely generic way in terms of
combinations of hyperbolic functions whose powers are computed from
generating functions involving either q-deformed Coxeter elements (\ref
{detpower}) or alternatively twisted q-deformed Coxeter elements (\ref
{detpower2}). The q-deformation appears to be vital in the construction
since it achieves that the properties of the two dual algebras are merged
together. It would be interesting to investigate whether it is possible at
all to construct generic formulae solely from non-deformed quantities as it
is possible in the simply laced case. However, it appears to us that the
q-deformation is vital to describe non-simply laced theories. Closely
related to this is the question of how to derive the q-deformed versions of
the fusing rules directly from the non-deformed versions. We have
demonstrated that the proposed scattering matrices fulfill all the
requirements of the generalized bootstrap equations. In particular, we
established the equivalence of the fusing rules and the generalized S-matrix
boostrap equations. Furthermore, we provide a simple criterion which allows
to exclude poles from the participation in the bootstrap.

It is intriguing that the combined bootstrap equation (\ref{Ravqq})
incorporates the information of \emph{all }individual fusing processes.
These equations do in fact not constitute anything new since they may always
be obtained from the individual fusing processes. They correspond to
particular graphs (see figure 5 and 6) of higher order.

The matrix $[K]_{q\bar{q}}$ plays a central role in several ways. The
components of its nullvectors constitute conserved quantities, e.g. the
particle masses. We show how these quantities are related to the fusing
rules. The properties of the matrix $[K]_{q\bar{q}}$ are further utilized in
order to formulate a matrix $M$ which serves to derive and prove a generic
integral representation for the scattering matrix. The same goal may be
achieved by exploiting the properties of the matrix $[\hat{K}]_{q\bar{q}}$
which is related to the twisted algebra and allows to define the matrix $%
\hat{N}$. We established the equality between these two matrices.

It is interesting to note that the properties of the blocks are reflected by
the polynomial (\ref{M}), such that we can carry out a one-to-one
identification between $\{x,y\}_{\theta }$ and $q^{x}\bar{q}^{y}$. In
addition we can also manipulate them in an identical way if we further
define $q^{-x}\bar{q}^{-y}=-q^{x}\bar{q}^{y}$ in analogy to $%
\{-x,-y\}_{\theta }=\{x,y\}_{\theta }^{-1}$ or choose $q$ and $\bar{q}$ to
be roots of unity. This means we can treat the whole bootstrap properties in
an entirely polynomial fashion.

From the matrix relation $\hat{N}=M$ one deduces immediately the equality $%
\mu (x,y)=\nu (x,y)$. However, it remains a challenge to develop a more
direct Lie algebraic understanding of the equation.

\noindent \textbf{Acknowledgments: } A.F. and C.K. are grateful to the
Deutsche Forschungsgemeinschaft (Sfb288) for partial financial support.
B.J.S. is grateful to the Studienstiftung des deutschen Volkes for financial 
\label{cc}support. We are grateful to P. Dorey, P. Mattsson and R. Weston
for bringing ref. \cite{FR} to our attention.


\begin{thebibliography}{99}
\bibitem{BPZ}  {\small A.A.~Belavin, A.M.~Polyakov and A.B.~Zamolodchikov, 
\emph{Nucl. Phys.} \textbf{B241} (1984) 333. }

\bibitem{SCH}  {\small B.~ Schroer, \emph{``A Trip to Scalingland'', V.
Brazilian Symposium on Theoretical Physics, Rio de Janeiro 1974, }ed. E.
Ferreira.}

\bibitem{Zper}  {\small A.B.~Zamolodchikov, \emph{Int. J. Mod. Phys. } 
\textbf{A1} (1989) 4235.}

\bibitem{PT}  {\small T.~Eguchi and S.-K.~Yang, \emph{Phys. Lett.} \textbf{%
B224} (1989) 373;}

{\small T.J.~Hollowood and P.~Mansfield, \emph{Phys. Lett.} \textbf{B226}
(1989) 73.}

\bibitem{ATFT}  {\small A.V.~ Mikhailov, M.A.~Olshanetsky and
A.M.~Perelomov, \emph{Commun. Math. Phys. } \textbf{79} (1981) 473;}

{\small G.~Wilson, \emph{Ergod. Th. Dyn. Syst}. \textbf{1} (1981) 361;}

{\small D.I.~Olive and N.~Turok, \emph{Nucl. Phys. }\textbf{B257} [FS14]
(1985) 277. }

\bibitem{TodaS}  {\small A.E.~Arinshtein, V.A.~Fateev and
A.B.~Zamolodchikov, \emph{Phys. Lett.} \textbf{B87} (1979) 389;}

{\small H.W.~Braden, E.~Corrigan, P.E.~Dorey and R.~Sasaki, \emph{Nucl.
Phys. } \textbf{B338} (1990) 689;}

{\small C.~Destri and H.J.~De Vega, \emph{Phys. Lett.} \textbf{B233 }(1989)
336, \emph{Nucl. Phys.} \textbf{B358} (1991) 251;}

{\small P.G.O.~Freund, T.R.~Klassen and E.~Melzer, \emph{Phys. Lett.} 
\textbf{B229} (1989) 243;}

{\small P.~Christe and G.~Mussardo,} {\small \emph{Nucl. Phys.} \textbf{B330}
(1990) 465, \emph{Int. J. Mod. Phys.} \textbf{A5} (1990) 4581.}

{\small G.~Mussardo,} {\small \emph{Phys. Rep.} \textbf{218} (1993) 215.}

\bibitem{PD}  {\small P.E.~Dorey, \emph{Nucl. Phys. } \textbf{B358} (1991)
654; \emph{Nucl. Phys. } \textbf{B374} (1992) 741.}

\bibitem{FO}  {\small A.~Fring and D.I.~Olive, \emph{\ Nucl. Phys.} \textbf{%
\ B379} (1992) 429.}

\bibitem{FLO}  {\small M.D.~Freeman, \emph{Phys. Lett.} \textbf{B261} (1991)
57;}

{\small A.~Fring, H.C.~Liao and D.I.~Olive, \emph{Phys. Lett.} \textbf{B266}
(1991) 82.}

\bibitem{G2}  {\small G.W.~Delius, M.T.~Grisaru and D.~Zanon,\emph{\ Nucl.
Phys.} \textbf{B382} (1992) 365;}

{\small H}.{\small G.~Kausch and G.M.T.~Watts,\emph{\ Nucl. Phys. } \textbf{%
B386} (1992) 166;}

{\small G.M.T.~Watts and R.A.~Weston, \emph{Phys. Lett.} \textbf{B289}
(1992) 61.}

\bibitem{nons}  {\small E.~Corrigan, P.E.~Dorey and R.~Sasaki, \emph{Nucl.
Phys. } \textbf{B408} (1993) 579.}

\bibitem{Donons}  {\small P.E.~Dorey, \emph{Phys. Lett.} \textbf{B312}
(1993) 291.}

\bibitem{Khast}  {\small S.P.~Khastgir, \emph{Nucl. Phys.} \textbf{B499}
(1997) 650.}

\bibitem{Oota}  {\small T.~Oota, \emph{\ Nucl. Phys.} \textbf{B504} (1997)
738.}

\bibitem{DT}  {\small P.E.~Dorey, \emph{``Hidden geometrical structures in
integrable models'' in ``Integrable Quantum Field Theories'',} ed.
L.~Bonora, et al (Plenum, 1993). }

\bibitem{OM}  {\small D.I.~Olive and C.~Montonon \emph{\ Phys. Lett. } 
\textbf{B72} (1977) 117.}

\bibitem{CP}  {\small V.~Chari and A.~Pressley \emph{Commun. Math. Phys. } 
\textbf{181} (1996) 265.}

\bibitem{Kac}  {\small V.G.~Kac, \emph{``Infinite Dimensional Lie Algebras''
(CUP, Cambridge,1990).}}

\bibitem{Springer}  {\small T. A.~Springer, \emph{Inv. Math.} \textbf{25}
(1974) 159.}

\bibitem{FF}  {\small M.~Karowski and P.~Weisz,\emph{\ Nucl. Phys.} \textbf{%
B139} (1978) 445.}

\bibitem{TBAZam}  {\small Al.B.~Zamolodchikov, \emph{Nucl. Phys. } \textbf{%
B342} (1990) 695.}

\bibitem{FKS}  {\small A.~Fring, C.~Korff and B.J.~Schulz, \emph{\ Nucl.
Phys.} \textbf{B549} (1999) 579.}

\bibitem{Rava}  {\small F.~Ravanini, A.~Valleriani and R.~Tateo, \emph{Int.
J. Mod. Phys.} \textbf{A8} (1993) 1707.}

\bibitem{TO}  {\small D.I.~Olive and N.~Turok, \emph{Nucl. Phys. }\textbf{%
B215} [FS7] (1983) 470. }

\bibitem{KO}  {\small M.A.C.~Kneipp and D.I.~Olive, \emph{Commun. Math.
Phys. } \textbf{177} (1996) 561.}

\bibitem{FR}  {\small E.~Frenkel and N.~Reshetikhin, ``\emph{Deformations of
W-algebras associated to simple Lie algebras'' q-alg/9708006.}}
\end{thebibliography}
\end{document}